\documentclass[a4,12pt,english]{book}
\usepackage{epsfig,amsmath,subeqnarray,caption2}
 \usepackage{natbib}
 \usepackage{color}
 \usepackage{chapterbib}
\usepackage{fancyhdr}
\usepackage{float}
\usepackage{graphicx}
\DeclareGraphicsExtensions{.pdf}
\graphicspath{{figures/}}

\newcommand{\var}[1]{\ensuremath{#1}}
\newcommand{\vect}[1]{\ensuremath{\mbox{\boldmath$\mathrm{#1}$}}}
\newcommand{\vectdot}[1]{\ensuremath{\mbox{\boldmath$\dot{\mathrm{#1}}$}}}

\newcommand{\df}[2]{\ensuremath{\frac{d #1}{d #2}}}
\newcommand{\ddf}[2]{\ensuremath{\frac{d^2 #1}{d #2^2}}}
\newcommand{\citec}[1]{\citeauthor{#1}~\citeyear{#1}}

\newcommand{\captionfonts}{\small}

\makeatletter
  \long\def\@makecaption#1#2{
  \vskip\abovecaptionskip
  \sbox\@tempboxa{{\captionfonts #1: #2}}
  \ifdim \wd\@tempboxa >\hsize
    {\captionfonts {\bf #1:} #2\par}
  \else
    \hbox to\hsize{\hfil\box\@tempboxa\hfil}
  \fi
  \vskip\belowcaptionskip}
\makeatother

\let\origdoublepage\cleardoublepage 
\newcommand{\clearemptydoublepage}{%
  \clearpage
  {\pagestyle{empty}\origdoublepage}%
}
\let\cleardoublepage\clearemptydoublepage

\newcommand{\fig}[1]{Fig.\ \ref{#1}}
\newcommand{\Fig}[1]{Figure \ref{#1}}

\begin{document}
\ifx\href\undefined\else\hypersetup{linktocpage=true}\fi 

\begin{figure}[h]
\vspace{-4.0cm}\hspace{-4cm}\includegraphics[width=21.6cm,height=27.9cm]{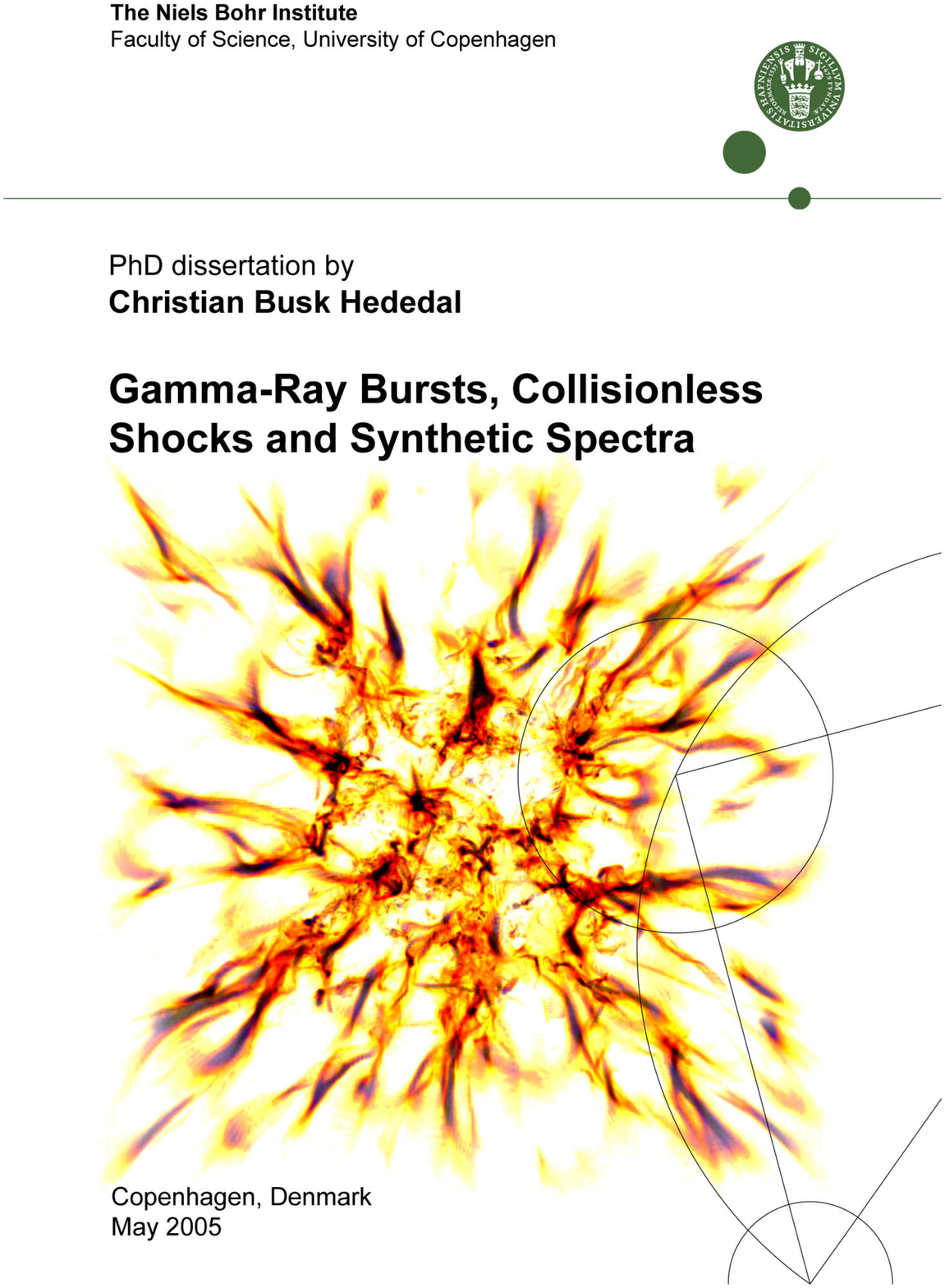}
\end{figure}

 \frontmatter
\addcontentsline{toc}{chapter}{Preface}
\chapter*{Preface}
\thispagestyle{empty}
This thesis is submitted in partial fulfillment of the PhD degree at
the Niels Bohr Insitute in Denmark. The thesis concludes three years
of work, of which two years were dedicated to scientific work and one
year to lecturing and various other duties. The PhD project was
carried out under supervision of Professor \AA{}ke Nordlund, in the
Computational Astrophysics group at the Niels Bohr institute in
Copenhagen, Denmark. As part of the PhD education, I spent 4
months at the National Space Science and Technology Center (NSSTC)
in Huntsville Alabama, USA. Here I collaborated with the gamma-ray
group under Dr. Gerald J. Fishman and in particular with Dr.
Ken-Ichi Nishikawa.

The backbone of the thesis consists of three papers published in ApJL (the
Astrophysical Journal Letters), plus recent work on large-scale 2D
plasma simulations and generation of synthetic spectra that has not
yet been submitted as scientific papers. The three papers are
included as chapters here, with minor extensions compared to the published versions.
Co-author statements are attached at the end of the thesis.

The aim has been to write a thesis that both students and
experienced scientist may benefit from reading. This is a
difficult undertaking and thus you may find some parts either too trivial
or too complex. In the latter case, do not worry: Collisionless
plasma shocks are indeed complicated non-linear systems, and really
a contradiction in terms.

I hope that you will enjoy and appreciate the work I
present in this thesis.

\bigskip
\bigskip
\bigskip
\bigskip
\smallskip
\noindent Christian Hededal\\
Copenhagen, May 2005

\bigskip
\bigskip
\noindent{\footnotesize\begin{tabular}{l}
\hline
\footnotesize The illustration on the cover is a ray-traced image of data, taken directly from our \\
\footnotesize simulations. It shows plasma filaments that are generated by the Weibel two-stream\\
\footnotesize instability, as a relativistic plasma is travelling directly towards the observer.
\end{tabular}}

\addcontentsline{toc}{chapter}{Abstract}\chapter*{Abstract}

In this thesis I present the results of three-dimensional,
self-consistent particle-in-cell simulations of collisionless
shocks.

The radiation from afterglows of gamma-ray bursts is generated in
collisionless plasma shocks between a relativistic outflow and a
quiescent circum-burst medium. The two main ingredients responsible
for the radiation are high-energy, non-thermal electrons
($N(\gamma)d\gamma\propto\gamma^{-p}$) and a strong magnetic field.
Fermi acceleration is normally believed to be responsible for the
acceleration of the electrons. Fermi acceleration has been employed
extensively in Monte Carlo simulations, where it operates in
conjunction with certain assumptions about the scattering of
particles and the structure of the magnetic field. The mechanism
has, however, not been conclusively demonstrated to occur in ab
initio particle simulations and also faces additional problems.
Furthermore, the requirement of a strong magnetic field in the shock
region indicates that the magnetic field is generated in situ in.

In this thesis, I argue that in order to make the right conclusions
about gamma-ray burst and afterglow parameters, it is crucial to
have a firm understanding of collisionless shocks: How are electrons
accelerated, what is the topology of the generated magnetic field,
and how do these two aspects affect the resulting radiation. Thus,
the main goal of the work I present in this thesis has been to
expand our knowledge about the microphysics of collisionless plasma
shocks. To accomplish this, a self-consistent, three-dimensional
particle-in-cell computational code has been utilized. The
simulation tool works from first principles by solving Maxwell's
equations for the electromagnetic field, consistently coupled to the
momentum equation for the charged particles.

In the experiments, I study the collision of two plasma populations
travelling at relativistic velocities. When the plasma populations
are initially unmagnetized or weakly magnetized, the Weibel
two-stream instability generates a magnetic field in the shock ramp
with strengths up to percents of equipartition with the plasma ions.
The nature of the magnetic field is predominantly transverse to the
plasma flow. The transverse coalescence scale is comparable to the
ion skin depth whereas the parallel scale extends up to hundreds of
ion skin depths. A spatial Fourier decomposition of the magnetic
field shows that the structures follow a power-law distribution with
negative slope.

The experiments also reveal a new non-thermal electron acceleration
mechanism, which differs substantially from Fermi acceleration.
Acceleration of electrons is directly related to the formation of
ion current channels by the non-linear Weibel two-stream
instability. This links particle acceleration closely together with
magnetic field generation in collisionless shocks. The resulting
electron spectrum consists of a thermal component and a non-thermal
component at high energies. In an experiment with a bulk Lorentz
factor of $\Gamma=15$, the non-thermal tail has the power-law index
$p=2.7$.

Finally, I have developed a tool that generates synthetic radiation
spectra from the experiments. The radiation is calculated directly,
by tracing a large number of electrons in the generated magnetic
field, and thus continuous the line of work from first principles.
Numerous tests show that the radiation tool successfully reproduces
synchrotron, bremsstrahlung and undulator radiation from small-angle
deflections. I then go on to perform a parameter study of
three-dimensional jitter radiation. Using the tool on
particle-in-cell experiments of collisionless shocks I find that the
radiation spectrum from particles in a randomized magnetic field is
not fully consistent with radiation from particles in
shock-generated magnetic field, even when the two have the same
statistical properties.

In experiments where magnetic field generation and particle
acceleration arise as natural consequences of the Weibel two-stream
instability, the resulting radiation spectrum is consistent with
observations. In simulations of a collisionless shock that
propagates with bulk Lorentz factor $\Gamma=15$ through the
interstellar medium, I find that the radiation spectrum peaks around
$10^{12}$Hz. Above this frequency, the spectrum follows a power-low
 $F\propto\nu^{-\beta}$ with $\beta=0.7$. Below the peak frequency, the
spectrum follows a power law  $F\propto\nu^{\alpha}$ with
$\alpha\simeq2/3$. This is steeper than the standard synchrotron
value of $1/3$ and more compatible with observations.

I conclude that strong magnetic field generation
($\epsilon_B\sim0.01-0.1$), non-thermal particle acceleration, and
the emission of radiation with properties that are consistent with
GRB afterglow observations are all unavoidable consequences of the
collision between two relativistic plasma shells.

\tableofcontents
\addcontentsline{toc}{chapter}{List of Figures}\listoffigures
\addcontentsline{toc}{chapter}{Acknowledgement}
\chapter*{Acknowledgement}
I would like to extend my gratitude to:

\begin{description}
\item Professor \AA{}ke Nordlund for being my supervisor and a great source of inspiration and motivation to
students. \vspace{-.2cm}
\item The University of Copenhagen and the Niels Bohr Institute for three years of financial support of my PhD studies. \vspace{-.2cm}
\item My girlfriend Pernille - Science and family can be hard to mix but you have made it very
easy for me. \vspace{-.2cm}
\item My family for supporting me throughout my entire life - even when I signed up for becoming an Astronaut. \vspace{-.2cm}
\item Dr. Ken-Ichi Nishikawa and Rosa Sanchez for helping me in every way during my stay in Huntsville
and for inviting me into your home. \vspace{-.2cm}
\item Dr. Gerald J. Fishman for inviting me to the National Space Science and Technology Center at
Marshal Space Flight Center, Huntsville and for letting me drive the
ATV. That was great fun!  And thanks to the rest of the BATSE team
for your friendly attitude towards a young student, far away from
home. \vspace{-.2cm}
\item Jacob Trier Frederiksen and Troels Haugb\o{}lle for collaboration, discussions and much
more. \vspace{-.2cm}
\item Dr. Therese Moretto and Dr. Michael Hesse for triggering a profound
interest in plasma physics. I also thank you for supporting me and
taking great care of me during my 6 months at NASA Goddard Space
Flight Center. And thanks to Dr. Michael Hesse for generously
providing the original particle-in-cell code. \vspace{-.2cm}
\item The Danish Center for Scientific Computing for access to almost unlimited computing time and resources. \vspace{-.2cm}
\item The students and staff at the NBI for coffee and fruitful
discussions. \vspace{-.2cm}
\item "Drengene" for beers and fun.
\end{description}

\addcontentsline{toc}{chapter}{Units and Conventions}
\chapter*{Units and Conventions}
The field of gamma-ray bursts is multi-disciplinary in the sense
that it covers the electromagnetic spectrum from radio to gamma-ray,
length scales from microphysical to cosmological and velocities from
newtonian to highly relativistic. It thus connects many branches of
physics and astrophysics, with all their traditions with regard to
formalisms, phenomenologies and units. Some people are almost
religious about units. I'm not, as long as the use of units is
consistent. I have chosen to use SI-units throughout the thesis. To
those readers that are mostly familiar with the use of Gaussian cgs
units in electrodynamics, the following conversion table may be of
help:

\begin{center}
\begin{tabular}{c c}
SI & cgs\\ \hline
Replace & by\\
\hline
$\epsilon_0$ & $1/(4\pi)$\\
$\mu_0$ & $4\pi/c^2$\\
$\vect{B}$ & $\vect{B}/c$\\
 \hline
\end{tabular}
\end{center}

Throughout the thesis I use $\gamma$ for Lorentz of individual
particles and $\Gamma$ for bulk flows, e.g.\ jet Lorentz factor. On
many occasions, I use the terms "parallel" and "perpendicular". If
nothing else is stated, this refers a direction relative to
propagation direction of the shock.

Since the units in PIC codes are often re-scaled (and our code is no
exception) it is normal to measure time and length in terms of the
typical plasma parameters that govern the physical processes in a
plasma. Time is often given in units of one over the electron plasma
frequency $\omega_{pe}=(n_e q^2/(m_e \epsilon_0))^{1/2}$, where
$n_e$ is the electron plasma density, $q$ is the unit charge, $m_e$
is the electron mass, and $\epsilon_0$ is the electric vacuum
permittivity. Lengths are typically given in units of skin depths
$\delta=c/\omega_{pe}$ where c is the speed of light in vacuum.
 If not otherwise stated, the units in the figures should be
taken as arbitrary units. However, on some occasions I make an
effort to scale the results into real-space values. In this case,
the units are clearly marked on the axis.

\setcounter{chapter}{0}

 \mainmatter
 \pagestyle{fancy}
\chapter{Introduction}\label{sec:intro}
\section{The ``early'' history of gamma-ray bursts}
\begin{quote}
{\it Gamma-ray bursts are the largest explosions we know of in the
universe after the Big Bang}

 {\it If you have seen one gamma-ray
burst... you have seen one gamma-ray burst!}
\end{quote}
These two statements are among the most repeated in the history of
gamma-ray bursts. They very nicely cover what the fuss about
gamma-ray bursts is all about. With their extreme brightness,
gamma-ray bursts have the potential of being used as lighthouses,
shining from the far and dark ages of the universe. At the same
time, gamma-ray bursts apparently come in a large number of colors
and flavors, and thus their origin is still a puzzle after many
years in scientific focus.

The existence of gamma-ray bursts (GRBs) came into human knowledge
at the end of the 1960s. As a product of the nuclear arms race, the
USA had launched a series of gamma-ray nuclear blast wave detectors
--- the VELA satellites. The aim of the VELA satellites was to make
sure that the USSR did not break the nuclear test ban treaties with
secret nuclear tests in the upper atmosphere and in space. Testing
the satellites, \cite{bib:Klebesadel} found unidentified spikes in
the data. It was easily realized that the signals were not from
nuclear tests. Using the timing offset from several satellites, it
was possible to make a crude triangulation and place the origin of
the gamma-rays outside our Solar system. Distributed randomly in the
sky, the positions indicated that the bursts were either from an
extended galactic halo, or were an extra-galactic phenomenon.

In 1991, the {\it Compton Gamma Ray Observatory} (CGRO) was
launched, carrying the {\it Burst and Transient Source Experiment}
(BATSE). Several thousand detections during the 1990s, isotropically
distributed over the sky \citep{bib:meegan}, still left two
possibilities  for the origin of the GRBs. Either they were
cosmological \citep{bib:paczynski} or they were from a very extended
spherical halo of the Milky way \citep{bib:lamb}. Which of the two
was determined in 1997, following the launch of the BeppoSAX
satellite. BeppoSAX was able to rapidly locate the position of GRB
970228 (970228 for 1997, February 28). This triggered a
multiwavelength campaign resulting in the detection of an x-ray
afterglow \citep{bib:costa}, and an optical afterglow
\citep{bib:paradijs} within the error-box. Using the Hubble Space
Telescope, the origin of the burst was found to lie in a galaxy at
cosmological distance \citep{bib:sahu}.

The spectrum of the host galaxy to GRB 970228 showed prominent
emission lines \citep{bib:bloom}. From these, the redshift of the
galaxy was found to be $z=0.695$. With the given flux and enormous
distance, the energy of the GRB was estimated to be as large as
$10^{47}$ J ($10^{54}$ erg), many thousand times stronger than any
previous known type of astrophysical explosion. Moreover, the
duration over which the energy was released was apparently only a
few seconds. The great variability in the burst suggested that this
huge amount of energy was released within a volume a few 1000 km in
radius. \cite{bib:ruderman} had realized that such a scenario would
inevitably lead to a {\it compactness problem}. The fireball would
be extremely optically thick with respect to pair production and
this would not allow us to observe the high-energy, non-thermal
photon tail. This problem was solved by suggesting that the emitting
surface was ejected with a highly relativistic bulk velocity
\citep{bib:goodman,bib:paczynski1986,bib:piran}. Fireball expansion
speeds comparable to the speed of light where later observationally
confirmed from changes in radio scintillation of GRB 970508.
\cite{bib:goodman1997} and \cite{bib:waxman1998} estimated the size
of the fireball to $10^{15}\mathrm{m}$, only four weeks after the
trigger.

\begin{figure}[htb]
\begin{center}
\epsfig{figure=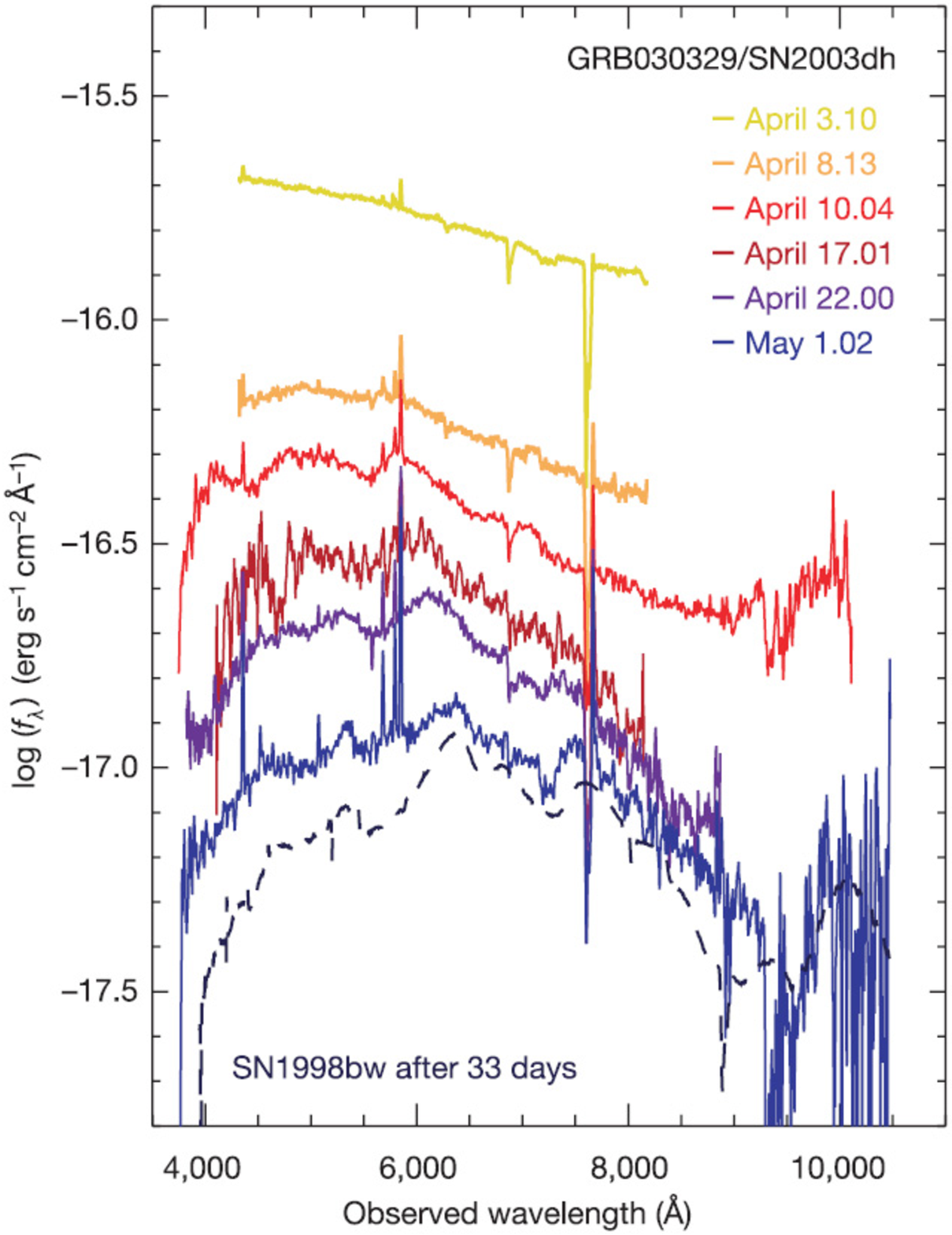,width=0.41\textwidth}
\epsfig{figure=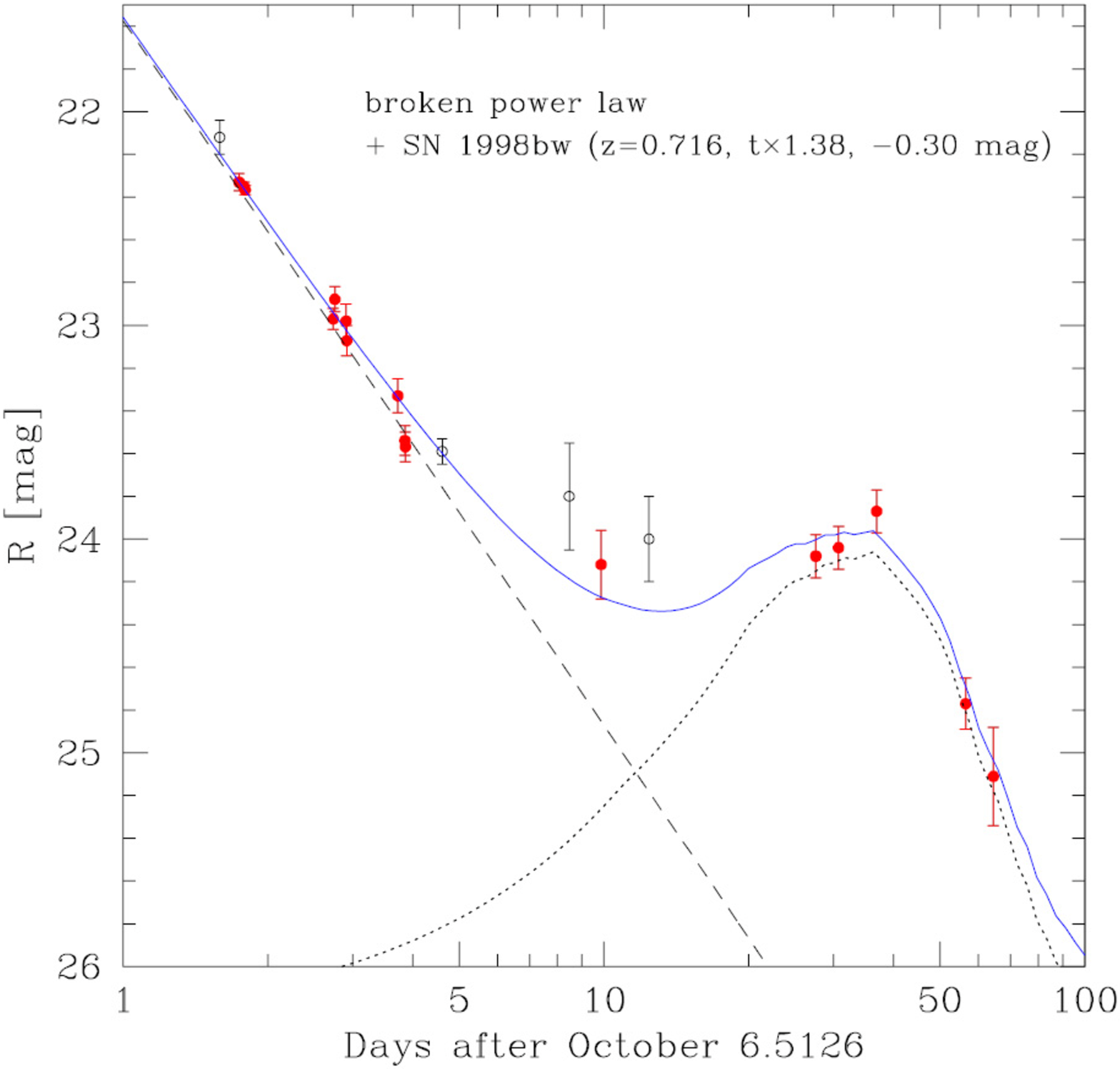,width=0.55\textwidth}
 \caption[Supernova-GRB connection]{{\it Left}:
Spectral evolution of the combined optical flux density of GRB
030329, the associated SN2003dh, and its host galaxy ({\it colored
lines}) compared to the spectrum of SN1998bw ({\it dashed line}).
{\it Right}: R-band light curve for GRB041006. A typical GRB
power-law decay of the optical afterglow is shown as the dashed line
and the SN1998bw light curve extended by a factor of 1.38 is shown
as the dotted curve. Together they fit the observations.
 From \cite{bib:hjorth2003} ({\it left}) and \cite{bib:stanek2005} ({\it right}).} \label{bib:snbump}
\end{center}
\end{figure}

In 1998, supernova 1998sw was found within the error box of GRB
980425 \citep{bib:galama1998, bib:kulkarni1998} and in 2003, the
Supernova-GRB connection was unambiguous established by the
discovery of a clear supernova light-curve bump and spectral
signature in the optical afterglow of GRB 030329
\citep{bib:hjorth2003, bib:stanek2003} (see Fig.\ \ref{bib:snbump}).

\section{The general picture}
After roughly 40 years with GRBs in the scientific spotlight, we are
converging towards a working theory for GRBs.

There appears to be a general consensus in the scientific community
about the fireball internal-external shock model, in which the
gamma-ray burst and the subsequent afterglow radiation is created by
dissipation of collisionless plasma shocks. Independent of the true
nature of the progenitor, the extremely large amount of energy
deposited in a very small volume inevitably creates a highly
energetic outflow that will interact with the surrounding medium
\citep{bib:shemi1990}.

For a detailed discussion, several good review papers exist (e.g.,
\citec{bib:fishman1995}, \citec{bib:piran1999},
\citec{bib:paradijs2000}, \citec{bib:meszaros2001},
\citec{bib:mesz2002}, \citec{bib:zhang2004}, and
\citec{bib:piran2005}).

 To explain the origin and variability of the prompt gamma
emission, it was suggested that the progenitor may expel multiple
plasma shells with different energies. The shells heat up in shocks
when they overtake each other
\citep{bib:rees1994,bib:paczynski1994}. At later times, an external
shock forms as the ejecta blasts through the external medium. The
external medium can either be the interstellar medium or a
progenitor wind. This shock heats up the external plasma and creates
the afterglow \citep{bib:rees1992,bib:meszaros1993}.

As the fireball expands into the external medium it sweeps up
external matter and is expected to eventually approach the self
similar Blandford-McKee solution \citep{bib:blandfordmckee}. This is
a blast wave solution analogues to the non-relativistic Sedov-Taylor
solution \citep{bib:granot2002,bib:sari1995}. For an extensive
review on the GRB blast wave physics, see \cite{bib:piran1999}.

\cite{bib:rees1992} and \cite{bib:meszaros1993} suggested
synchrotron radiation as the main radiation mechanism. The
synchrotron radiation assumption naturally implied that the spectrum
would soften and fade as a power law in time, and that an optical
and radio afterglow would be present at later times
\citep{bib:paczynski1993,bib:katz1994}. The radiation from both
internal and external shocks is fairly well fitted by synchrotron
and inverse Compton radiation from a high-energy non-thermal
electron population in a strong magnetic field. Below (section
\ref{sec:intro-acc} and \ref{sec:intro-magn}) I discuss the
non-thermal acceleration and the possible origin of the magnetic
field.

\begin{figure}[htb]
\begin{center}
\epsfig{figure=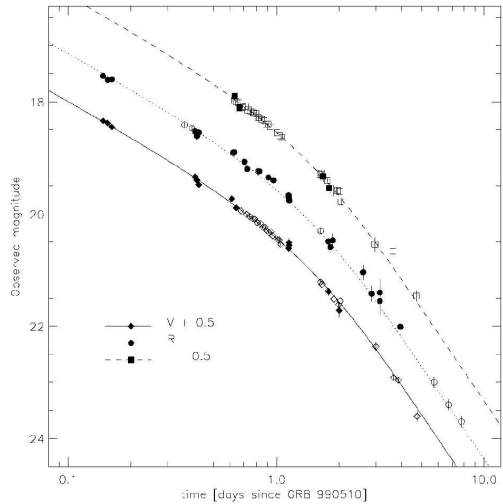,width=0.5\textwidth}
 \caption[Jet break of GRB 990510.]{The jet break of GRB 990510.
 The break is interpreted as the limit where the relativistic jet
is decelerated enough so that the relativistic beaming angle
($~1/\Gamma$) becomes larger than the jet opening angle $\theta_j$.
When this happens, the beaming angle covers an increasingly larger
area outside the jet and the temporal decay will appear faster. From
\cite{bib:harrison1999}.} \label{fig:jetbreak}
\end{center}
\end{figure}

It is now well established that the relativistic gamma-ray burst
ejecta are collimated. A collimated outflow is indeed a more
compelling scenario since the required energy release from the
progenitor is greatly reduced. If the emission were isotropic, the
cosmological distances would imply that some bursts emit more than
one solar mass in gamma rays. Such energies are hard to produce
instantaneously in any stellar model. An observational fact that
supports the collimation model is an achromatic break in the power
law slope of the light curves. For many afterglows this happens days
to weeks after the burst \citep{bib:kulkarni1999,bib:harrison1999}
(see Fig.\ \ref{fig:jetbreak}). Such a break was suggested and
interpreted as the limit where the relativistic jet is decelerated
enough so that the relativistic beaming angle ($~1/\Gamma$) becomes
larger than the jet opening angle $\theta_j$
\citep{bib:rhoads1997,bib:rhoads1999,bib:panaitescu1999,bib:sari1999b}.
When this happens, the beaming angle covers an increasingly larger
area outside the jet and the temporal decay will appear faster.

One of the big unanswered questions concerns the jet structure.
Calculating the energy budget for a burst requires crucial knowledge
of the angular shape of the jet. The simplest structure is a jet
where the internal energy, density and bulk Lorentz factor are
constant throughout the jet cone
\citep{bib:rhoads1999,bib:sari1999b}. A more advanced model is the
universal structured jet where the jet parameters vary smoothly with
the angle measured from the jet symmetry axis
\citep{bib:lipunov2001,bib:rossi2002,bib:zhang2002}. Simulations by
\cite{bib:zhang2004} of the jet break-out from a massive Wolf-Rayet
star show that the jet consists of at least two components (a highly
relativistic thin jet and a less relativistic cocoon jet).

\begin{figure}[htb]
\begin{center}
\epsfig{figure=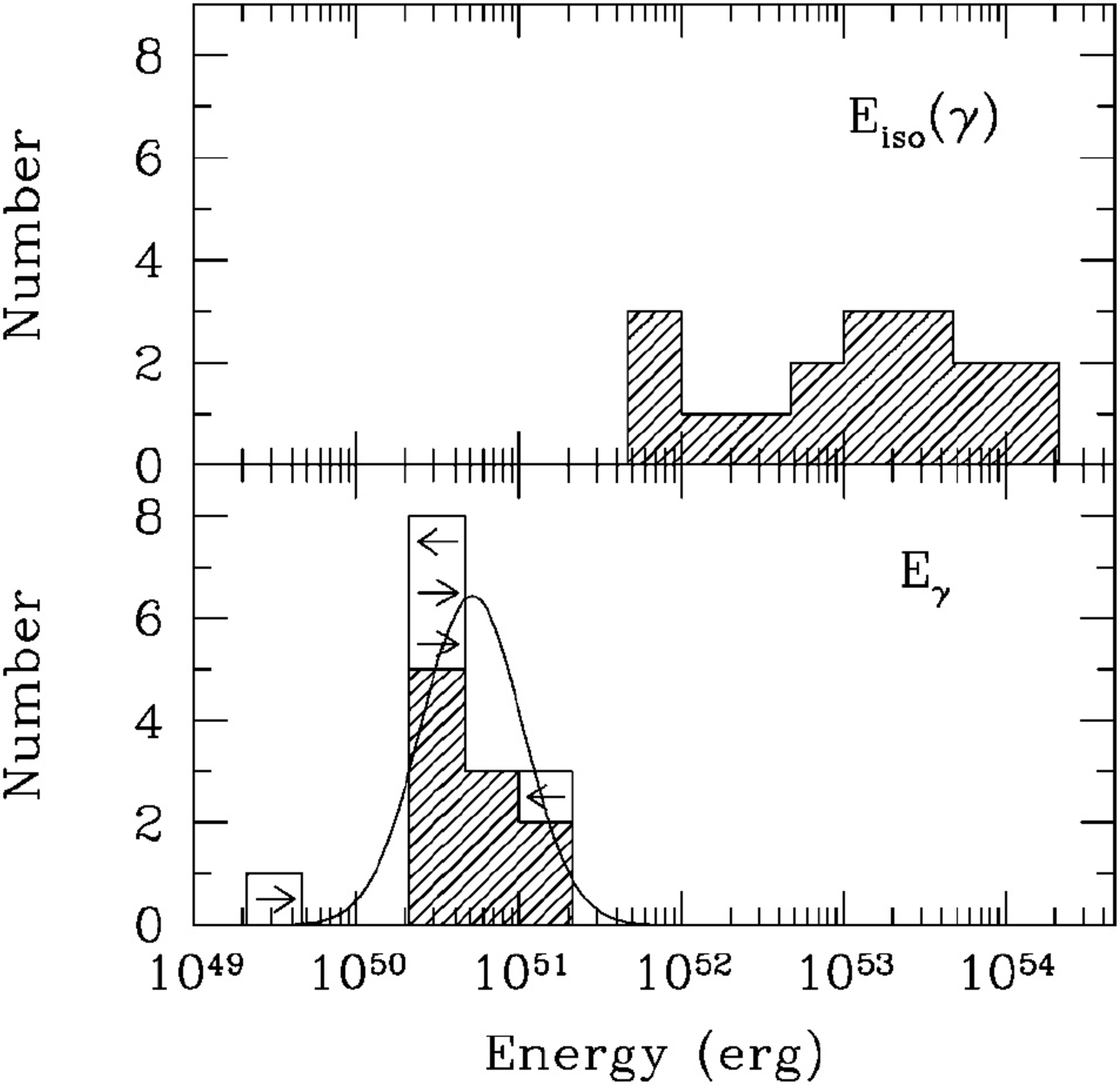,width=0.45\textwidth}
\epsfig{figure=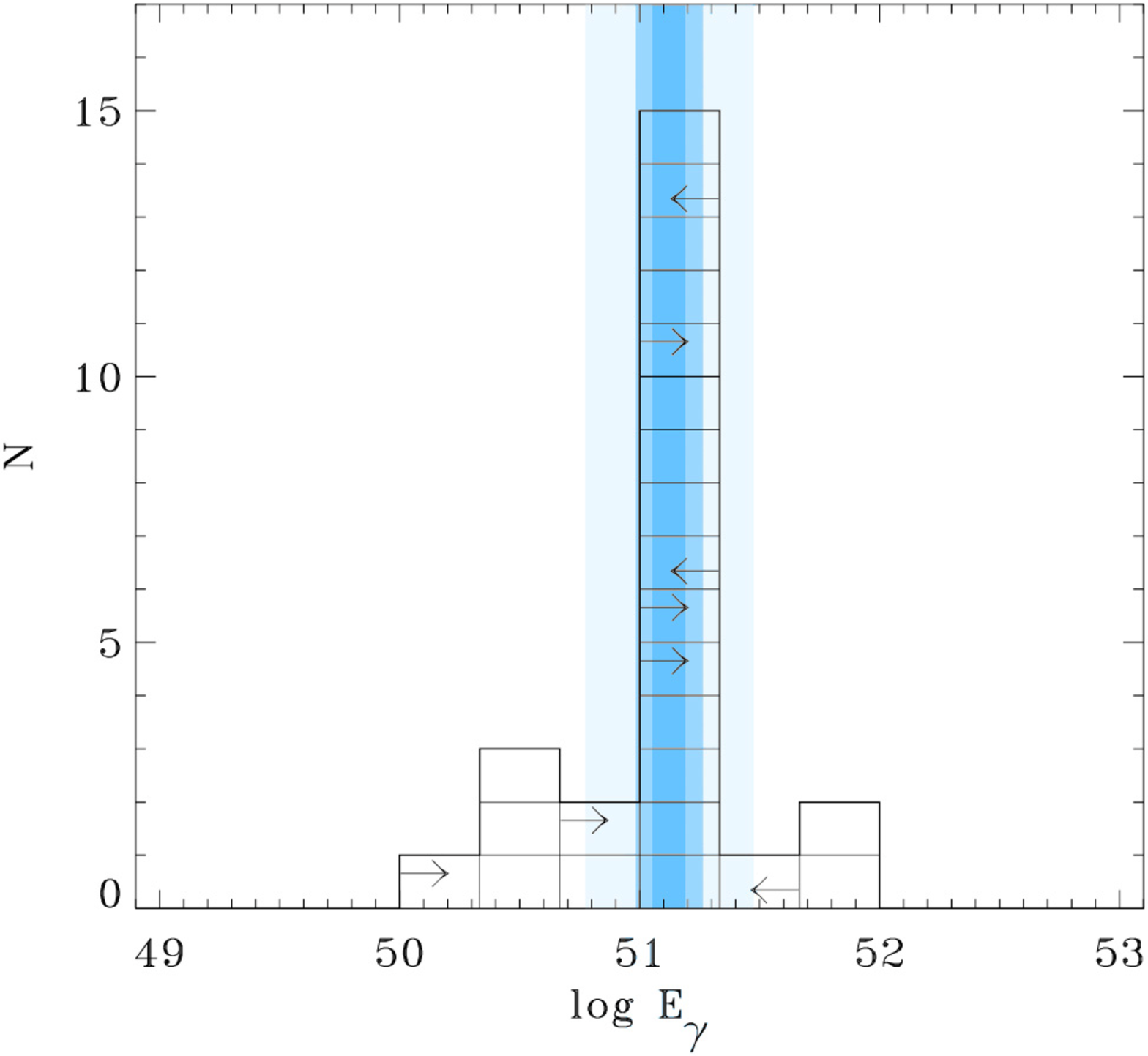,width=0.48\textwidth}
 \caption[Beaming angle corrected energies.]{Correcting the total
 emitted gamma-ray energy corrected for beaming angle.
 From \cite{bib:frail2001} {\it left} and \cite{bib:bloom2003} ({\it right}).} \label{fig:frail}
\end{center}
\end{figure}
Correcting for the jet geometry, the total required burst energy
drops from $10^{47}\ \mathrm{J}$ to around $10^{44}\ \mathrm{J}$,
not too far from the supernova output. More remarkably, in both the
uniform structured jet and the universal structured jet model, the
gamma-ray energy releases for many bursts are narrowly clustered
around $5\times10^{43}\ \mathrm{J}$
\citep{bib:frail2001,bib:bloom2003} (see Fig.\ \ref{fig:frail}). In
the uniform structured jet model, the different light curve break
times are explained by different opening angles. The universal
structured jet explains the break time spread by differences in the
angle between the line of sight and the jet symmetry axis
\citep{bib:rossi2002,bib:zhang2002}.

\begin{figure}[htb]
\begin{center}
\epsfig{figure=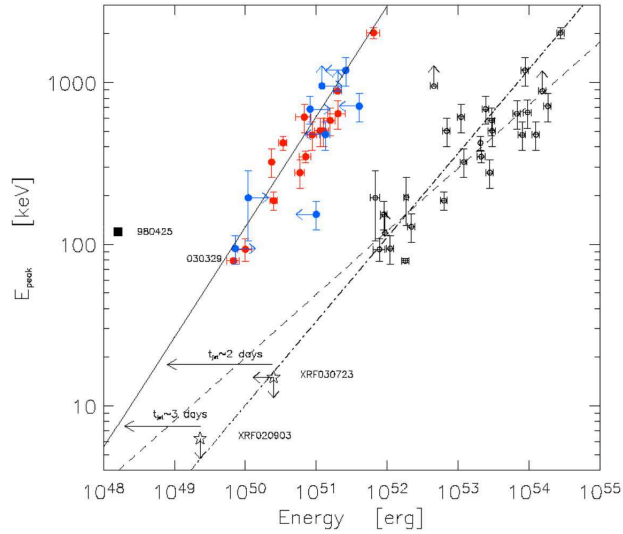,width=0.6\textwidth}
 \caption[The Ghirlanda and Amati relations]{The \cite{bib:amati}
 and \cite{bib:ghirlanda} relations. The Amati relation links the
 total isotropic equivalent gamma-ray energy to the peak energy.
 The Ghirlanda relation ({\it solid line}) links the
 beaming corrected total energy to the $\nu F_\nu$ peak energy in the cosmological rest frame of the burst. From \cite{bib:ghirlanda}.} \label{fig:ghirlanda}
\end{center}
\end{figure}
Another striking correlation is the Amati correlation.
\cite{bib:amati} found a correlation between the peak in $\nu
F_{\nu}$ and the total isotropic equivalent burst energy. An even
tighter correlation was found between the typical photon energy and
the beaming corrected gamma-ray output during the burst
\citep{bib:ghirlanda} (see Fig.\ \ref{fig:ghirlanda}). These
correlations have so far been purely empirical, with no viable
physical explanation. Recently however, \cite{bib:ryde2005}
suggested a hybrid model consisting of a strong thermal component
 accompanied by a non-thermal component of similar strength.
\cite{bib:ryde2005} made fits of 25 strong bursts and compared the
hybrid model to the commonly used Band function
\citep{bib:band1993}. The result showed almost equally good fits for
the two models except for ten of the burst where the hybrid model is
marginally better.

In Ryde's hybrid model, the peak of the burst is determined
primarily by the temperature and is less sensitive on $\Gamma$. In
this case, the Amati/Ghirlanda correlations have a natural
explanation, since for a thermal emitter the luminosity and the
temperature are correlated. \cite{bib:rees2005} show that a
correlation close to the observed one arises naturally under certain
assumptions.

Thus, it would seem that GRBs are becoming excellent standard
candles for probing the dark ages of the universe.

\section{Particle acceleration in collisionless shocks}\label{sec:intro-acc}
One of the key ingredients in generating what is believed to be
non-thermal synchrotron radiation from GRB afterglows is a
non-thermal, high-energetic electron population. Placing an ensemble
of electrons with a power-law energy distribution function
$dN(E)\propto E^{-p}dE$ in a homogenous magnetic field will result
in a synchrotron radiation spectrum with a power-law segment
$F(\nu)\propto\nu^{-(p-1)/2}$ (e.g. \citec{bib:landau} and
\citec{bib:rybicki}).

Many afterglow and Cosmic Ray models assume that electrons are
accelerated in collisionless shocks by diffusive Fermi acceleration
\citep{1949PhRv...75.1169F}. In first-order Fermi acceleration, the
particles are accelerated as they repeatedly cross the shock
transition jump. In the shock rest frame, the incoming upstream
particles are stochastically deflected by a magnetic field as they
pass the shock region. In this process, a fraction of the particles
are kicked to higher energy. These high-energy particles then run
into the upstream region and a fraction of the particles are
reflected into the shock again for further acceleration. This
iterative process continues in a competitive game between energy
gain and escape of particles.

The theory of diffusive shock acceleration predicts that for
non-relativistic shocks, the resulting particle distribution
function converges to a power-law with the slope
\begin{equation}
p=\frac{v_u+2v_d}{v_u-v_d}\ \ \ \ \ \mathrm{or}\ \ \ \ \ s=\frac{3v_u}{v_u-v_d}
\end{equation}
(e.g.\ \citec{bib:axford1977}, \citec{bib:bell1978} and
 \citec{bib:blandford1978}). Here $v_u$  and $v_d$ are the upstream and
downstream plasma bulk velocities. The notation $s=p+2$ is more
common in the literature of particle acceleration
($d^3N(p)/dp^3\propto p^{-s}$ where p is the particle momenta).

In the relativistic limit, the Fermi acceleration formalism becomes
more complicated. The non-relativistic derivation is based on the
diffusion equation \citep{bib:kirk1987}, derived under the
assumption that the particle distribution function is approximately
isotropic in the local plasma rest frame. But for relativistic
shocks, the bulk flow is comparable to the particle velocities and
then the particle angular distribution function becomes highly
anisotropic near the shock. In this case, strong magnetic
fluctuations downstream of the shock are essential
\citep{bib:achterberg2001}.
 \cite{bib:kirk1987} and \cite{bib:heavens1988} investigated the relativistic problem and found the distribution slopes
\begin{equation}
p=\frac{R+2}{R-1}\ \ \ \ \ \mathrm{or}\ \ \ \ \ s=\frac{3R}{R-1},
\end{equation}
where $R$ is the shock compression factor. Both analytical
\citep{bib:kirk2000} and Monte-Carlo
\citep{bib:bednarz1998,bib:achterberg2001} find that $p$ converges
at 2.23 for $\Gamma\to\infty$ ($\Gamma$ is here the bulk Lorentz
factor).

The major force of relativistic first-order Fermi acceleration is
that it predicts indices very close to the ones inferred from
observations. Estimates from a number of GRB afterglows yield
$p=2.2\pm0.2$ \citep{bib:waxman1997a,bib:berger2003}. Good agreement
and predictions are, however, not the same as a scientific proof.
Afterglows exist that have a much larger variety in $p$. E.g.
\cite{bib:campana2005} find $p=1.3$ in the very early afterglow.
Here I emphasize some of the problems that the Fermi acceleration
scenario in GRB afterglows is still facing:

\begin{description}
\item[Problem 1] It is very important to stress that Fermi acceleration in collisionless shocks is still not understood from
first principles. The foundation of the acceleration mechanism is
based on the test-particle approximation. It is assumed that the
particles scatter on electromagnetic waves but the model does not
self-consistently account for the generation of these waves. Nor
does it account for the back-reaction that the high-energy particle
distribution have on the electromagnetic field. Acceleration that
results from currents and charge separation near the shock must be
probed with a full kinetic approach (e.g.\ particle-in-cell codes).
See section \ref{sec:introacc} and Chapter \ref{sec:bperp}.

\item[Problem 2] In all derivations of the relativistic
Fermi acceleration mechanism, the downstream magnetic field is
required to be strongly turbulent on scales smaller than the typical
gyro-radius (e.g.\ \citec{bib:ostrowski2002}). This is, however, in
conflict with the GRB afterglow synchrotron interpretation where the
high-energy particles are expected to gyrate in circular orbits in a
magnetic field with variation scale length much longer that the
gyro-radii. Either the ansatz of strong downstream turbulence must
be relaxed, with the result that the high-energy particle
distribution function is not nearly as universal and possible not
power-law at all
\citep{bib:ostrowski2002,bib:Niemiec,bib:baring2005}. Or, the
radiation model must be altered to include jitter radiation
\citep{bib:Medvedev_jitter}. See also Chapter \ref{sec:rad}.
\item[Problem
3] Relativistic Fermi acceleration requires a pre-acceleration
mechanism that injects electrons into the iterative acceleration
process. The pre-acceleration mechanism is not well-known. How large
a fraction of the electrons that are accelerated greatly affects our
estimates of the total GRB energy \citep{bib:eichler2005}. We define
this fraction as $f$. Moreover, according to \cite{bib:baring},
agreeable synchrotron and/or inverse Compton fits are only
attainable when the electron population has a significant
non-thermal component. This is in disagreement with the Fermi
process where electrons are injected from a dominant thermal pool.
The lack of a dominant thermal pool also raises a question of how
the electromagnetic turbulence is sustained in the shock-region.
\item[Problem 4] This problem is
connected to problem 3. In the closest and best studied mildly
relativistic shock in the Crab Nebula, most of the electrons radiate
below the expected injection energy and this means that $f\ll1$
\citep{bib:eichler2005}. The "low" energy electrons have a power-law
distribution spectrum $1.3\ge p\ge 1.1$ \citep{bib:weiler1978}. This
is much lower than what is expected from test particle simulations.
The high-energy electrons, however, are more consistent with slope
expected from first order Fermi acceleration $p\simeq2.2$.
\item[Problem 5] If the standard Fermi diffusive shock
acceleration theory is correct, one expects an X-ray halo around the
shock. This is because higher energy electrons are expected to
diffuse further ahead of the shock, so the halo should become more
extensive at X-ray wavelengths. \cite{bib:long2003} have
investigated high resolution Chandra images of the close by
Supernova remnant SN 1006. They fail to detect such a halo. Instead
they see a sharp jump in emissivity at the shock. They conclude that
either Fermi acceleration is absent in this shock, or some kind of
shock instability must be operating in shocks that can create or
amplify a magnetic field with a factor significantly larger than
that given by the fluid compression, resulting in greater contrast
between upstream and downstream emission \citep{bib:long2003}.
\end{description}

\subsection{Particle acceleration in PIC simulations}\label{sec:introacc}
Clearly, further advances in our understanding of particle
acceleration in collisionless shocks require a full, self-consistent
kinetic treatment. This may be provided by particle-in-cell (PIC)
simulations (see Chapter \ref{sec:pic}). Unfortunately, PIC code
simulations are very computationally demanding. Therefore, many PIC
simulations up to date are one-dimensional. Here, I briefly review
the state of particle acceleration in PIC code simulations.

One of the first reports of non-thermal acceleration in PIC
simulations of relativistic collisionless plasma shocks came from
\cite{bib:hoshino1992}. In their (one-dimensional) simulations,
plasma, carrying a magnetic field, is injected at the leftmost
boundary. At the rightmost boundary, the plasma flow is reflected
and thus collides with itself. In a plasma consisting purely of
electron/positron pairs, they find that the acceleration is mainly
thermal. When protons are present the positrons are strongly
accelerated. The acceleration is driven by resonant absorbtion of
magnetosonic waves, excited by energy dissipated from the gyrating
ions. \cite{bib:hoshino1992} found that the positrons could be
accelerated to a power-law distribution with slope $p=1.5-2.5$ and
that the spectrum extended up to $\Gamma m_ic^2$. The reason why
only positrons and not electrons are participating in the
acceleration has to do with the polarization of the magnetosonic
waves.

Another mechanism was examined with PIC simulations by
\cite{bib:Dieckmann2000,bib:Dieckmann2004}. Based on theoretical
work by \cite{bib:galeev1995},
\cite{bib:Dieckmann2000,bib:Dieckmann2004} used PIC simulations of
counter-streaming proton beams on a cold plasma background in a
transverse magnetic field. They found traces of non-thermal particle
acceleration up to mildly relativistic energies, offering an
explanation to the injection problem. One should note, however, that
their simulations were limited to one spatial dimension and the
setup in itself is based on many assumptions about the initial
conditions. It is not clear if the ion beam is injected on a
quasi-neutral plasma background. If this is indeed the case, there
is an excess of correlated positive charges in the simulations. This
might trigger unrealistic instabilities. The size and duration of
the simulations are rather limited and it would be interesting to
see the same simulations carried out in three-dimensions.

\begin{figure}[htb]
\begin{center}
\epsfig{figure=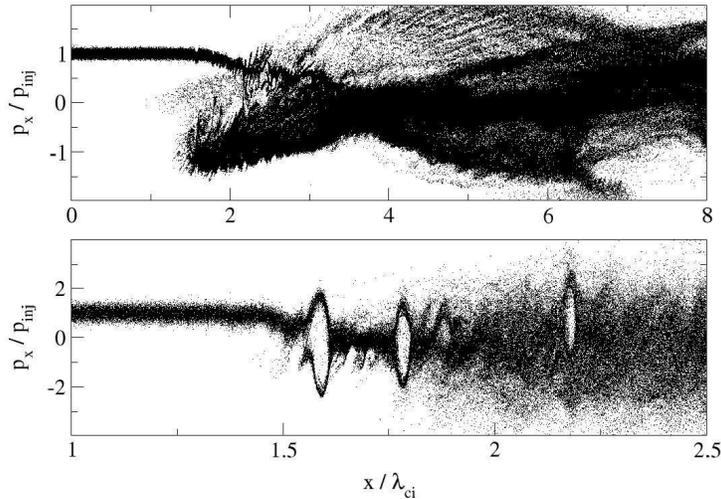,width=0.7\textwidth}
 \caption[Beaming angle corrected energies.]{
Phase space plot of ions ({\it top panel}) and electrons ({\it
bottom panel}). The surfatron mechanism accelerate electrons by
 trapping in nonlinear wave modes linked to the existence of solitary electron
phase space holes. The result is based on 1D PIC simulations. From
\cite{bib:Schmitz2002}.} \label{fig:loophole}
\end{center}
\end{figure}

A very well studied acceleration mechanism is the so called
surfatron \citep{bib:Katsouleas1983,bib:dawson1983}. The surfatron
is a mechanism in which a particle is accelerated while it is
trapped by a propagating large-scale perpendicular electrostatic
wave. The waves are driven via the Buneman instability that arises
when electron and ion beams drift at different velocities
\citep{bib:buneman1958,bib:buneman1959}. The acceleration of
particles by trapping in electrostatic waves has been simulated with
PIC simulations by \cite{bib:hoshino2002} and \cite{bib:shimada2004}
and has been linked to phase space holes or loopholes
\citep{bib:Schmitz2002} (see Fig.\ \ref{fig:loophole}).

It should be noted that several independent simulations have
indicated that the shock-fronts of collisionless plasma shocks,
propagating in an ambient magnetic field, show great time
variability (e.g. \cite{bib:lembege2004} and references therein).
These simulations include both PIC simulations and hybrid
simulations. In the latter, the ions are treated as particles and
the electrons as a massless fluid. In one-dimensional PIC
simulations by \cite{bib:Lee2004} with parameters aimed at supernova
shocks, the shock structure was found to be cyclically reforming on
ion cyclotron timescales. In the reformation, \cite{bib:Lee2004}
found both electron loophole acceleration but also a suprathermal
population of ions that could potentially explain the injection of
high-energy ions into a Fermi acceleration scenario. The
acceleration of electrons is interesting in the GRB context, and the
acceleration of ions is interesting in regard to ultra high-energy
cosmic rays.

All the simulations reviewed above have two things in common, 1)
they are one-dimensional and 2) they assume that a rather strong
transverse magnetic field is present. It would indeed be very
interesting to explore whether effects such as electron trapping by
electrostatic solitary waves will survive in three-dimensions, or if
other effects become dominant, rendering the previous results
artifacts of the one-dimensionality.

Specifically, I would like to express my reservation concerning the
surfatron and loophole acceleration in relativistic collisionless
shocks for the following reasons. The main driving mechanism is the
excitation of electrostatic waves as a result of the Buneman
two-stream instability. However, at relativistic velocities, the
Weibel two-stream instability has a larger growth rate than the
Buneman instability and will dominate the shock region
\citep{bib:califano2002}. But the Weibel two-stream instability
cannot be represented  correctly in simulations with only one
spatial dimension. \cite{bib:shimada2004} have shown that the
generation of loopholes is suppressed when the plasma frequency to
cyclotron frequency ratio
($\omega_{pe}/\omega_{ce}\propto\sqrt{n}/B$) is less than 10
($\omega_{pe}\equiv(n q^2/m_e\epsilon_0)^{1/2}$ is the electron
plasma frequency and $\omega_{ce}\equiv qB/m_e$ is the electron
cyclotron frequency). \cite{bib:hededal2005} found in
three-dimensional simulations that for $\omega_{pe}/\omega_{ce}>10$,
the initial ordered ambient magnetic field becomes curled and even
locally reversed because of the Weibel instability.
\cite{bib:hededal2005} did find non-thermal acceleration, but for
other reasons (see Chapter \ref{sec:bperp}). So for
$\omega_{pe}/\omega_{ce}<10$ the solitary electrostatic surfatron
acceleration mechanism is suppressed and for
$\omega_{pe}/\omega_{ce}>10$ the Weibel two-stream instability is
present, which will distort the electrostatic wave generation.
Finally, \cite{bib:hededal2005} found that in the interstellar
medium where $\omega_{pe}/\omega_{ce}\simeq1000$, the Weibel
two-stream instability evolves unhindered, with no signs of
loopholes or surfatrons (although \cite{bib:hededal2005} do point
out the importance of larger simulations).

Clearly, the time has come for 2D or even 3D simulations to provide
a more self-consistent explanation for the origin of the ambient
magnetic field and non-thermal particle acceleration. Evidence from
3D PIC-simulations is gathering that suggests that particle
acceleration and magnetic field generation are two highly connected
features of collisionless plasma shocks
\citep{bib:frederiksen2002,bib:silva,bib:frederiksen2004,bib:nishikawa,bib:nishikawa2004,bib:hededal2004,bib:hededal2005}.
I save the discussion of these 3D PIC simulations of particle
acceleration and the connection to field generation to section
\ref{sec:magn_pic} below.

\section{Magnetic fields in gamma-ray bursts}\label{sec:intro-magn}
A second crucial ingredient in generation of radiation in GRB
afterglows is the presence of a strong magnetic field. For a review
on the role of magnetic fields in GRBs, see \cite{bib:piran2005b}.

The general working assumption is that the energy that resides in
the magnetic field and in the non-thermal electrons may be
parameterized by the equipartition parameters $\epsilon_B$ and
$\epsilon_e$, and that the non-thermal electrons follow a power-law
distribution with slope $p$ (where $\epsilon_B$ and $\epsilon_e$ are
defined as the fractions of the total internal energy of the shock
that are deposited in magnetic energy and kinetic energy of the
electrons, respectively). It is generally assumed that these
parameters are constant through the shock and even throughout the
duration of the afterglow. The values are observationally determined
by localizing certain characteristic break frequencies in the
spectra. From low to high frequencies these are the synchrotron
self-absorbtion frequency, the synchrotron frequency of the typical
electron, and the self absorption frequency
\citep{bib:sari1998,bib:piran1999,bib:piran2005b,bib:piran2005}.
Regarding the mangetic field, the typical value of the equipartition
parameter in the afterglow is $\epsilon_B=0.0001-0.1$
\citep{bib:waxman1997a,bib:wijers1999,bib:Panaitescu+Kumar,bib:yost2003}.
This value may be translated to a magnetic field of the order of
$10^{-4}\ $T (1 G) in the afterglow shock. In the interstellar
medium the typical magnetic field strength is of the order a few
$10^{-10}$ T (few $\mu$G). According to the relativistic
Rankine-Hugoniot plasma shock jump conditions \citep{bib:taub1948},
shock compression can only give a factor of $4\Gamma_{shock}$ and
this is clearly far from enough to match the interpretations from
observations \citep{1999ApJ...511..852G}. This leaves us with two
possibilities for the origin of the magnetic field. One possibility
is that the magnetic field is generated or amplified in the shock by
microphysical instabilities and one is that the magnetic field is
carried with the outflow from the progenitor. A magnetic field that
originates from the progenitor and is frozen into the ejecta might
account for the magnetic field in the internal shocks. But as the
plasma shell expands, the magnetic field is diluted and dissipated
to well below the anticipated values (e.g.
\citec{bib:medvedevloeb}). Additionally, there exists a question of
how to transport a magnetic from the ejecta and into the shocked ISM
(all though theoretically it may happen via the Rayleigh-Taylor
instability). Hence, the magnetic field responsible for the
afterglow is most likely to be generated {\it in situ} in the shock.

Under collision of two relativistic plasma populations, the particle
phase space is extremely anisotropic. Naively one could argue that
in the absence of particle collisions, the two plasma populations
would stream right through each other. The anisotropy is, however,
unstable to several plasma instabilities, including the
electrostatic Buneman two-stream instability and the electromagnetic
Weibel two-stream instability. At relativistic shocks, the latter
has the largest growth rate and will dominate
\citep{bib:califano2002,bib:hededal2005}. \cite{1999ApJ...511..852G}
and \cite{bib:medvedevloeb} suggested that the Weibel two-stream
instability could generate a strong magnetic field in the shock
region. Since many of the results that I present in this thesis are
connected to the Weibel two-stream instability, it is worthwhile to
explain the nature of the instability in some detail. The following
description is based on papers by \cite{bib:Weibel},
\cite{bib:freid1959}, \cite{bib:medvedevloeb}, and
\cite{bib:Wiersma2004}.

\begin{figure}[htb]
\begin{center}
\epsfig{figure=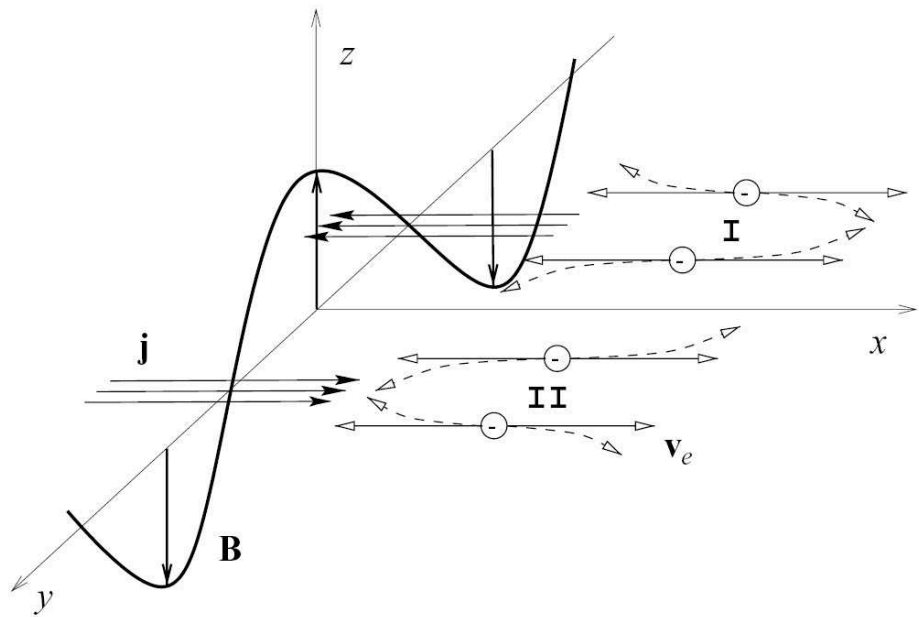,width=0.9\textwidth}
 \caption[Schematic of the Weibel two-stream instability]{Schematic of the Weibel two-stream instability.
Counter-streaming electrons are collected into opposite directed
current channels by a magnetic perturbation. The current channels
amplify the initial perturbation.
 From \cite{bib:medvedevloeb}.} \label{fig:weibel_medv}
\end{center}
\end{figure}

\cite{bib:Weibel} suggested that if an isotropic plasma population
is anisotropically perturbed, relaxation will lead to a growing
transverse magnetic field, even in the absence of an external
electromagnetic field. The same year, \cite{bib:freid1959} gave a
physical interpretation, where the anisotropic perturbation was
described as an actual two-stream configuration. Since there will
always be infinitesimal magnetic perturbations in a plasma, Fried
suggested that deflection of the electrons (by the Lorentz force) in
such a fluctuating magnetic field will create currents that can
amplify the initial magnetic perturbation. Figure
\ref{fig:weibel_medv} shows a schematic drawing of this mechanism
\citep{bib:medvedevloeb}. In the center of mass rest frame, two
oppositely directed electron beams collide at $x=0$ (with bulk
velocities $v_{1x}=-v_e$ and $v_{2x}=v_e$ so that the net current is
zero). The ions are treated as a homogenous non-interacting
background, to ensure charge neutrality.
 At x=0, a magnetic perturbation is initially present
with $B_z=B_0\cos{ky}$. This perturbation deflects left streaming
 and right streaming electrons into anti-parallel currents
with a distance comparable to the wavelength of the magnetic
perturbation. According to Ampere's law these currents represent a
curl in the magnetic field. This curl amplifies the initial
perturbation, which, in turn, collects even more electrons into the
current channels. This positive feedback results in an instability
where magnetic fluctuations grow with a rate
$\sigma=\omega_{pe}v_e/c$ \citep{bib:freid1959}. Again,
$\omega_{pe}$ is the electron plasma frequency defined above.

The relativistic generalization is not trivial. The problem have
been investigated by \cite{bib:YoonDavidson} who find the maximum
growth rate to be $\sigma=\omega_{pe}/\Gamma^{1/2}$ where $\Gamma$
is the bulk relativistic Lorentz factor. The result can be
understood intuitively by evaluating the non-relativistic expression
in the limit where $v_e\to c$ and $\omega_{pe}\to(n q^2/\Gamma
m_e\epsilon_0)^{1/2}$.

\cite{bib:medvedevloeb} estimated that the instability would
saturate at $\epsilon_B\sim10^{-5}-10^{-4}$ if only the electrons
participate in the instability, and $\epsilon_B\le10^{-1}$ if also
the ions take part in the instability. \cite{bib:Wiersma2004} found
from a linear analysis that the two-stream instability in an
electron-proton plasma shock has an early end where $\epsilon_B\le
m_e/m_i$ and that the wavelength of the most efficient mode for
magnetic field generation equals the electron skin depth. Numerical
simulations of the instability have shown that this limitation not
true for the non-linear stage.

\subsection{Magnetic field generation in PIC simulations}\label{sec:magn_pic}

To investigate the non-linear stage of the Weibel two-stream
instability, self-consistent kinetic particle-in-cell (PIC)
simulations are necessary. In Chapter \ref{sec:pic} I briefly
describe the theory behind PIC codes with their advantages and
disadvantages.

The first PIC simulations with direct focus on the Weibel
instability in an astrophysical context were performed by
\cite{bib:Kazimura}. They were interested in the plasma wind
interaction in millisecond binary pulsars. In two-dimensional
simulations of size $6.6\times106.6$ electron skin depths, they
investigated the collision of two mildly relativistic pair-plasma
winds ($v_0=0.5c$). They found that up to five percent of the
kinetic energy was converted into magnetic fields.

\begin{figure}[h!]
\begin{center}
\epsfig{figure=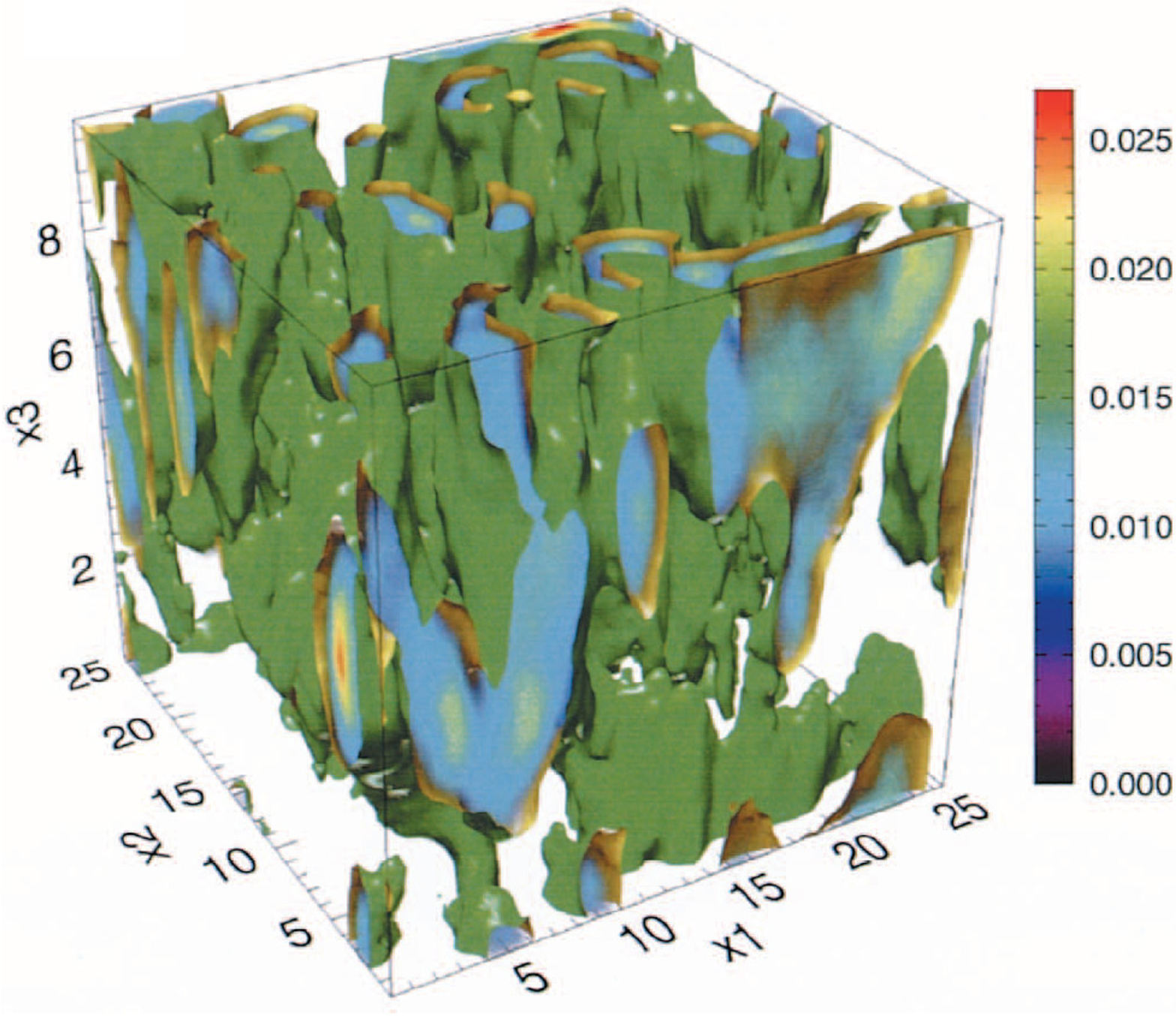,width=0.4\textwidth}
\epsfig{figure=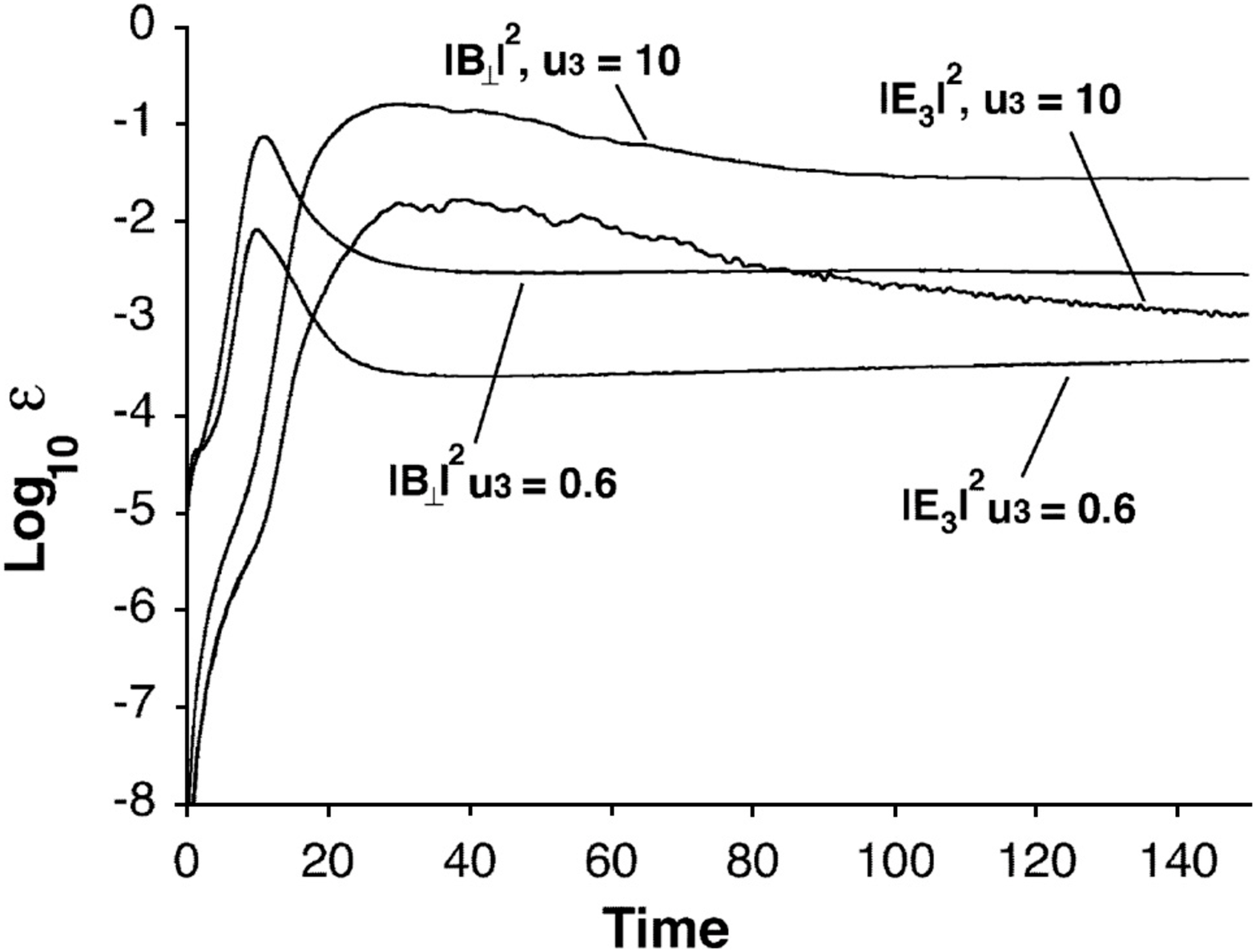,width=0.48\textwidth} \caption[Magnetic
field generation by the Weibel two-stream instability in
pair-plasma]{The structure of current channels generated by the
Weibel two-stream instability under the collision of two pair-plasma
shells ({\it left panel}). The corresponding amount of generated
magnetic end electric field is close to equipartition ({\it right
panel}). From \cite{bib:silva}.} \label{fig:silva}
\end{center}
\end{figure}
The collision of pair plasma shells was investigated with
three-dimensional simulations by \cite{bib:silva}. In a simulation
box with size $25.6\times25.6\times10.0$ electron skin depths, they
explored the collision of pair-plasma shells with relativistic
Lorentz factors ranging from $\Gamma\sim1$ to $\Gamma=10$. They
found that the generated magnetic field reached maximum of
$\epsilon_B\sim0.1$ and afterward relaxed to
$\epsilon_B\sim0.0001-0.01$ (see Fig.\ \ref{fig:silva}). I emphasize
that in these simulations the boundaries in the flow direction are
periodic. It is not clear what this implies, but it can potentially
introduce non-physical, stabilizing feed-back, since each
current-filament is feeding itself.

\begin{figure}[!htbp]
\begin{center}
\epsfig{figure=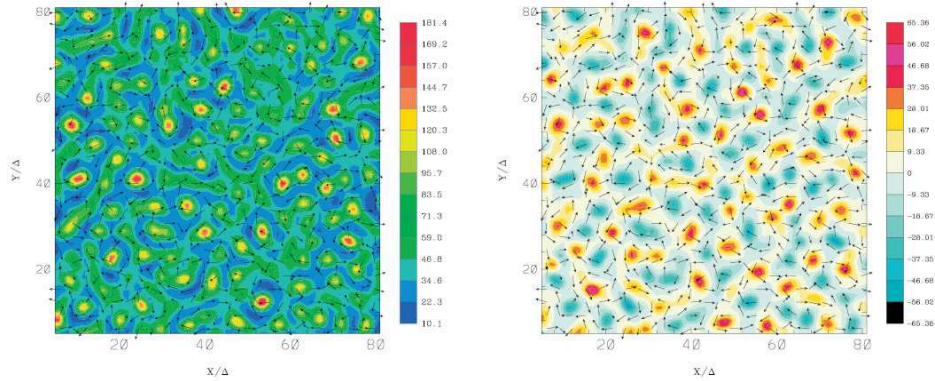,width=.9\textwidth} \caption[Transverse
slice of the particle density during the Weibel instability]{The
electron density ({\it left panel}) and current density ({\it right
panel}). The arrows show the induced magnetic field. From
\cite{bib:nishikawa2004}.} \label{fig:nishikawa}
\end{center}
\end{figure}

\cite{bib:nishikawa} and \cite{bib:nishikawa2004} have performed
three-dimensional simulations of both pair-plasma and
electron-proton shocks in a numerical box with size
$17.7\times17.7\times66.7$ electron skin depths (corresponding to
$4\times4\times14.9$ ion skin depths for ion-electron mass ratio
$m_i/m_e=20$) (see Fig.\ \ref{fig:nishikawa}). With these
simulations they were able to follow primarily the linear stage of
the instability at the front of the shock ramp (the ambient and jet
electron populations are not thermalized to a single population in
the simulations). With the simulations
\citeauthor{bib:nishikawa2004} confirmed the growth rate predicted
by \cite{bib:Weibel}, \cite{bib:freid1959} and
\cite{bib:medvedevloeb}. They also found signs of electron
acceleration connected to the instability, but whether the nature of
this acceleration is non-thermal or merely a thermalization is not
clear.

\begin{figure}[!htp]
\begin{center}
\epsfig{figure=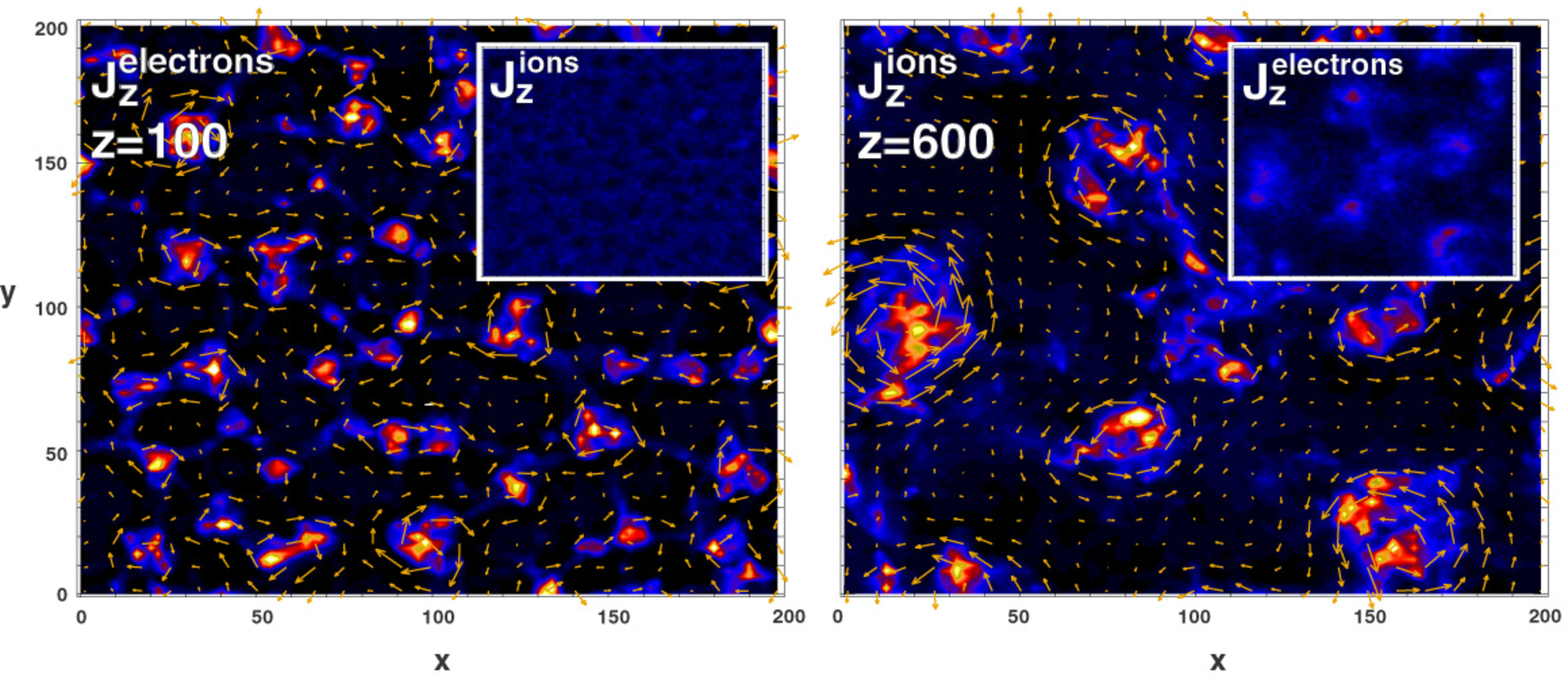,width=0.9\textwidth} \caption[Contour
plot of electron and ion current density in the two-stream
region]{The left hand side panel shows the longitudinal electron
current density through a transverse cut early in the shock. The
right hand side panel shows the ion current deeper in the shock. The
arrows represent the transverse magnetic field. From
\cite{bib:frederiksen2004}.} \label{fig:Slice_intro}
\end{center}
\end{figure}

\cite{bib:frederiksen2002,bib:frederiksen2004} and
\cite{bib:hededal2004} have investigated the non-linear evolution
the two-stream instability in electron-proton shocks. Details are
given in Chapter \ref{sec:weibel} and Chapter \ref{sec:acc} of this
thesis. Here I give a short summary. \cite{bib:frederiksen2004}
performed simulations of electron-proton shocks with $\Gamma=3$. The
size of the simulation box was $40\times40\times160$ electron skin
depths (corresponding to $10\times10\times40$ ion skin depths for
ion-electron mass ratio $m_i/m_e=16$). The results of these
simulations showed how Weibel generated ion current filaments were
collected into increasingly larger patterns in the non-linear stage.
This phenomenon has been investigated both analytically and
numerically in great details by \cite{bib:medv_jit2005}.
\cite{bib:frederiksen2004} also found that the ion current filaments
are partially Debye shielded by shock-heated electrons. The
electrons thus act as insulators for the ion-current filaments,
making these rather robust (see Fig.\ \ref{fig:Slice_intro}). The
results showed that a magnetic field was generated
($\epsilon_B\sim0.05$) and sustained for many ion skin depths. This
answers a concern raised by \cite{bib:Gruzinov1999}, who estimated
that the magnetic field can only be sustained for roughly one
electron skin depth.

\begin{figure}[!htbp]
\begin{center}
\epsfig{figure=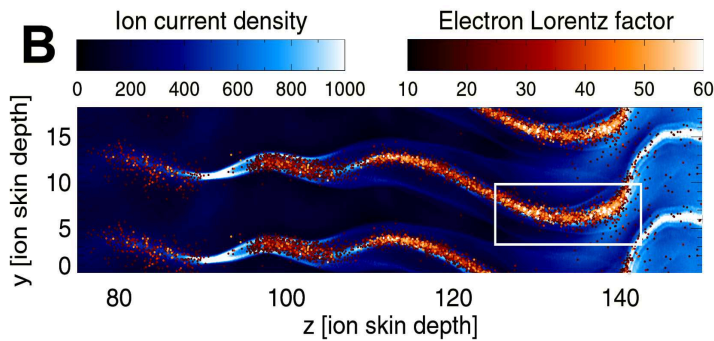,width=.7\textwidth}
\caption[Accelerated particles and ion current channels]{ The
high-energy electrons (red colored dots) are found where the
ion-current filaments are strongest (blue colors). From
\cite{bib:hededal2004}.} \label{fig:hededal_intro}
\end{center}
\end{figure}

\cite{bib:hededal2004} found that not only can the Weibel two-stream
instability account the generation of a strong magnetic field, but
it appears that non-thermal electron acceleration is a natural
consequence of the process. They found that high-energy electrons
are spatially connected to the ion current filaments (see Fig.\
\ref{fig:hededal_intro} and Chapter \ref{sec:acc}). The acceleration
and deceleration of electrons is local and instantaneous. This is in
contrast to the recursive process of Fermi acceleration.

\section{Summary and Thesis Outline}

The radiation from GRB afterglows is produced in relativistic,
collisionless plasma shocks by two key ingredients, namely 1) a
population of highly energetic, non-thermal electrons and 2) a
strong electromagnetic field. All our knowledge of GRBs are based on
this radiation. Nevertheless, the questions of how the electrons are
accelerated, and what the exact generation mechanism and nature of
the electromagnetic field in the shock is, have not yet been
answered in a self-consistent way and are still open questions.

Even though magnetic field generation by the Weibel two-stream
instability seems necessary (and also unavoidable) in collisionless
shocks in GRB afterglows, it have not yet been possible to verify or
discard it from observations. One way to investigate on the magnetic
morphology in these shocks is to measure the polarization of
afterglows, although this also relies heavily on the jet structure.
\cite{bib:lazatti2004} found that polarization of GRB 020813 is not
very well fitted with a homogenous jet with shock generated field. I
stress that to draw any conclusions from observations about the
magnetic field and particle acceleration, it is vital to have a firm
understanding of collisionless shocks, and the generation of
radiation in these.

 \bigskip
  It is the goal of this thesis to shed light
on the microphysics of collisionless plasma shocks, mainly in the
context of gamma-ray burst afterglows. Using a particle-in-cell code
that works from first principles, the aim is to obtain insight in,
and explain the origin of, the magnetic field in these shocks, the
nature of the field, and how it may influence the radiation emitted
from GRB afterglows.

The thesis is divided into \ref{sec:concl} chapters:

\begin{itemize}
\item In Chapter \ref{sec:pic} I describe the particle-in-cell code
that have been used and how radiative cooling has been implemented.

\item Chapter \ref{sec:weibel}  presents the results of simulations of the
non-linear evolution of the Weibel two-stream instability. This
chapter is based on the paper by Frederiksen, Hededal, Haugb\o{}lle
and Nordlund (2004).

\item In Chapter \ref{sec:2d3d} I compare two- and three-dimensional
simulations, and present results of large-scale shock simulations.
This work has not yet been submitted to a scientific journal.

\item Chapter \ref{sec:bperp} investigates the microphysics of
collisionless plasma shocks in the presence of an ambient magnetic
field. This chapter is based on the paper by \cite{bib:hededal2005}.

\item In Chapter \ref{sec:acc} I present a new particle acceleration
mechanism that differs from Fermi acceleration. This paper is based
on the paper by Hededal, Haugb\o{}lle, Frederiksen and Nordlund
(2004).

\item In Chapter \ref{sec:rad} I describe the development and test of a
new and powerful numerical tool, which may be used to create
radiation spectra directly from PIC simulations. This work has not
yet been submitted to a scientific journal.

\item Chapter \ref{chap:photonplasma} describes the foundations
of a next generation particle-in-cell code that includes photons and
various scattering processes. The chapter includes some preliminary
test results with Compton scattering.

\item In Chapter \ref{sec:concl} I collect all the pieces from the
thesis in a summary and conclusions. Here I also discuss the future
of the line of work that I have presented in this thesis.

\end{itemize}

\chapter{The Particle--In--Cell code}\label{sec:pic}
In this Chapter I briefly describe the kinetic particle-in-cell
code, and justify why it is important compared to a fluid
description. I also discuss some of the limitations of a kinetic
numerical description. Finally, I describe in some detail the
derivation, implementation and testing of a radiative cooling
mechanism in the code.

\section{Kinetic or fluid description of collisionless shocks.}
The mean free path for a $90^\circ$ Coulomb deflection of an
electron moving with relativistic momentum $\Gamma v_e m_e$ in a
plasma (with density $n$) is
\begin{equation}
\lambda_c=\frac{1}{n\sigma_c}=\frac{16\pi\epsilon_0^2\sqrt{2}\gamma^2m_e^2v_e^4}{n
e^4}.
\end{equation}
See Appendix \ref{app:cross_section} for details.
 For a relativistic
GRB jet that expands into the interstellar medium (ISM), we may use
this expression to estimate the typical mean free path for Coulomb
collisions between the jet and ISM particles. With an ISM density
$n\simeq10^6\mathrm{m}^{-3}$ and a jet bulk Lorentz factor
$\gamma=5$ ($v\simeq c$), the mean free path for Coulomb collisions
is $10^{24}\mathrm{m}$. This is of the order of a billion times the
expected size of the fireball. Thus one might naively expect a
relativistic jet to expand unhindered through the ISM. This is
however, in direct contrast with observations, where gamma-ray burst
afterglows are described by synchrotron emission from decelerating
relativistic shells that collides with, and heats, an external
medium. Lack of interactions would pose serious problems in
explaining the particle acceleration and origin of the magnetic
field, which is needed to produce the observed synchrotron radiation
(e.g.\ \citec{bib:waxman1997} and \citec{bib:sari1998}). Not
surprisingly, the collisional interaction agent must be found in the
microphysical processes at play between particles and
electromagnetic fields \citep{bib:Sagdeev}.

From this discussion it is clear that a treatment of the jet/ISM
interaction needs to be established. For this purpose we need a
theoretical framework. The use of the Magneto-Hydro-Dynamic
equations (MHD) in this context is discarded by several arguments:
\begin{itemize}
\item The low collision rate cannot provide the equilibration of
ion and electron energies sufficiently fast for the plasma to behave
as a fluid. This is the case even for the "low energy" shocks
associated with supernova remnants
\citep{bib:draine1993,bib:vink2004}. Observations are consistent
with an energetically important population of accelerated particles
superimposed on a low energy background population
\cite{bib:gruzinov2001}.
\item MHD shocks are stable and do not generate magnetic fields. In MHD shocks, magnetic fields are only compressed, with a resulting
field strength that is orders of magnitudes smaller than what is
required by the synchrotron model of GRB afterglows
 \citep{bib:gruzinov2001}.
\end{itemize}

\section{From first principles}
The solutions that we are seeking fall into the kinetic and highly
non-linear regime of plasma physics. In this case, we need to work
from first principles, by solving the Maxwell equations with source
terms for the electromagnetic fields, together with the relativistic
equation of motion for charged particles
\begin{subeqnarray}
\nabla\cdot\vect{E}&=&\frac{\rho}{\epsilon_0}\\
\nabla\times\vect{B}-\frac{d\vect{E}}{c^2dt}&=&\mu_0\vect{J}\label{eq:maxwell}\\
-\nabla\times\vect{E}&=&\frac{d\vect{B}}{dt}\\
\nabla\cdot\vect{B}&=&0
\end{subeqnarray}
and
\begin{equation}
m\frac{d(\gamma\vect{v})}{dt}=q(\vect{E}+\vect{v}\times\vect{B})\label{eq:lorentz_pic}.
\end{equation}
Here $\epsilon_0$ and $\mu_0$ are the constants of electric
permittivity and magnetic permeability of vacuum, with
$c^2\mu_0\epsilon_0=1$. $m$ and $q$ are the mass and charge of a
particle of a given species, $\vect{v}$ is the velocity vector and
$\gamma\equiv(1-v^2/c^2)^{-1/2}$ is the relativistic Lorentz factor.
The source terms, $\vect{J}$ and $\rho$, in the Maxwell equations
are determined by the particles in the simulations.

\begin{figure}[!t]
\begin{center}
\epsfig{figure=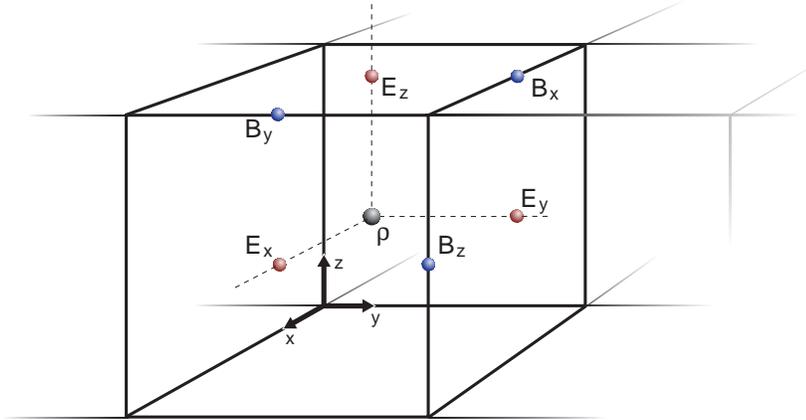,width=.8\textwidth} \caption[The
staggered mesh] {The staggered mesh. In a grid-cell, the particle
densities are cell centered, the electric field and current source
terms are face centered and the magnetic field components are edge
centered. Distributing the field variables on a staggered mesh gives
a factor 2 increase in resolution. In 3+1 dimensions this is a
factor 16 in computing time.} \label{fig:stagger}
\end{center}
\end{figure}

We wish to find a general solution to the coupled differential
equations of Eq.\ \ref{eq:maxwell} and Eq.\ \ref{eq:lorentz_pic}.
For a given set of initial/boundary conditions with, say, $10^{25}$
particles, this is not analytically possible. Numerically, however,
solving a scaled-down version of the same problem is doable, with
particle-in-cell (PIC) codes (e.g. \citec{bib:birdsall}). Much like
in a real plasma, a PIC code integrates the trajectories of a large
number of charged particles in both external and self-induced
electromagnetic fields. Some limitations exist in this approach.
Some of the main differences between a PIC simulated plasma and a
real plasma are:
\begin{itemize}
\item The number densities of real space are many orders of magnitudes larger than what can be fitted in a computer: A $10\times10\times10
\ \mathrm{m}^3$ cube of the ISM contains approximately $10^9$
charged particles, and even this is barely computational affordable
today. Therefore, in the simulations, each particle is a {\it
macro-particle} that represents a large number of real-plasma
charges. Each macro-particle keeps the same charge to mass ratio as
the individual particles it is made of.
\item Even though continuous in space and momentum space, the particles positions are discretized in time.
\item The electromagnetic fields are discretized is space as well as time. The Maxwell equations are integrated on a fixed numerical
grid and the interactions with the particles in a given grid cell
are done via interpolations from grid to particle positions and vice
versa (hence the name particle-in-cell). The electromagnetic field
components and source-terms are staggered and distributed on a 3D
Yee lattice \citep{bib:Yee}. This gives a resolution improvement
that corresponds to a factor 16 in computing time (Fig.\
\ref{fig:stagger}).

\item Many plasma processes evolve on time scales that are proportional to the plasma frequency $\tau\propto\omega_p^{-1}$ and on length scales
that are proportional to the skin depth $\delta\equiv c/\omega_p$.
Therefore, a large spatial and temporal span exists in plasma
processes that are dominated by respectively ions and electrons. To
comply with the limitations in computational resources it is
convenient to compress the dynamical ranges by reducing the ion
(proton) to electron mass ratio $m_i/m_e$ from the real value (1836)
to 15-30. This is clearly an approximation but we have performed
tests that have shown that the results show good convergence for
mass ratios above 15-30. For lower mass ratio, the results are still
qualitatively correct.

\item The maximum temporal and spatial scale-lengths in PIC simulations are limited because it is important to resolve microphysical
plasma oscillations. Since the electron plasma frequency
$\omega_{pe}$ is often the limiting factor, we normalize time with
respect to the oscillation period $\omega_{pe}^{-1}$ and the space
with respect to the electron skin depth $c/w_{pe}$. The plasma
frequency is defined as $\omega_{pe}\equiv(n_e
q^2/m_e\epsilon_0)^{-1/2}$, and thus the plasma density $n_e$
determines the re-scaling.
\end{itemize}

Despite these constraints, the PIC code representation of a plasma
is far more fundamental than the MHD approximation. Still, PIC
simulations are computationally challenging and fully
three-dimensional experiments have only become practically possible
within the last few years.

The PIC-code implementation I use is based on a non-relativistic
code developed by Dr. Michael Hesse. The code was initially
developed for simulating reconnection topologies in the context of
space weather \citep{bib:HesseKuzenova}. The code was later made
relativistic by \cite{bib:trier_master} as part of a masters
project. Since then it has been used mainly for numerical plasma
shock experiments related to GRB afterglows
\citep{bib:frederiksen2002,bib:frederiksen2004,bib:hededal2004,bib:hededal2005}.
As part of the current PhD project, radiative cooling has recently
been included to the code. This is important for investigation of
particle acceleration and generation of radiative spectra. A new PIC
code, which includes photons and several scattering mechanisms, is
under development (see Chapter \ref{chap:photonplasma}).

\section{Cooling of an accelerated charge}
In this section I derive and describe the implementation of
radiative cooling in the PIC-code. Radiative cooling is essential
for highly relativistic particle dynamics and especially for
experiments aimed at investigating particle acceleration. Before I
describe the derivation and implementation an expression for the
energy radiated from an accelerated charged particle into the
PIC-code, I briefly revive the concept of retarded time and space.

\subsection{Retarded time and position}\label{sec:retard} 
\begin{figure}[h]
\begin{center}
\epsfig{figure=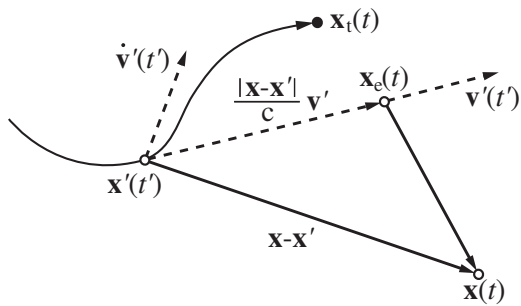,width=0.5\linewidth} \caption[Definition
of retarded time and position] {Definition of the retardation of a
particles position. From an observers point $\vect{r}$ we see the
particle at the position $\vect{r}_0(t')$ where it was at the {\it
retarded time}.} \label{fig:retarded}
\end{center}
\end{figure}

Let a particle be in the
position $\vect{r}_0(t)$ at time $t$ (Fig.\ \ref{fig:retarded}). At
the same time, we observe the electric field from the particle in
the position $\vect{r}$. However, because of the finite propagation
velocity of light, we observe the particle at an earlier position
$\vect{r}_0(t')$ where it was at the retarded time $t'=t-\delta
t'=t-R(t')/c$. Here $R(t')=\left|\vect{r}-\vect{r}_0(t')\right|$ is
the distance from the charge (at the retarded time $t'$) to the
observer point.

\subsection{Radiated power}
 To skip a trivial, but rather long derivation, we adopt the expression from
\cite{bib:jackson} for the retarded electric field from a charged
particle moving with instant velocity \vect{\beta} under
acceleration \vectdot{\beta},
\begin{equation}
\vect{E}=\underbrace{\frac{q}{4\pi\epsilon_0}\left[
\frac{\vect{n}-\vect{\beta}}{\gamma^2\left(1-\vect{n\cdot
\beta}\right)^3 R^2}\right]_\mathrm{ret}}_{\mathrm{velocity\ field}}
+\underbrace{\frac{q}{4\pi\epsilon_0 c}\left[\frac{\vect{n} \times
\{(\vect{n}-\vect{\beta})\times\vectdot{\beta}\}}
{\left(1-\vect{n\cdot\beta}\right)^3 R}\right]_\mathrm{ret}}_
{\mathrm{acceleration\ field}}\label{eq:e_rad_field}.
\end{equation}
Here, $\vect{n}\equiv\vect{R}(t')/\left|\vect{R}(t')\right|$ is a
unit vector that points from the particles retarded position towards
the observer.
 The first term on the right hand side, containing the velocity field, is
the Coulomb field from a charge moving without influence from
external forces. The second term is a correction term that arises if
the charge is subject to acceleration. Since the velocity-dependent
field is falling of as $R^{-2}$ while the acceleration-dependent
field falls off as $R^{-1}$, the latter becomes dominant when
observing the charge at large distances ($R\gg1$). This term is
therefore often referred to as the radiation term. The corresponding
magnetic field is given by
\begin{equation}
\vect{B}=\left[\frac{\vect{n}\times\vect{E}}{c}\right]_\mathrm{ret}
\end{equation}.

The energy $W$ per unit area $dA$ per unit time $dt$ that is
radiated from the accelerated particle is given by the Poynting flux
\vect{S}
\begin{equation}
\vect{S}\equiv\frac{d^2W}{dtdA}=\frac{\vect{E}\times\vect{B}}{\mu_0}=
\frac{\left|E\right|^2}{\mu_0c}\vect{n}
\end{equation}
from which we define the energy per unit time, received through a
unit solid angle element $d\Omega$ about \vect{n}
\begin{equation}
\df{P(t)}{\Omega}=R^2\left[\vect{S\cdot
n}\right]_\mathrm{ret}\label{eq:dpdomega}.
\end{equation}
A note of caution must be added here. The Poynting vector is related
to the observer time \var{t} but we are interested in the radiated
power measured at the particle's retarded time \var{t'}. Thus, to
get the total {\it emitted} power we multiply with the correction
term $dt/dt'$ (see Chapter \ref{sec:rad} for details)
\begin{eqnarray}
\df{P_{rad}}{\Omega}&=&R^2\left(\vect{S\cdot n}\right)\df{t}{t'}
=R^2\vect{S\cdot n}\left(1-\vect{n\cdot\beta}\right)\nonumber\\
&=&\frac{\mu_0 q^2 c}{16\pi^2}\frac{\left[\vect{n}\times
\left\{\left(\vect{n}-\vect{\beta}\right)\times\vectdot{\beta}
\right\}\right]^2}{\left(1-\vect{n\cdot\beta}\right)^5}\label{eq:P_rad-solid}.
\end{eqnarray}
Rather tedious vector algebra and integration over all directions
\vect{n} through the solid angle $d\Omega$ gives us the total power
radiated by the particle in a time interval $dt'$ (see Appendix
\ref{app:dpdomega_integration} for details)
\begin{subeqnarray}
P_{rad}&=&\frac{\mu_0q^2c}{6\pi}\gamma^4
\left(\dot{\beta}^2+\gamma^2
|\vect{\beta\cdot\dot{\beta}}|^2\right)\label{eq:p_rad_a}\\
&=&\frac{\mu_0q^2c}{6\pi}
\gamma^6\dot{\beta}^2\left(1-\beta^2\sin^2\theta\right)\label{eq:p_rad_b},
\end{subeqnarray}
where $\theta$ is the angle between the particles velocity vector
\vect{\beta} and its acceleration vector \vectdot{\beta}. When
$\sin\theta=1\Leftrightarrow\vect{\beta\perp\dot{\beta}}$, we
recognize the solution for synchrotron radiation (magnetic
bremsstrahlung). In the other limit,
$\sin\theta=0\Leftrightarrow\vect{\beta\parallel\dot{\beta}}$, we
recover the result for bremsstrahlung. The results above may be
found in many textbooks \citep{bib:landau,bib:jackson, bib:rybicki}.

In Chapter \ref{sec:rad} we deal with radiation from relativistic
particles in more details.

\subsection{Implementing the radiative cooling}
Before implementing Eq.\ \ref{eq:p_rad_a} into the PIC-code, it is
fruitful to derive the expression for a particles energy as it
looses momentum to radiation.

From energy conservation we demand that the energy radiated from a
particle must equally correspond to a loss in the particles kinetic
energy $E_{kin}=m c^2(\gamma-1)$
\begin{eqnarray}
\df{E_{kin}}{t}&=&-P_{rad}\nonumber\\
\Rightarrow\df{\gamma(t)}{t}&=&-\frac{\mu_0q^2}{6\pi m c}\gamma^6
\left(\dot{\beta}^2\left(1-\beta^2\right)+
|\vect{\beta\cdot\dot{\beta}}|^2\right)\label{eq:energyloss},
\end{eqnarray}
which is the relativistic counterpart to the Larmor formula for
radiated power. In order to continue we need an expression for
\vectdot{\beta}. The only force acting on the particle is the
Lorentz force (we deal with the radiation reaction force in Section
\ref{sect:reactionforce})
\begin{eqnarray}
m\df{(\gamma\vect{v})}{t}&=&q(\vect{E}+\vect{v}\times\vect{B})\\
\Rightarrow\gamma\df{\vect{v}}{t}+\frac{\vect{v}\gamma^3}{c^2}
\left(\vect{v\cdot}\df{\vect{v}}{t}\right)&=&\frac{q}{m}
(\vect{E}+\vect{v}\times\vect{B})\label{eq:lorentz}.
\end{eqnarray}
Here we have expanded the right hand side and used the relation
\begin{equation}
\df{\gamma}{t}=\df{(1-\vect{\beta\cdot \beta})^{-1/2}}{t}=
\frac{\gamma^3}{c^2}\vect{v\cdot}\df{\vect{v}}{t}.
\end{equation}

\begin{figure}[!t]
\begin{center}
\epsfig{figure=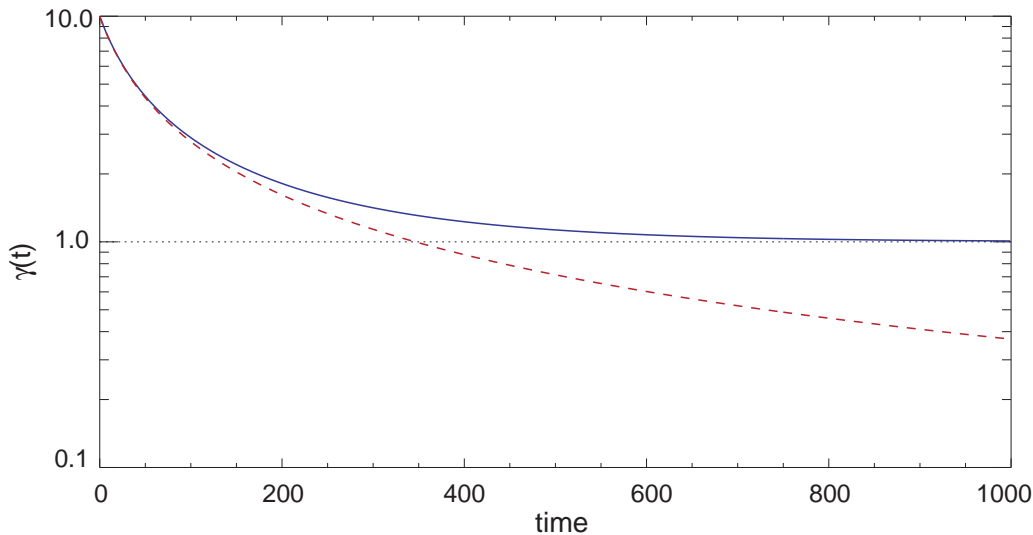,width=\textwidth}
\caption[$\gamma(t)$ under radiative cooling.] {The two versions of
$\gamma(t)$ -- Eq.\ \ref{eq:gamma_t} and Eq.\ \ref{eq:gamma_t_vc} --
compared. The main difference is that in deriving Eq.\
\ref{eq:gamma_t_vc} $v\sim c$ was assumed instead of the correct
$v=c\sqrt{1-\gamma^{-2}}$ in Eq.\ \ref{eq:gamma_t}. Thus, Eq.\
\ref{eq:gamma_t_vc} becomes incorrect as $\gamma(t)\to1$. The units
on the time axis is given in terms of the initial orbital period.}
\label{fig:gamma_t_compare}
\end{center}
\end{figure}

Unfortunately, we cannot provide a general solution to Eq.\
\ref{eq:energyloss} and Eq.\ \ref{eq:lorentz}. We may, however,
simplify the equation system by looking at a particle moving with
$\gamma(t=0)=\gamma_0$ at an angle $\alpha$ to a homogeneous
magnetic field ($\vect{E}=0$). In this case, the Lorentz force is
always perpendicular to the velocity vector and the term \vect{v
\cdot}\vectdot{v} vanishes. Equation \ref{eq:energyloss} and Eq.\
\ref{eq:lorentz} then reduce to
\begin{equation}
\df{\gamma(t)}{t}=\frac{\mu_0 q^4}{6\pi  m^3c}\sin^2(\alpha)
B^2(\gamma^2(t)-1) \label{eq:cool_diff_eq},
\end{equation}
where we have used that $v^2(t)=c^2(1-\gamma^{-2}(t))$. This
differential equation has the formal solution
\begin{equation}
\gamma(t)=\frac{2(\gamma_0-1)}
{(\gamma_0+1)\exp\left[{\frac{\mu_0q^4}{3\pi m^3
c}\sin^2(\alpha)B^2t}\right]-\gamma_0+1}\label{eq:gamma_t}.
\end{equation}
If we in Eq.\ \ref{eq:cool_diff_eq} assume that
$\gamma\gg1\Rightarrow v\sim c$, the solution becomes simpler, but
less valid for low $\gamma$ (see fig.\ \ref{fig:gamma_t_compare})
\begin{equation}
\gamma(t)=\frac{\gamma_0}{1+\frac{\mu_0q^4}{6\pi m^3
c}\sin^2(\alpha) B^2\gamma_0 t}\label{eq:gamma_t_vc}.
\end{equation}
In Section \ref{sect:reactionforce} I describe the implementation of
the radiative damping from Eq.\ \ref{eq:p_rad_a} into the PIC-code
and find good agreement with Eq.\ \ref{eq:gamma_t}.

Eq.\ \ref{eq:gamma_t_vc} gives us an indication of when momentum
losses to radiation becomes an important issue. If a physical
process of interest occurs over a time span $T$, radiation may be
neglected whenever
\begin{eqnarray}
\kappa T\ll1 \label{eq:negl_rad},
\end{eqnarray}
where
\begin{eqnarray}
\kappa\equiv\frac{\mu_0q^4}{6\pi m^3 c}\sin^2(\alpha)B^2\gamma_0.
\end{eqnarray}
We may also estimate the time $t_{1/2}$ it takes for a particle to
loose half of its energy
\begin{eqnarray}
m c^2(\gamma(t_{1/2})-1)=1/2 m c^2 (\gamma_0-1)\nonumber\\
\Rightarrow t_{1/2}=\frac{6\pi m^3
c}{\mu_0q^4B^2\gamma_0}\frac{\gamma_0-1}{\gamma_0+2} \
\mathop{\longrightarrow}_{\gamma_0\gg1}\ \frac{6\pi m^3 c}
{\mu_0q^4B^2\gamma_0}.
\end{eqnarray}
\subsection{Implementing the reaction force}\label{sect:reactionforce}
When a particle emits radiation, it looses both energy and momentum
to the emitted photons. Thus, in theory, we need to add an extra
force term in Eq.\ \ref{eq:lorentz}. However, as stated by
\cite{bib:jackson} {\it "a complete satisfactory classical treatment
of the reactive effects of radiation does not exist"} (and it is
noteworthy that Jackson only discuss the issue in the very last
Chapter). One approach is the {\it Abraham-Lorentz equation of
motion}
\begin{equation}
m\left(\df{(\gamma\vect{v})}{t}-\ddf{(\gamma\vect{v})}{t}\right)
=\vect{F}_{\mathrm Lorentz},
\end{equation}
but this solution is limited to periodic motions and have runaway
solutions. A consistent, but rather extensive, implementation of the
reaction force in a PIC code have been described by
\cite{bib:Noguchi2004}.

Instead we derive a more empirical solution to the problem that is
suitable for a numerical implementation.
\begin{figure}[!th]
\begin{center}
\epsfig{figure=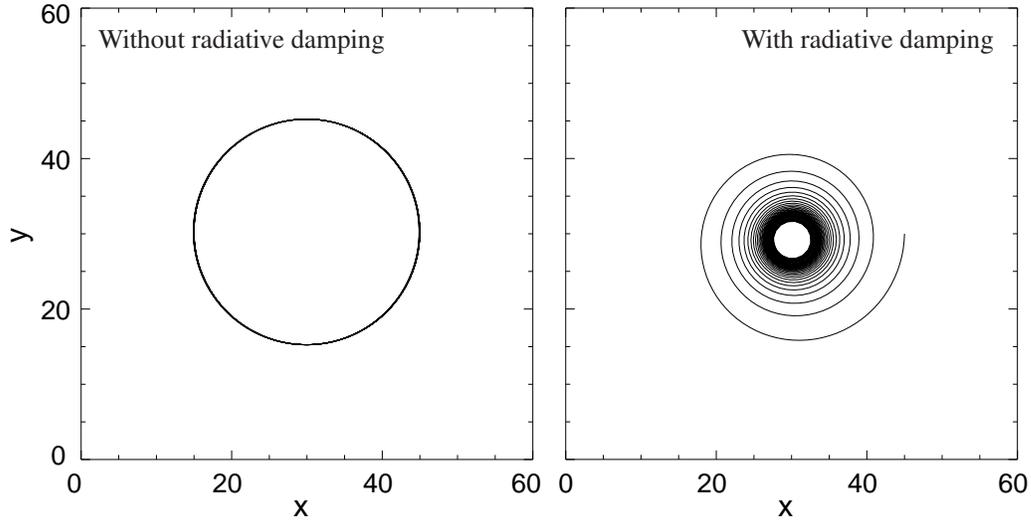,width=\textwidth} \caption[Radiative damping
of a relativistic particle in a homogeneous magnetic field] {The
trajectory of a relativistic particle in a transverse homogeneous
magnetic field. The particles motion is integrated by the PIC code.
The left panel shows the orbit without damping whereas the right
panel shows the orbit of the particle as it looses momentum to
radiation according to Eq.\ \ref{eq:u_rad}. Initially $\gamma_0=60$,
$\kappa=0.005$ and the orbit is integrated over over the time
interval $T=[0,1000]$.} \label{fig:damp_orbit}
\end{center}
\end{figure}
We recall the expression for the particles kinetic energy lost to
radiation
\begin{equation}
m c^2\df{(\gamma(t)-1)}{t}=-P_{rad}\label{eq:en_loss}.
\end{equation}
We are looking for a way to modify the particles momentum vector
$\vect{u}=\gamma\vect{v}$ that we may implement in the PIC-code. We
allow ourself to make a discrete representation of Eq.\
\ref{eq:en_loss}
\begin{eqnarray}
\frac{\tilde\gamma-\gamma}{\Delta t}&=&-\frac{P_{rad}}{m c^2}\nonumber\\
\Rightarrow\tilde\gamma&=&\gamma-\frac{P_{rad}}{m c^2}\Delta t,
\end{eqnarray}
where $\gamma$ is the particle Lorentz factor from the code and $\tilde\gamma$
is the radiation corrected Lorentz factor. The corresponding radiation
corrected momentum is $|\tilde{\vect{u}}|=c\sqrt{\tilde\gamma^2+1}$.

\begin{figure}[!h]
\begin{center}
\epsfig{figure=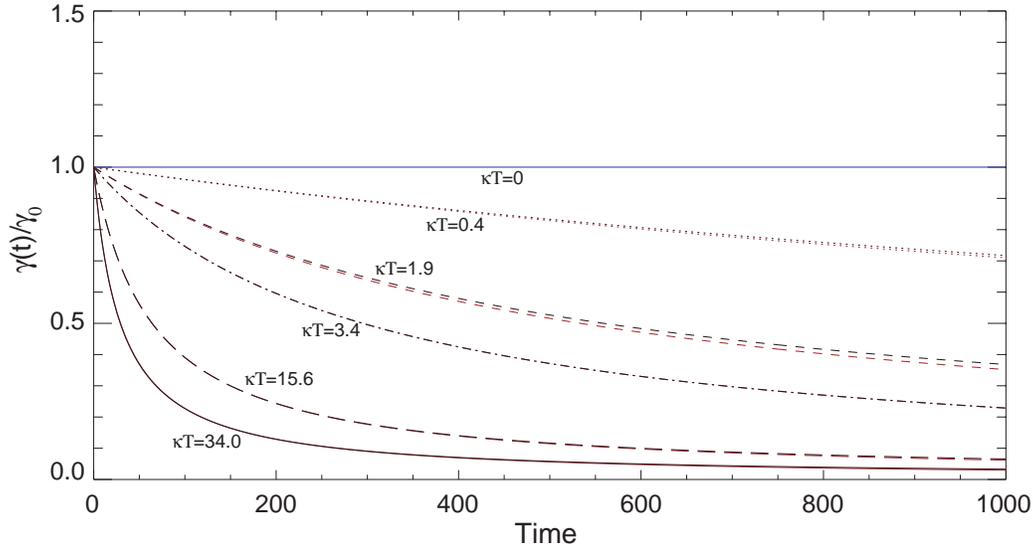,width=\textwidth} \caption[Cooling of
particles] {A comparison between cooling times in the theoretical
approach of Eq.\ \ref{eq:gamma_t} ({\it red lines}) and the PIC-code
implementation of Eq.\ \ref{eq:u_rad} ({\it black lines}). The small
discrepancies between the two approaches are mainly caused by
numerical integration inaccuracy. The time is in arbitrary units.}
\label{fig:cooling}
\end{center}
\end{figure}

We know that in an observers frame, the radiation from a
relativistic particle is concentrated around the direction of its
velocity vector (e.g.\ \citec{bib:landau}). Therefore, we can make
the assumption that the radiation reaction force is directed
opposite to the momentum vector. In this case, the change in
momentum occurs only along the momentum vector \vect{u} and we have
\begin{eqnarray}
\frac{\tilde{\vect{u}}}{|\tilde{\vect{u}}|}&=&\frac{\vect{u}}{|\vect{u}|}\nonumber\\
\Rightarrow\tilde{\vect{u}}&=&\vect{u}\frac{|\tilde{\vect{u}}|}{|\vect{u}|}
=\vect{u}\frac{\sqrt{[\gamma-P_{rad}\Delta t/(m c^2)]^2+1}}
{\sqrt{\gamma^2+1}}\label{eq:u_rad},
\end{eqnarray}
where $P_{rad}$ is given by Eq.\ \ref{eq:p_rad_a}. We have
implemented Eq.\ \ref{eq:u_rad} in the PIC code. To verify the
correctness of the method, we test against the result from the
analytical approach, Eq.\ \ref{eq:gamma_t}. We make a simulation run
with $\vect{E}=0$ and a homogeneous magnetic field $\vect{B}$. A
single particle is setup with a given velocity perpendicular to the
magnetic field ($\sin^2\alpha=1$). The integrated path of the
particle may be seen in Fig.\ \ref{fig:damp_orbit}. The figure shows
that as the particle cools to lower gamma it approaches a state
where cooling becomes negligible and the orbit becomes semi stable
($\kappa T\ll 1$).

A comparison of the PIC-code cooling rate against Eq.\
\ref{eq:gamma_t} may be found Fig.\ \ref{fig:cooling}. Five
different combinations of $\gamma_0$ and $B_0$  has been tested.
Good agreement is found between the two approaches. The minor
discrepancies that do appear are mainly caused by integrated
interpolation errors in the second order scheme used in the
PIC-code.

\chapter{The Weibel two-stream instability}\label{sec:weibel}



In this chapter I present results from three-dimensional particle
simulations of collisionless shock formation, with relativistic
counter-streaming ion-electron plasmas \citep{bib:frederiksen2004}.
Particles are followed over many skin depths downstream of the
shock.  Open boundaries allow the experiments to be continued for
several particle crossing times. The experiments confirm the
generation of strong magnetic and electric fields by a Weibel-like
kinetic streaming instability, and demonstrate that the
electromagnetic fields propagate far downstream of the shock.  The
magnetic fields are predominantly transversal, and are associated
with merging ion current channels.  The total magnetic energy grows
as the ion channels merge, and as the magnetic field patterns
propagate down stream.  The electron populations are quickly
thermalized, while the ion populations retain distinct bulk speeds
in shielded ion channels and thermalize much more slowly. The
results help us to reveal processes of importance in collisionless
shocks, and may help to explain the origin of the magnetic fields
responsible for afterglow synchrotron/jitter radiation from
Gamma-Ray Bursts.

\section{Introduction}

The existence of a strong magnetic field in the shocked external
medium is required in order to explain the observed radiation in
Gamma-Ray Burst afterglows as synchrotron radiation
\citep[e.g.][]{bib:Panaitescu+Kumar}.  Nearly collisionless shocks,
with synchrotron-type radiation present, are also common in many
other astrophysical contexts, such as in super-nova shocks, and in
jets from active galactic nuclei.  At least in the context of
Gamma-Ray Burst afterglows the observed synchrotron radiation
requires the presence of a stronger magnetic field than can easily
be explained by just compression of a magnetic field already present
in the external medium.

\citet{bib:medvedevloeb} showed through a linear kinetic treatment
how a two-stream magnetic instability -- a generalization of the
Weibel instability \citep{bib:Weibel,bib:YoonDavidson} -- can
generate a strong magnetic field  ($\epsilon_B$, defined as the
ratio of magnetic energy to total kinetic energy, is
$10^{-5}$-$10^{-1}$ of equipartition value) in collisionless shock
fronts \citep[see also discussion in][]{bib:RossiRees}. We note in
passing that this instability is well-known in other plasma physics
disciplines, e.g. laser-plasma interactions
\citep{bib:YangGallantAronsLangdon,bib:califano1}, and has been
applied in the context of pulsar winds by \citet{bib:Kazimura}.

Using three-dimensional particle-in-cell simulations to study
relativistic collisionless shocks (where an external plasma impacts
the shock region with a bulk Lorentz factor $\Gamma = 5-10$),
\cite{bib:frederiksen2002}, \cite{bib:nishikawa}, and
\cite{bib:silva} investigated the generation  of magnetic fields by
the two-stream instability. In these first studies the growth of the
transverse scales of the magnetic field was limited by the small
sizes of the computational domains.  The durations of the
\cite{bib:nishikawa} experiments were less than particle travel
times through the experiments, while \cite{bib:silva} used periodic
boundary conditions in the direction of streaming. Further,
\cite{bib:frederiksen2002} and \cite{bib:nishikawa} used
electron-ion ($e^-p$) plasmas, while experiments reported upon by
\cite{bib:silva} were done with $e^-e^+$ pair plasmas.

Here, we report on 3D particle-in-cell simulations of
relativistically counter-streaming $e^-p$ plasmas. Open boundaries
are used in the streaming direction, and experiment durations are
several particle crossing times. Our results can help to reveal the
most important processes in collisionless shocks, and help to
explain the observed afterglow synchrotron radiation from Gamma-Ray
Bursts. We focus on the earliest development in shock formation and
field generation. Late stages in shock formation will be addressed
in successive work.

\section{Simulations}
Experiments were performed using a self-consistent 3D3V (three
spatial and three velocity dimensions) electromagnetic
particle-in-cell code originally developed for simulating
reconnection topologies \citep{bib:HesseKuzenova}, redeveloped by
the present authors to obey special relativity and to be second
order accurate in both space and time.

The code solves Maxwell's equations for the electromagnetic field
with continuous sources, with fields and field source terms defined
on a staggered 3D Yee-lattice \citep{bib:Yee}. The sources in
Maxwell's equations are formed by weighted averaging of particle
data to the field grid, using quadratic spline interpolation.
Particle velocities and positions are defined in continuous
(${\bf{r}},\gamma{\bf{v}}$)-space, and particles obey the
relativistic equations of motion.

The grid size used in the main experiment was
$(x,y,z)=200\times200\times800$, with 25 particles per cell, for a
total of $8\times10^8$ particles, with ion to electron mass ratio
$m_{i}/m_{e} = 16$. To adequately resolve a significant number of
electron and ion skin-depths ($\delta_e$ and $\delta_i$), the box
size was chosen such that $L_{x,y} = 10\delta_i \sim 40\delta_e$ and
$L_z \sim 40 \delta_i \sim 160\delta_e$. Varying aspect and mass
ratios were used in complementary experiments.

\begin{figure*}[!th]
\begin{center}
\epsfig{figure=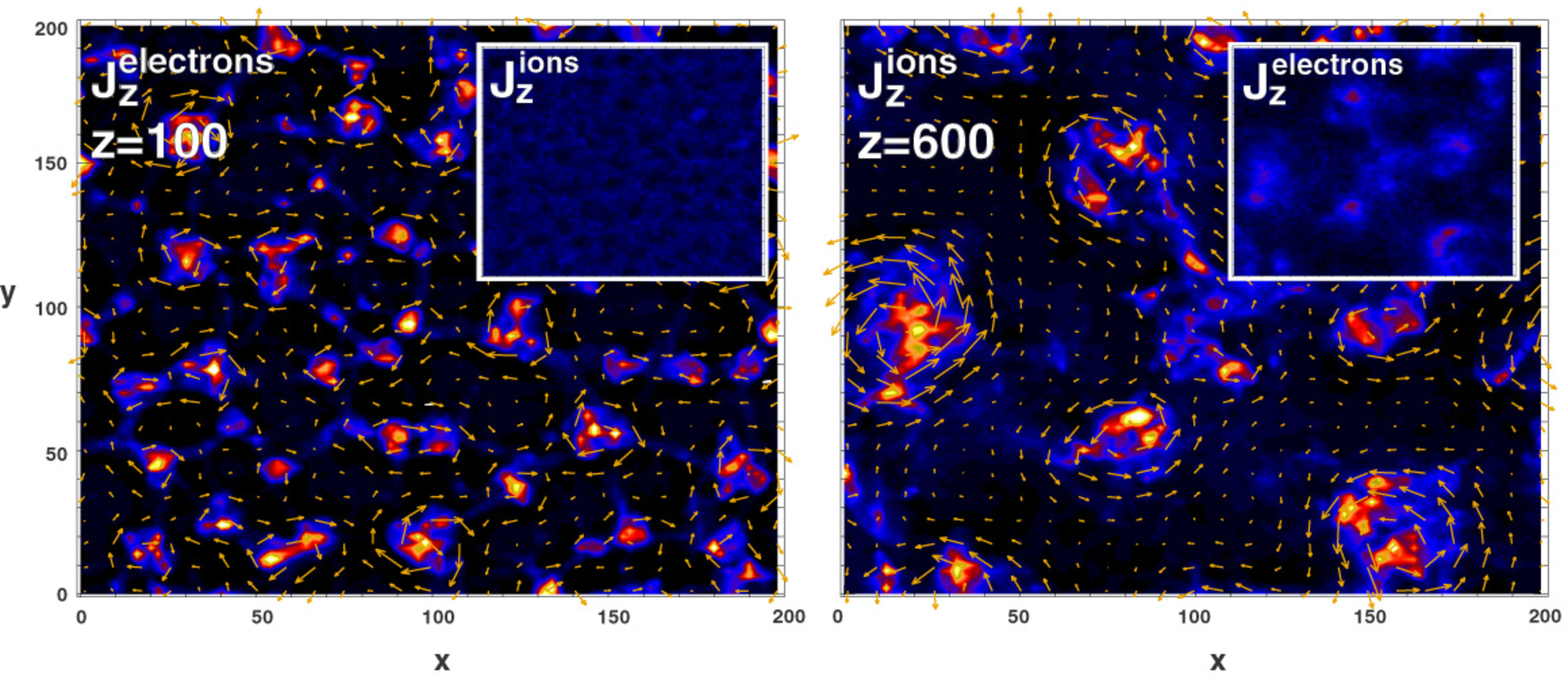,width=\textwidth} \caption[Contour plot
of electron and ion current density in the two-stream region]{The
left hand side panel shows the longitudinal electron current density
through a transverse cut at $z=100$, with a small inset showing the
ion current in the same plane.  The right hand side panel shows the
ion current at $z=600=30\delta_i$, with the small inset now instead
showing the electron current. The arrows represent the transverse
magnetic field. Both panels are from time $t =1200$
($240\omega_{pe}^{-1}$).}

\label{fig:Slice}
\end{center}
\end{figure*}

Two counter-streaming -- initially quasi-neutral and cold -- plasma
populations are simulated. At the two-stream interface (smoothed
around $z=80$) a plasma ($z<80$) streaming in the positive
z-direction, with a bulk Lorentz factor $\Gamma=3$, hits another
plasma ($z\ge80$) at rest in our reference frame. The latter plasma
is denser than the former by a factor of 3. Experiments have been
run with both initially sharp and initially smooth transitions, with
essentially the same results. The long simulation time allows the
shock to gradually converge towards self-consistent jump conditions.
Periodic boundaries are imposed in the $x$-- and $y$--directions,
while the boundaries at $z=0$ and $z=800$ are open, with layers
absorbing transverse electromagnetic waves.  Inflow conditions at
$z=0$ are fixed, with incoming particles supplied at a constant rate
and with uniform speed.  At $z=800$ there is free outflow of
particles. The maximum experiment duration is 480 $\omega_{pe}^{-1}$
(where $\omega_{pe}$ is the electron plasma frequency), sufficient
for propagating $\Gamma \approx 3$ particles 2.8 times through the
box.

\section{Results and Discussions}

The extended size and duration of these experiments make it possible
to follow the two-stream instability through several stages of
development; first exponential growth, then non-linear saturation,
followed by pattern growth and downstream advection.  We identify
the mechanisms responsible for these stages below.

\subsection{Magnetic Field Generation, Pattern Growth \\ and Field Transport} \label{field_generation}
Encountering the shock front the incoming electrons are rapidly
(being lighter than the ions) deflected by field fluctuations
growing as a result of the two-stream instability
\citep{bib:medvedevloeb}. The initial perturbations grow non-linear
as the deflected electrons collect into first caustic surfaces and
then current channels (\fig{fig:Slice}). Both streaming and rest
frame electrons are deflected, by arguments of symmetry.

In accordance with Ampere's law the current channels are surrounded
by approximately cylindrical magnetic fields (illustrated by arrows
in \fig{fig:Slice}), causing mutual attraction between the current
channels.  The current channels thus merge in a race where larger
electron channels consume smaller, neighbouring channels. In this
manner, the transverse magnetic field grows in strength and scale
downstream. This continues until the fields grow strong enough to
deflect the much heavier ions into the magnetic voids between the
electron channels. The ion channels are then subjected to the same
growth mechanism as the electrons. When ion channels grow
sufficiently powerful, they begin to experience Debye shielding by
the electrons, which by then have been significantly heated by
scattering on the growing electromagnetic field structures. The two
electron populations, initially separated in $\gamma{\bf{v}}$-space,
merge to a single population in approximately $20\delta_e$
($z=80$--$200$) as seen in \fig{fig:acc}. The same trend is seen for
the ions -- albeit at a rate slower in proportion to $m_i/m_e$.

\begin{figure}[!t]
\begin{center}
\epsfig{figure=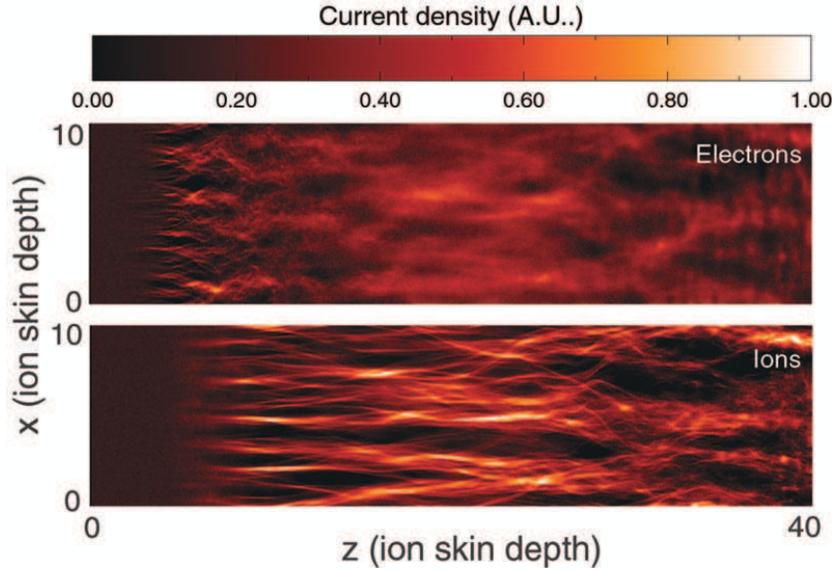,width=.8\textwidth} \caption[Current
filaments in the shock region.]{ Electron (top) and ion (bottom)
currents, averaged over the $y$-direction, at time $t=1200$
($240\omega_{pe}^{-1}$). However, due to the finite propagation time
of the particles, these two figures can be interpreted as earlier
and later times.} \label{fig:jiz}
\end{center}
\end{figure}
\begin{figure}[!ht]
\begin{center}
\epsfig{figure=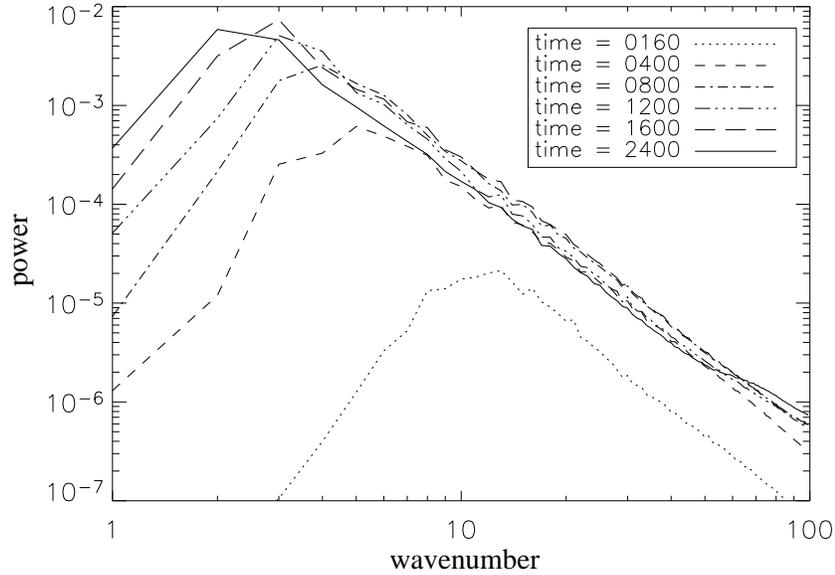,width=12cm} \caption[Power spectrum of
the magnetic field generated in a $\Gamma=3$ collisionless
shock]{Power spectrum of ${\mathbf B}_{\perp}$ for $z = 250$ at
different times.} \label{fig:power}
\end{center}
\end{figure}

The Debye shielding quenches the electron channels, while at the
same time supporting the ion-channels; the large random velocities
of the electron population allow the concentrated ion channels to
keep sustaining strong magnetic fields. Figure \ref{fig:Slice} shows
the highly concentrated ion currents, the more diffuse -- and
shielding -- electron currents, and the resulting magnetic field.
The electron and ion channels are further illustrated in
\fig{fig:jiz}. Note the limited $z$-extent of the electron current
channels, while the ion current channels extend throughout the
length of the box, merging to form larger scales downstream. Because
of the longitudinal current channels the magnetic field is
predominantly transversal; we find $|B_z|/|B_{tot}| \sim 10^{-1} -
10^{-2}$.

\begin{figure}[!t]
\begin{center}
\epsfig{figure=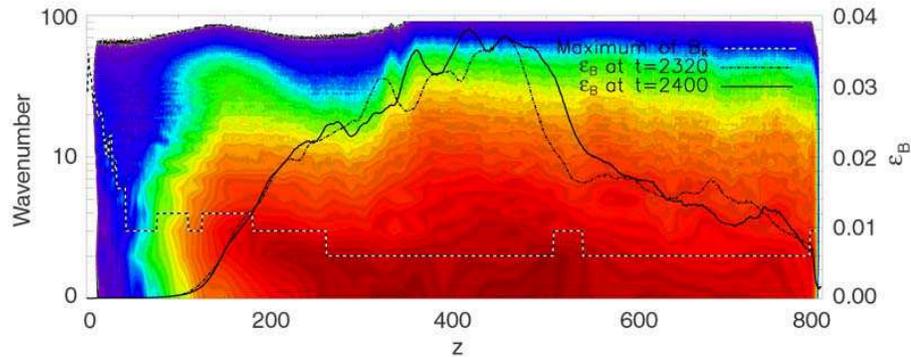,width=12cm} \caption[$\epsilon_B$ and
function of shock depth in a $\Gamma=3$ shock]{Relative
electromagnetic energy density $\epsilon_{B}$. The contour color
plot shows the power in the transverse magnetic field through the
box distributed on spatial Fourier modes at $t=2400$
($480\omega_{pe}^{-1}$), with the dotted line marking the wavenumber
with maximum power. Superposed is the spatial distribution of
$\epsilon_{B}$, averaged across the beam, at $t=2320$
($464\omega_{pe}^{-1}$) (dashed-dotted) and $t=2400$
($240\omega_{pe}^{-1}$)(full drawn), highlighting how EM-fields are
advected down through the box. } \label{fig:epsb}
\end{center}
\end{figure}

\Fig{fig:power} shows the temporal development of the transverse
magnetic field scales around $z=250$. The power spectra follow
power-laws, with the largest scales growing with time. The dominant
scales at these $z$ are of the order $\delta_i$ at early times.
Later they become comparable to $L_{x,y}$. \Fig{fig:epsb} captures
this scaling behavior as a function of depth for $t=2400$
($480\omega_{pe}^{-1}$).

\begin{figure}[!t]
\begin{center}
\epsfig{figure=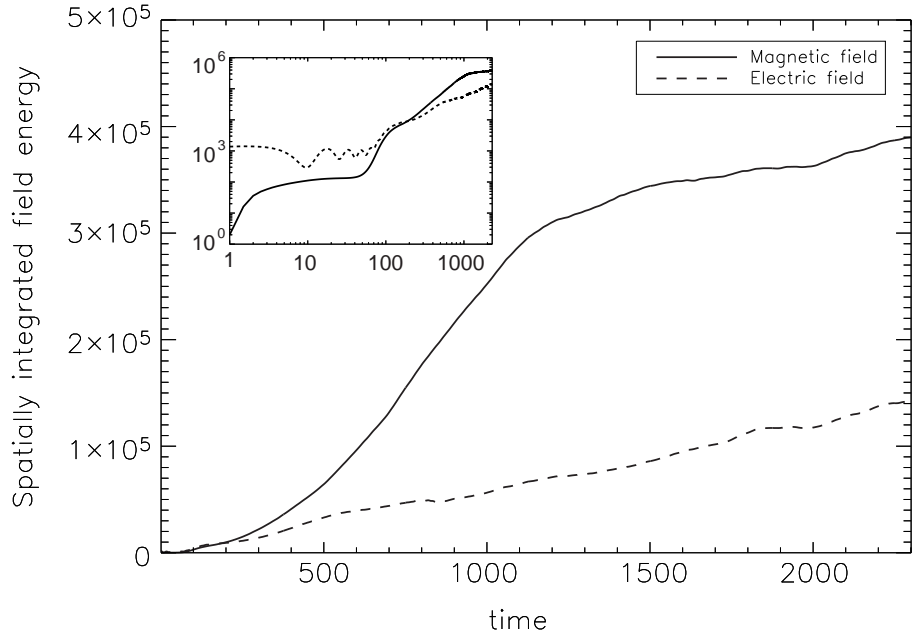,width=12cm} \caption[The total energy in
magnetic and electic field as function of time]{ Total magnetic
(full drawn) and electric (dashed) energy in the box as a function
of time. The inset shows a log-log plot of the same data. }
\label{fig:B_energy}
\end{center}
\end{figure}

The time evolutions of the electric and magnetic field energies are
shown in \fig{fig:B_energy}. Seeded by fluctuations in the fields,
mass and charge density,  the two-stream instability initially grows
super-linearly ($t=80-100$ = $16-20\omega_{pe}^{-1}$), reflecting
approximate exponential growth in a small sub-volume. Subsequently
the total magnetic energy grows more linearly, reflecting
essentially the increasing volume filling factor as the non-linearly
saturated magnetic field structures are advected downstream.

At $t\approx 1100$ the slope drops off, because of advection of the
generated fields out of  the box.  The continued slow growth, for $t
> 1100$ ($220\omega_{pe}^{-1}$), reflects the increase of the pattern size with time (cf.\
\fig{fig:power}). A larger pattern size corresponds to, on the
average, a larger mean magnetic energy, since the total electric
current is split up into fewer but stronger ion current channels.
The magnetic energy scales with the square of the electric current,
which in turn grows in inverse proportion to the number of current
channels.  The net effect is that the mean magnetic energy increases
accordingly.

The magnetic energy density keeps growing throughout our experiment,
even though the duration of the experiment (480 $\omega_{pe}^{-1}$)
significantly exceeds the particle crossing time, and also exceeds
the advection time of the magnetic field structures through the box.
This is in contrast to the results reported by \citet{bib:silva},
where the magnetic energy density decays after about 10-30
$\omega_{pe}^{-1}$. It is indeed obvious from the preceding
discussion that the ion-electron asymmetry is essential for the
survival of the current channels.

From the requirement that the total plasma momentum should be
conserved, the (electro)magnetic field produced by the two-stream
instability acquires part of the z-momentum lost by the two-stream
population in the shock; this opens the possibility that magnetic
field structures created in the shock migrate downstream of the
shock and thus carry away some of the momentum impinging on the
shock.

Our experiments show that this does indeed happen; the continuous
injection of momentum transports the generated field structures
downstream at an accelerated advection speed. The dragging of field
structures through the dense plasma acts as to transfer momentum
between the in-streaming and the shocked plasmas.

\subsection{Thermalization and Plasma Heating}
\begin{figure}[!t]
\begin{center}
\epsfig{figure=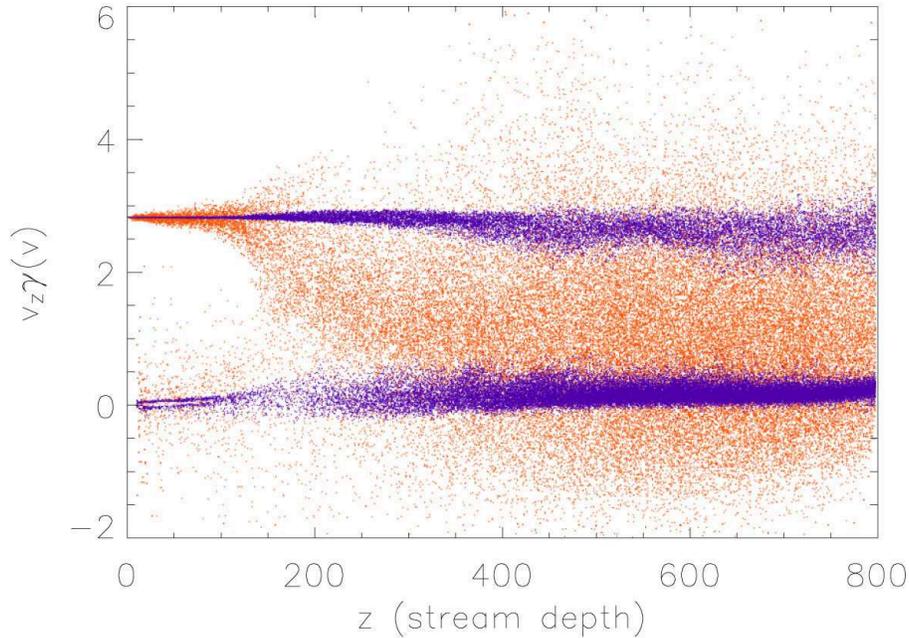,width=12cm}
 \caption[Scatter plot of electrons and ions in a $\Gamma=3$ shock]{Thermalization and longitudinal acceleration, illustrated
by scatter plots of the electron (orange) and ion (blue)
populations. Note the back-scattered electron population
($v_z\gamma(v) < 0$). } \label{fig:acc}
\end{center}
\end{figure}

At late times the entering electrons are effectively scattered and
thermalized: The magnetic field isotropizes the velocity
distribution whereas the electric field generated by the
$e^{-}$--$p$ charge separation acts to thermalize the populations.
\Fig{fig:acc} shows that this happens over the $\sim$ 20 electron
skin depths from around $z=80$ -- $200$. The ions are expected to
also thermalize, given sufficient space and time. This fact leaves
the massive ion bulk momentum constituting a vast energy reservoir
for further electron heating and acceleration. Also seen in
\fig{fig:acc}, the ions beams stay clearly separated in phase space,
and are only slowly broadened (and heated).

We do not see indications of a super-thermal tail in the heated
electron distributions, and there is thus no sign of second order
Fermi-acceleration in the experiment presented in this chapter.
\cite{bib:nishikawa} and \cite{bib:silva} reported acceleration of
particles in experiments similar to the current experiment, except
for more limited sizes and durations, and the use of an $e^-e^+$
plasma \citep{bib:silva}. On closer examination of the published
results it appears that there is no actual disagreement regarding
the absence of accelerated particles. Whence, \cite{bib:nishikawa}
refer to transversal velocities of the order of $0.2 c$ (their Fig.\
3b), at a time where our experiment shows similar transversal
velocities (cf. \fig{fig:acc}) that later develop a purely thermal
spectrum. \cite{bib:silva} refer to transversal velocity amplitudes
up to about $0.8 c$ (their Fig.\ 4), or $v\gamma\sim 2$, with a
shape of the distribution function that appears to be compatible
with thermal. In comparison, the electron distribution illustrated
by the scatter plot in \fig{fig:acc} covers a similar interval of
$v\gamma$, with distribution functions that are close to
(Lorentz-boosted) relativistic Maxwellians.  Thus, there is so far
no compelling evidence for non-thermal particle acceleration in
experiments with no imposed external magnetic field. Thermalization
is a more likely cause of the increases in transversal velocities.

\cite{bib:frederiksen2002} reported evidence for particle
acceleration, with electron gamma up to $\sim100$, in experiments
with an external magnetic field present in the up-stream plasma.
This is indeed a more promising scenario for particle acceleration
experiments \citep[although in the experiments by][results with an
external magnetic field were similar to those
without]{bib:nishikawa}. \Fig{fig:acc} shows the presence of a
population of back-scattered electrons ($v_z\gamma < 0$).  In the
presence of an external magnetic field in the in-streaming plasma,
this possibly facilitates Fermi acceleration in the shock.

\section{Conclusions}

The experiment reported upon here illustrates a number of
fundamental properties of relativistic, collisionless shocks:

1. Even in the absence of a magnetic field in the up-stream plasma,
a small scale, fluctuating, and predominantly transversal magnetic
field is unavoidably generated by a two-stream instability
reminiscent of the Weibel-instability. In the current experiment the
magnetic energy density reaches a few percent of the energy density
of the in-coming beam.

2. In the case of an $e^-p$ plasma the electrons are rapidly
thermalized, while the ions form current channels that are the
sources of deeply penetrating magnetic field structures.  The
channels merge in the downstream direction, with a corresponding
increase of the average magnetic energy with shock depth.  This is
expected to continue as long as a surplus of bulk relative momentum
remains in the counter-streaming plasmas.

3. The generated magnetic field patterns are advected downstream at
speeds intermediate between the in-coming plasma and the rest frame
plasma. The electromagnetic field structures thus provide scattering
centers that interact with both the fast, in-coming plasma, and with
the plasma that is initially at rest.  As a result the electron
populations of both components quickly thermalize and form a single,
Lorentz-boosted thermal electron population.  The two ion
populations merge much more slowly, with only gradually increasing
ion temperatures.

4. The observed strong turbulence in the field structures at the
shocked streaming interface provides a promising environment for
particle acceleration.

We emphasize that quantification of the interdependence and
development of $\epsilon_U$ and $\epsilon_B$ is accessible by means
of such experiments as reported upon here.

Rather than devising abstract scalar parameters $\epsilon_B$ and
$\epsilon_U$, that may be expected to depend on shock depth, media
densities etc., a better approach is to compute synthetic radiation
spectra directly from the models (see Chapter \ref{sec:rad}), and
then apply scaling laws to predict what would be observed from
corresponding, real supernova remnants and Gamma-Ray Burst afterglow
shocks.

\chapter{Large-scale Two-Dimensional Simulations}\label{sec:2d3d}
\section{Introduction}
Large-scale simulations of collisionless electron-proton plasma
shocks that are covering the full shock ramp are crucial for our
interpretation of the radiation that we receive from relativistic
jets (e.g.\ gamma-ray bursts). It is important to understand the
nature and interdependencies of the shock "parameters" $\epsilon_e$
and $\epsilon_b$. The complexity and non-linearity of collisionless
shocks give PIC code simulations a central role in the process of
seeking further progress in our understanding of these shocks.

 Even though the continuously increasing computational
power has reached a level where we may now begin to explore the
full, three-dimensional collisionless shock problem, it is still
only barely possible. The 3D simulations that were presented in
Chapter \ref{sec:weibel} take from weeks to months to run on a
modern supercomputer (parallized using 8-32 processors and 100 Gb of
memory). Still it is not possible to resolve the full shock
transitions region for an electron-proton collisionless plasma
shock. Some of the great unanswered questions that cannot be
targeted yet, because of this limitation in computer power, are for
example: How large do the structures that are created by merging of
current filaments grow in the non-linear phase of the Weibel
two-stream instability? How many ion skin depths is the shock
transition region? What is the fate of the generated magnetic
structures in the downstream region where the ions are thermalized
and the instability has saturated? What is the ion and electron
distribution function behind the shock ramp? Does the Fermi
acceleration mechanism work in collisionless shocks? Etc.

\section{Simulations and results}

To probe for answers to some of these questions, it is tempting to
return to only two spatial dimensions. To test whether 2D
simulations can account for the Weibel two-stream instability, we
have made a 2D3V (two spatial and three velocity dimensions --
sometimes referred to as $2\frac{1}{2}$D) repetition of the
simulation described in Chapter \ref{sec:weibel}. We use the exact
same plasma densities and velocities as in the 3D experiment: In the
shock rest frame, ISM particles are injected with $\Gamma=3$ into a
plasma, initially at rest, with electron plasma frequency
$\omega_{pe}=0.2\ c/\Delta$ (simulation units). The major difference
is that the simulation box is two-dimensional with $800\times4000$
grid zones instead of the 3D $200\times200\times800$. The 2D box
size corresponds to $40\times200$ ion skin depths. Compared to the
3D box, the extra grid zones in the $z$-direction is used to cover
320 grid zones upstream of the 3D box and and 2880 grid zones
downstream of the 3D box.
\begin{figure}[htb]
\begin{center}
\epsfig{figure=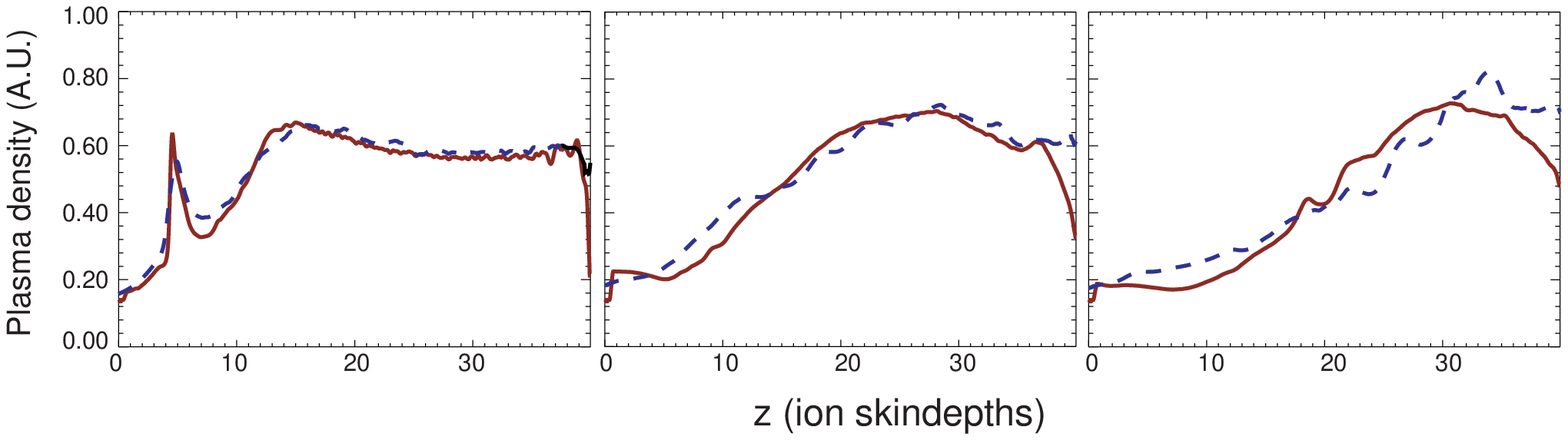,width=\textwidth}
\epsfig{figure=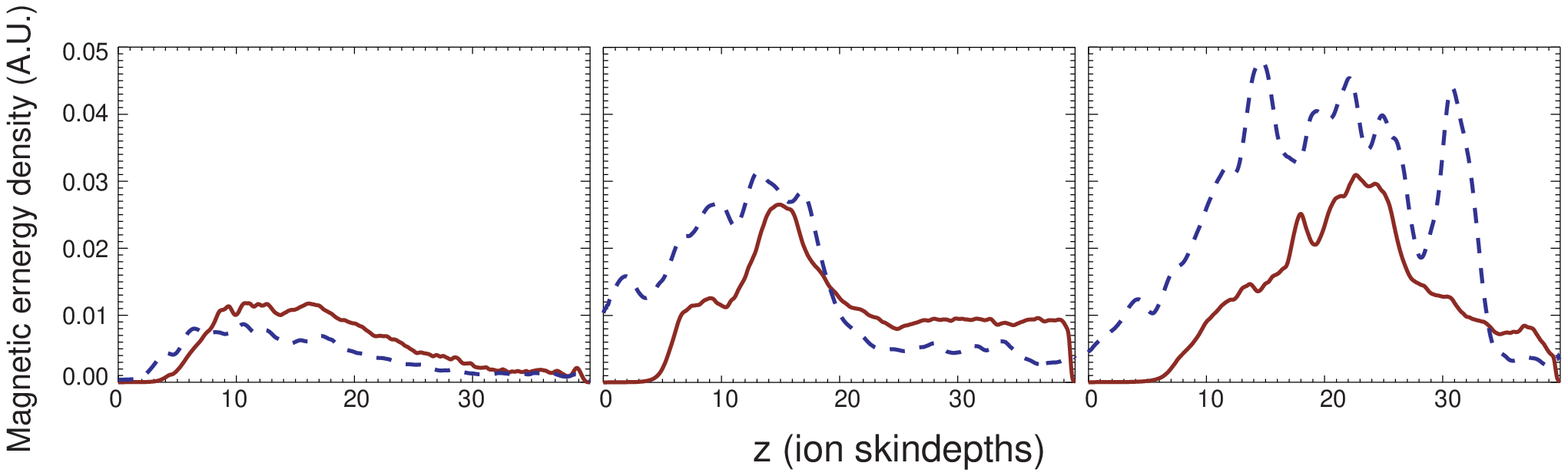,width=\textwidth}
 \caption[Density and magnetic energy in 2D and 3D simulations]{A comparison of
 density ({\it top})
 and magnetic field generation ({\it bottom}) between 2D ({\it dashed blue}) and 3D ({\it solid red}).
The three columns corresponds to snapshots taken at time $t=40$,
$80$ and $120\ \omega_{pi}^{-1}$.} \label{fig:2ddensity}
\end{center}
\end{figure}

Figure \ref{fig:2ddensity} shows the evolution of the plasma density
({\it top panel}) and magnetic energy ({\it bottom panel}) in the 2D
and 3D simulation. A good agreement between the two simulations is
found, at least with respect to the evolution of the density. The
minor differences are mainly caused by the artificial reflection of
thermal particles at the leftmost boundary, and the drain of
particles at the rightmost boundary in the 3D simulations. For the
generation of magnetic field, the situation is somewhat different.
In the very early stage, the field generated in the 3D simulation is
stronger than in 2D, because of the extra transverse dimension that
the 3D instability can collect particles from. However, at later
times the growth in magnetic field in the 2D run exceeds the 3D
case. This is primarily caused by two effects: 1) in the 2D box,
upscattered particles (defined as shocked particles that are
traveling upstream into the ISM) can generate a seed field further
upstream. Such a seed field enhances the growth of the instability.
This mechanism cannot be followed (yet) in the smaller 3D box. 2)
ion current channels can merge to larger transverse structures.
These also cannot be followed in the smaller 3D box. To a first
approximation, the two-dimensional simulation shows a promising
resemblance with the three-dimensional simulations, and potentially
can surpass them in some respects. Encouraged by these results, we
have continued the large-scale 2D simulations in order to follow the
long time evolution ($t_{max}=400\omega_{pi}^{-1}$). This duration
corresponds to two light crossing times of this larger box.

\begin{figure}[htb]
\begin{center}
\epsfig{figure=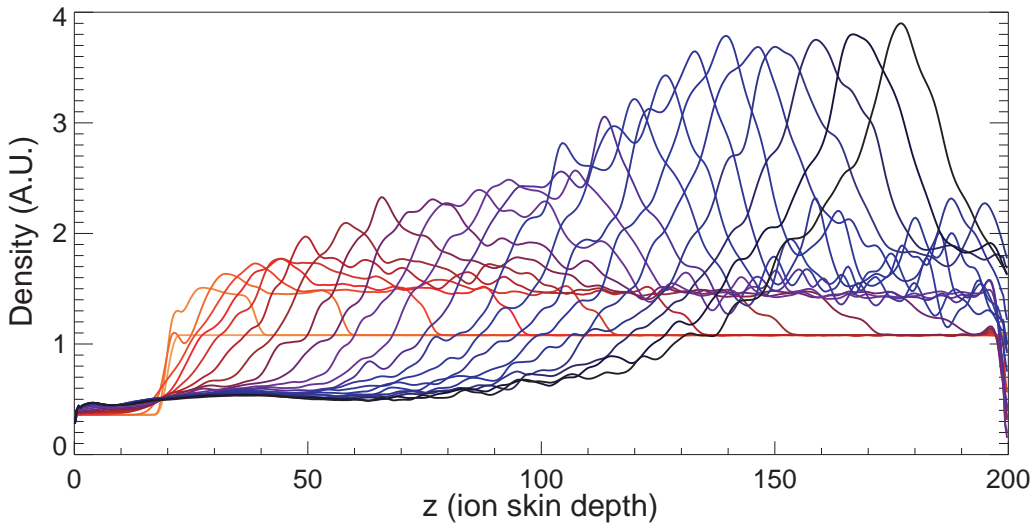,width=\textwidth}
 \caption[Evolution of the density profile from 2D simulations.]{
 The density profile at different time, sampled in the time range
 $t=0-400\omega_{pi}^{-1}$ with an interval of $20\omega_{pi}^{-1}$.
 The different time snapshots are color coded with red for early times through blue to black at late times.}
 \label{fig:2ddensity_plot}
\end{center}
\end{figure}

The evolution of the plasma density as function of shock depth is
shown in Fig.\ \ref{fig:2ddensity_plot} for different times. The
time-range is $t=0-400\omega_{pi}^{-1}$ with an interval of
$20\omega_{pi}^{-1}$.
 It is immediately clear that the initial setup is not consistent with
a steady state solution. This is not surprising, since initially
there are no electromagnetic fields to facilitate the momentum
transfer between the injected and quiescent material. As the
simulation evolves, the shock converges toward the real physical
solution. In three-dimensional simulations performed by us and other
groups, it has been neither temporally nor spatially possible to
follow the evolution to this stage. However, the 2D simulations that
we have performed cover a much larger range in space and time.

\begin{figure}[htb]
\begin{center}
\epsfig{figure=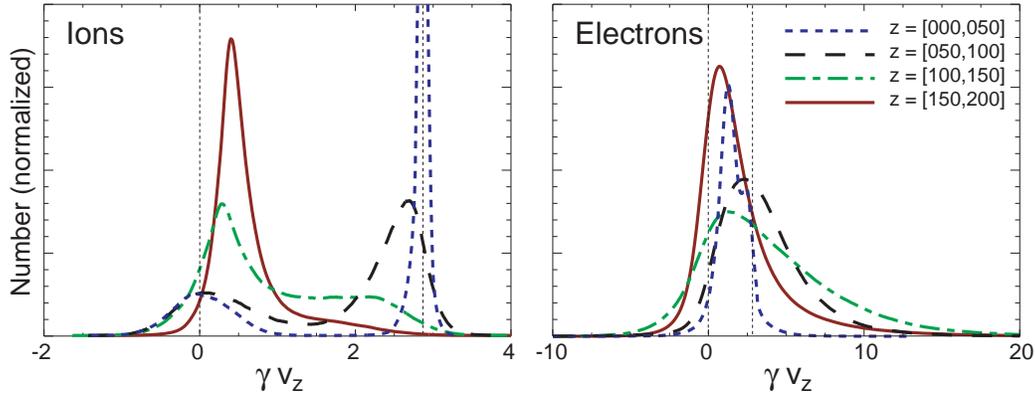,width=\textwidth}
 \caption[Ion and electron z-momentum distribution functions in a collisionless shock]{
The ion ({\it left}) and electron ({\it right}) distribution
functions of the momentum ($\gamma v_z$) along
 the shock propagation-direction. The particles are sampled at $t=400
 \omega_{pi}^{-1}$
 in four different segments of the shock ramp:
 $z=[0,50]$ ({\it blue, dotted}),
 $z=[50,100]$ ({\it black, dashed}),
 $z=[100,150]$ ({\it green, dot-dashed}) and
 $z=[150,200]$ ({\it red, solid}).
The (thin black dotted) vertical lines
 indicate $\gamma v_z=0$ and the injection momentum (corresponding to
 $\gamma=3$).
 Note that in the last segment ($z=[150,200]$ ion skin depths)
  the ions are very close to being one single population. This indicates that the full shock ramp is included in the box.
  } \label{fig:2dddist}
\end{center}
\end{figure}

At late times, the shock reaches a steady state where the value of
the main density peak is constant in time. The whole shock structure
is propagating with $0.35c$ in the downstream direction. Such a
steady state is expected when the computational domain covers the
full shock transition ramp. In this case, the injected ISM plasma
should have thermalized and merged with the shock plasma. Thus, only
a single plasma population should be present in the downstream
region. This is indeed the case, as seen in Fig.\ \ref{fig:2dddist}.
The figure shows the ion ({\it left}) and electron ({\it right})
distribution functions of the momentum along the shock propagation
direction ($\gamma v_z$), at time $t=400\omega_{pe}^{-1}$. The four
different curves correspond to different regions of the
computational domain ( $z=[0,50]$ , $z=[50,100]$, $z=[100,150]$, and
 $z=[150,200]$ ion skin depths). In the shock ramp ($z=[0,150]$ ion skin
 depths),
 where thermalization of the two ion populations (ISM and shocked) has not occurred yet,
 two distinct peaks are seen.
 In the last segment ($z=[150,200]$ ion skin depths), there exist only one single
 peak.
This means that the ions are fully thermalized. It is also
noteworthy that the merged ion population translates downstream with
$\gamma v_z\simeq0.4$, consistent with the estimate found from
figure \ref{fig:2ddensity_plot}. Even though the peak moves
relatively slowly, the particles that define the peak are moving
relativistically and are continuously getting replaced with "fresh"
particles.

\begin{figure}[htb]
\begin{center}
\epsfig{figure=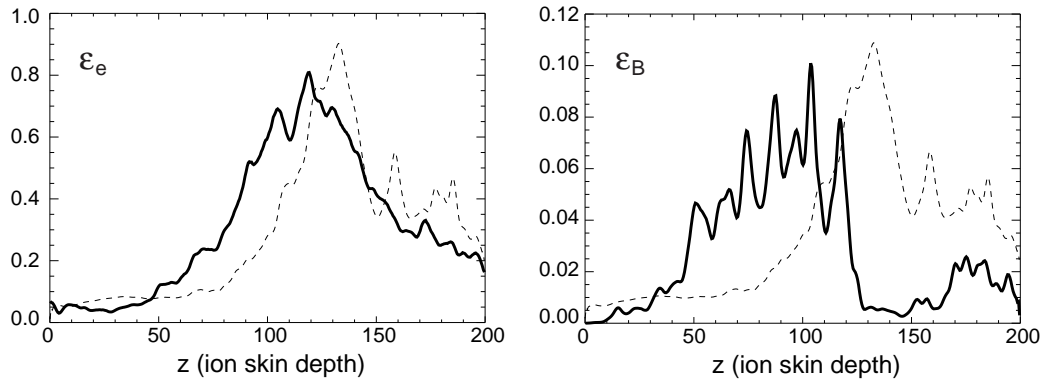,width=\textwidth}
 \caption[Electron and magnetic equipartition values ($\epsilon_e$ and $\epsilon_B$)]{Electron ({\it left}) and
 magnetic ({\it right}) equipartition values ($\epsilon_e$ and $\epsilon_B$) at time $t=300\omega_{pi}^{-1}$. The dotted line in both plots
 shows
 the shape of the total plasma density for comparison. The electrons reach up to 80\% of equipartition and the magnetic field up to 10\%.} \label{fig:2dequipart}
\end{center}
\end{figure}

Two shock parameters of great interest are the equipartition
parameters $\epsilon_e$ and $\epsilon_B$. They describe the amount
of energy that is deposited in the electrons and in the magnetic
field, relative to the kinetic energy of the ions. These parameters
are normally treated as free parameters, available to cover our
general ignorance of the details of the microphysics in
collisionless shocks. Fig.\ \ref{fig:2dequipart} shows $\epsilon_e$
({\it{}left}) and $\epsilon_B$ ({\it{}right}) as functions of the
shock depth. It is found that the electrons reach close to
equipartition with the ions. This, combined with the arguments
above, allows us to conclude that the simulations cover the full
shock. In the downstream region, equipartition has not yet been
reached. This is because since everything moves with velocities
close to the speed of light, we see earlier stages in the evolution
the further downstream we look. The $\epsilon_B$ plot shows us that
most of the field generation takes place in the foreshock. This is
not surprising, since the streaming instability in its nature
requires counter-streaming populations (anisotropy). Downstream, the
plasma populations are fully thermalized into on single population.

\begin{figure}[htb]
\begin{center}
\epsfig{figure=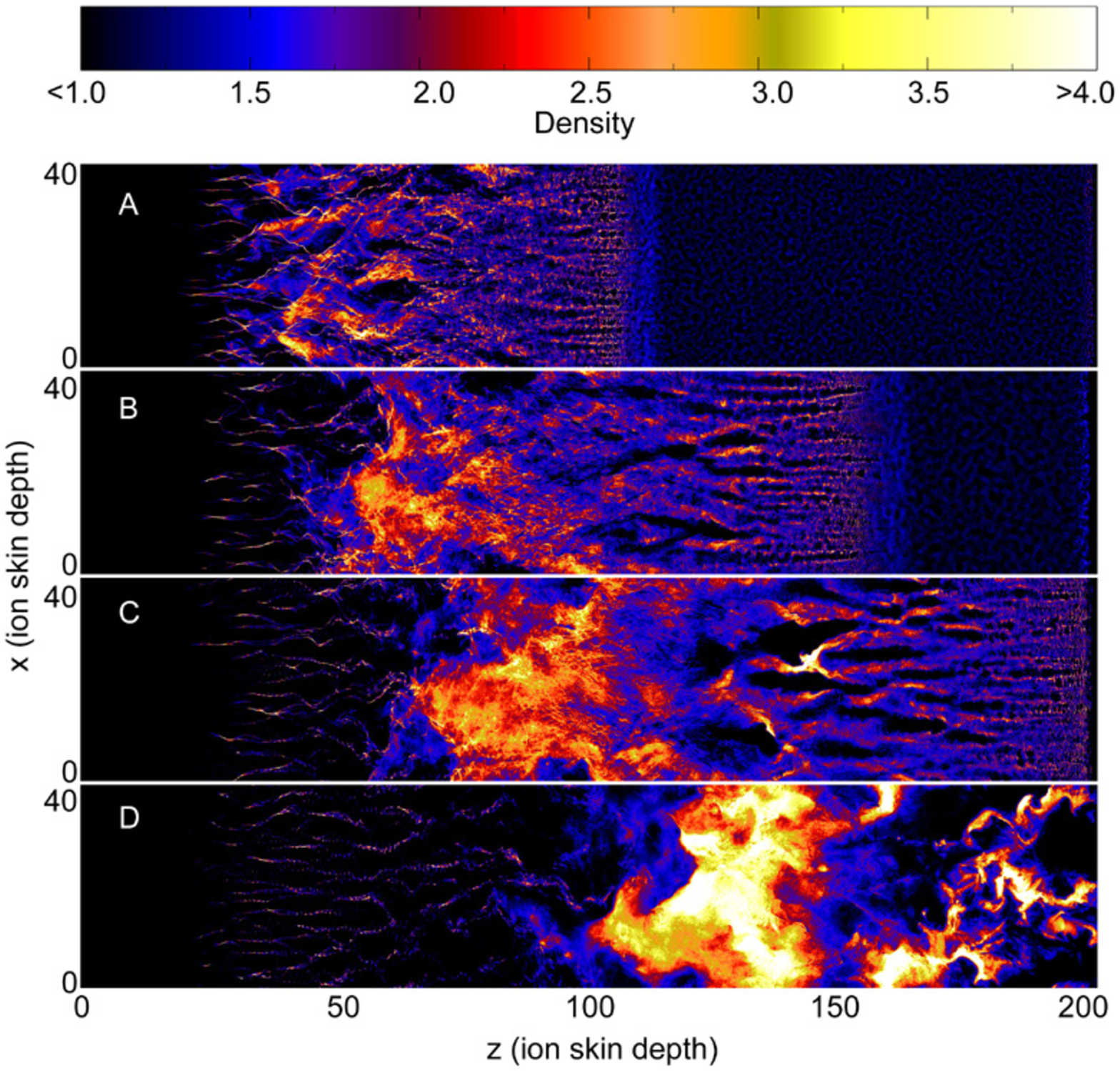,width=\textwidth}
 \caption[Density contour plot of the 2D Weibel instability.]{
 Plasma density contour plots at four different time levels A, B, C and D
 corresponding to $t=100,\ 150,\ 200\ \mathrm{and}\ 300\ \omega_{pi}^{-1}$} \label{fig:2ddens_cont}
\end{center}
\end{figure}

Finally, to give an impression of the extremely complex and
non-linear nature of the collisionless shock, we present contour
plots of the large-scale evolution of the shock in Figure
\ref{fig:2ddens_cont}. The figure shows the ion density structures
at four different time-steps ($t=100,150,200,$ and
$300\omega_{pi}^{-1}$). At early times we see that the Weibel
two-stream instability is occurring semi-symmetrically at the
contact discontinuities both upstream and downstream.

\section{Summary}
Even with the computational power available on supercomputers today,
it is not possible to make three-dimensional PIC simulations that
 cover a full collisionless shock, while also resolving the
microphysics. This has lead us to investigate whether
two-dimensional simulations can include the physical processes that
dominate these shocks (mainly the non-linear Weibel instability). We
find that this is the case as far as 3D results exist to compare
with. The density profile and generation of magnetic fields are
quite similar in the two cases. Encouraged by this, we have made
large-scale 2D PIC simulations of the formation of a collisionless
shock, for an ejecta propagating with a bulk lorentz factor
$\Gamma=3$. The simulation box covers $40\times200$ ion skin depths
(compared to the $10\times10\times40$ in the 3D case by
\citec{bib:frederiksen2004}, and $4\times4\times15$ in simulations
by \citec{bib:nishikawa2004}).

In the results we find that the 2D simulations cover a spatial
region just large enough for the ISM plasma (injected into a plasma
in the shock rest-frame) to thermalize and merge with the shocked
plasma. Even though the initial conditions are not setup correctly,
the system converges to a quasi-stationary state where at least
three different regions are identified: 1) An upstream foreshock
with great anisotropy, consisting of in-streaming ISM plasma and
upscattered shock particles. In this region, strong magnetic field
generation is taking place with $\epsilon_B=5-10\%$. 2) A dense, hot
thermalization region where the electrons and ions are close to
equipartition ($\epsilon_e=50-80\%$). In this region, the magnetic
field is relatively weak ($\epsilon_B\simeq1\%$) but still strong
enough to account for most GRB afterglow estimates. 3) A hot
downstream region where the magnetic field is of the order percents
of equipartition. This region is clumpy, with structure sizes of the
order $1-20$ ion skin depths (see Fig.\ \ref{fig:2ddens_cont}).

In these 2D simulations it is found that with an initial condition
where a dense plasma is at rest in the simulation frame, the
resulting shock rest frame was propagating with $0.35c$ in the
downstream direction. Based on these results, we plan to rerun the
simulations in a frame where the dense population initially travels
with $0.35c$ in the upstream direction. This would potentially allow
us to follow the shock for longer periods of time.

Finally we note that the upscattering of particles might be
important for Fermi acceleration. It is likely that these particles
can be deflected back into the shock region. To test whether Fermi
acceleration is taking place requires even larger simulations and
the question can only be answered by future work.

\chapter{Shocks in Ambient Magnetic Fields}\label{sec:bperp}

We now continue with an investigation of the Weibel instability in
collisionless shocks in the presence of an ambient magnetic field
\citep{bib:hededal2005}. This scenario is important especially in
the internal shock phase and in the very early afterglow, where it
is possible that a magnetic field, carried from the GRB progenitor,
still exist in the ejecta.

Plasma outflows from gamma-ray bursts, supernovae, and relativistic
jets, in general interact with the surrounding medium through
collisionless shocks. The microphysical details of such shocks are
still poorly understood, which, potentially, can introduce
uncertainties in the interpretation of observations. It is now well
established that the Weibel two-stream instability is capable of
generating strong electromagnetic fields in the transition region
between the jet and the ambient plasma. However, the parameter space
of collisionless shocks is vast and still remains unexplored. In
this chapter, we focus on how an ambient magnetic field affects the
evolution of the electron Weibel instability and the associated
shock. Using a particle-in-cell code, we have performed
three-dimensional numerical experiments on such shocks. We compare
simulations in which a jet is injected into an unmagnetized plasma
with simulations in which the jet is injected into a plasma with an
ambient magnetic field both parallel and perpendicular to the jet
flow. We find that there exists a threshold of the magnetic field
strengths, below which the Weibel two-stream instability dominates,
and we note that the interstellar medium magnetic field strength
lies well below this value. In the case of a strong magnetic field
parallel to the jet, the Weibel instability is quenched. In the
strong perpendicular case, ambient and jet electrons are strongly
accelerated because of the charge separation between deflected jet
electrons and less deflected jet ions. Also, the electromagnetic
topologies become highly non-linear and complex with the appearance
of anti-parallel field configurations.

\section{Introduction}
The collisionless plasma condition applies to many astrophysical
scenarios, including, the outflow from gamma-ray bursts (GRBs),
active galactic nuclei, and relativistic jets in general. The
complexity of kinetic effects and instabilities makes it difficult
to understand the nature of collisionless shocks. Only recently, the
increase in available computational power has made it possible to
investigate the full three-dimensional dynamics of collisionless
shocks.

In the context of GRB afterglows, observations indicate that
shock-compressed magnetic field from the interstellar medium (ISM)
is several orders of magnitude too weak to match observations.
Particle-in-cell (PIC) simulations have revealed that the Weibel
two-stream instability is capable of generating the required
electromagnetic field strength of the order of percents of
equipartition value \citep{bib:Kazimura, bib:medvedevloeb,
bib:nishikawa, bib:nishikawa2004,  bib:silva, bib:frederiksen2004}.
Furthermore, PIC simulations have shown that in situ non thermal
particle acceleration takes place in the shock transition region
\citep{bib:hoshino2002, bib:hededal2004, bib:saito2003}.
Three-dimensional simulations using $\sim 10^7$ electron-positron
pairs by \cite{bib:sakai} showed how complex magnetic topologies are
formed when injecting a mildly relativistic jet into a force-free
magnetic field with both parallel and perpendicular components. With
a two-dimensional analysis, \cite{bib:saito2003} found that an
ambient parallel magnetic field can quench the two-stream
instability in the weakly relativistic case. In this chapter, we use
three-dimensional PIC experiments to investigate how the two-stream
instability is affected by the presence of an ambient magnetic
field. Using up to $\sim10^9$ particles and $125\times125\times1200$
grid zones, we investigate the development of complex magnetic
topologies when injecting a fully relativistic jet (bulk Lorentz
factor $\Gamma=5$) into an ambient magnetized plasma. Using varying
field strengths, we focus on the case of a transverse magnetic field
and compare it with the case of a parallel magnetic field. The
simulations are mainly concerned with the electron dynamics since
processes involving the heavier ions evolve on much longer
timescales.

\section{The numerical experiments}
\begin{figure}[!ht]
\begin{center}
\epsfig{figure=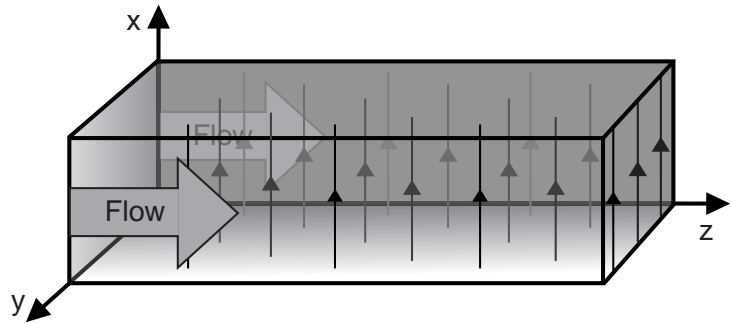,width=0.7\textwidth} \caption[Schematic
figure of the simulation setup for shocks with ambient magnetic
field ]{Schematic example of the simulation setup. Jet plasma is
homogeneously and continuously injected in the $z$-direction
throughout the $x-y$ plane at $z=0$. Inside, the box is populated by
a plasma population, initially at rest. In this specific example, an
ambient magnetic field is set up in the $x$-direction (perpendicular
case).} \label{fig:setup}
\end{center}
\end{figure}

We use the PIC code described by \cite{bib:frederiksen2004}. The
code works from first principles and evolves the equation of motion
for the particles together with Maxwell's equations.
In the simulation experiments, we inject an electron-proton plasma
(a jet) into an ambient plasma (the ISM) initially at rest (Fig.\
\ref{fig:setup}). The jet is moving with a relativistic velocity of
$0.98c$ along the $z$-direction, corresponding to Lorentz factor
$\gamma_{jet}=5$. The ion-to-electron mass ratio is set to $m_{\rm
i}/m_{\rm e}=20$. The jet plasma and the ambient plasma have the
same density, $n$, and the corresponding electron plasma rest-frame
frequency $\omega_{pe}\equiv [ne^2/(m_e\epsilon_0)]
=0.035\Delta_t^{-1}$ ($e$ is the unit charge, $\epsilon_0$ the
vacuum permittivity, and $\Delta_t$ the simulation time unit). We
choose this low value in order to properly resolve the microphysics.
Initially, the interface between the ambient and the injected plasma
is located at $z=3\lambda_e$, where $\lambda_e$ is the electron skin
depth defined as $\lambda_e\equiv c/\omega_{pe}=28.6\Delta_x$ ($c$
is the speed of light, and $\Delta_x$ the grid size). The time step
and grid size obey the Courant-Friedrichs-Levy condition
$\Delta_t=0.5\Delta_x/c$. Both plasma populations are, in their
respective rest frames, Maxwellian distributed with a thermal
electron velocity $v_{th}\simeq0.03c$. This temperature allows us to
numerically resolve the plasma Debye length with at least one grid
length.

We consider three different ambient magnetic configurations: no
magnetic field, a magnetic field parallel to the flow, and a
magnetic field perpendicular to the flow. The magnetic field is
initially setup to be homogeneous and at rest in the ambient plasma.
The experiments are carried out with 1 billion particles inside
$125\times125\times1200$ grid zones. In terms of electron skin
depths, this corresponds to $4.4\times 4.4\times 42 \lambda_e$. The
boundary conditions are periodic in the direction transverse to the
jet flow ($x, y$). In the parallel direction, jet particles are
continuously injected at the leftmost boundary ($z=0$). At the
leftmost and rightmost $z$ boundary, electromagnetic waves are
absorbed, and we allow particles to escape in order to avoid
unphysical feedback. The total energy throughout the simulations is
conserved with an error less than $1\%$.

\section{Results}
Initially, we ran simulations with no ambient magnetic field and
observed the growth of the Weibel two-stream instability also found
in previous work \citep{bib:Kazimura, bib:medvedevloeb,
bib:nishikawa,bib:nishikawa2004, bib:silva, bib:frederiksen2004}.
The Weibel two-stream instability works when magnetic perturbations
transverse to the flow collect streaming particles into current
bundles that in turn amplify the magnetic perturbations. In the
non-linear stage, we observe how current filaments merge into
increasingly larger patterns. The electromagnetic energy grows to
$\epsilon_B\simeq1\%$, where $\epsilon_B$ describes the amount of
total injected kinetic energy that is converted to magnetic energy.

\subsection{Parallel Magnetic Field}
\begin{figure}[!ht]
\begin{center}
\epsfig{figure=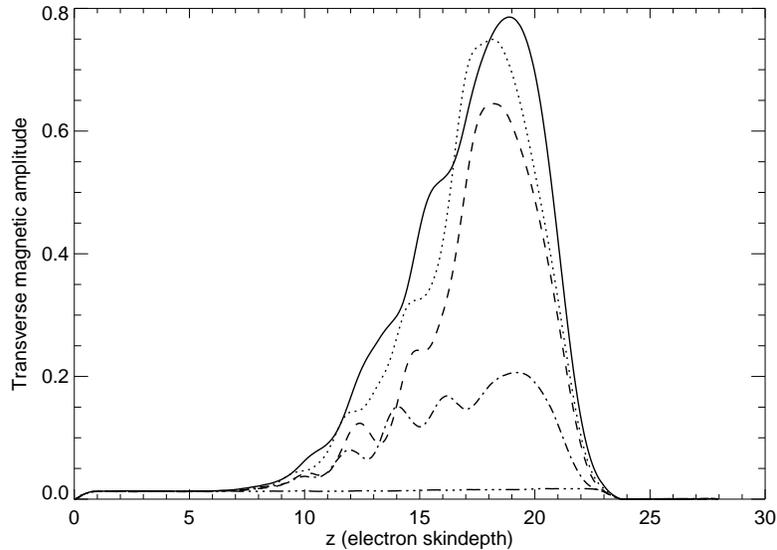,width=0.8\textwidth} \caption[Growth of
the Weibel instability in the presence of a parallel ambient
magnetic field]{ Growth of the Weibel two-stream instability for
different strengths of the parallel ambient magnetic field at time
$t = 21\omega_{pe}^{-1}$. Here we show the effectiveness of the
field generation, as measured by the average transverse magnetic
field amplitude as a function of $z$. The solid line corresponds to
$\omega_{pe}/\omega_{ce}=40$,
 the dotted line to 20, the dashed line to 10, the dot-dashed line
to 5, and the triple-dot-dashed line to 1. The case of no ambient
magnetic field is very similar to that of
$\omega_{pe}/\omega_{ce}=40$. The magnetic field amplitude is in
arbitrary units} \label{fig:b_par}
\end{center}
\end{figure}

In the presence of a strong magnetic field component parallel to the
flow, particles are not able to collect into bundles, since
transverse velocity components are deflected. We have performed five
runs, with parallel magnetic fields corresponding, respectively, to
$\omega_{pe}/\omega_{ce}= $40, 20, 10, 5, and 1, while keeping
$\omega_{pe}$ constant; $\omega_{ce}=eB/(\gamma_{jet} m_e)$ is the
jet electron gyrofrequency. The resulting field generation
efficiency can be seen in Fig.\ \ref{fig:b_par} at $t =
21\omega_{pe}^{-1}$ where the jet front has reached $z =
23\lambda_{e}$. In the case of $\omega_{pe}/\omega_{ce}=40$, the
Weibel instability overcomes the parallel field, and although
initially slightly suppressed, it eventually evolves as in the case
of no ambient magnetic field. Increasing the magnetic field to
$\omega_{pe}/\omega_{ce}=1$ effectively suppresses the instability.
Thus, for an ISM strength magnetic field
($\omega_{pe}/\omega_{ce}\simeq1500$) parallel to the plasma flow,
the Weibel two-stream instability evolves unhindered, and the
generated field will exceed the ISM field. We find from the
simulations that it would take a milligauss strength parallel
magnetic field to effectively quench the instability for a
$\gamma=5$ jet expanding in an environment with density similar to
the ISM.

\begin{figure*}[!ht]
\begin{center}
\epsfig{figure=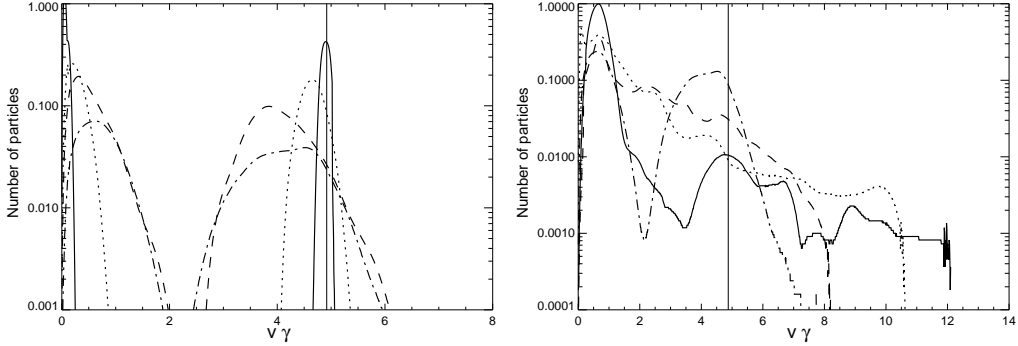,width=\textwidth} \caption[Electron
momentum distribution functions for shocks with parallel and
perpendicular ambient magnetic field]{Normalized electron momentum
distribution functions at time $30 \omega_{pe}^{-1}$. The left panel
is for runs with an ambient magnetic field parallel to the injected
plasma with $\omega_{pe}/\omega_{ce}=$ 1 ({\em solid line}), 20
({\em dotted line}), 40 ({\em dashed line}), and 1500 ({\em
dot-dashed line}). The right panel is for runs in which the initial
magnetic field is perpendicular to the inflow and
$\omega_{pe}/\omega_{ce}=$ 5 ({\em solid line}), 20 ({\em dotted
line}), 40 ({\em dashed line}) and 1500 ({\em dot-dashed line}). The
vertical line shows the injected momentum $\gamma=5$. The
distribution functions are for electrons with $z>15\lambda_e$.}
\label{fig:pdf}
\end{center}
\end{figure*}

The left panel of Fig.\ \ref{fig:pdf} shows the resulting electron
momentum distribution function for different values of
$\omega_{pe}/\omega_{ce}$
 at
$t=30\omega_{pe}^{-1}$. Since the presence of a strong parallel
magnetic field suppresses the generation of a transverse magnetic
field, there exists no mechanism that can heat the electrons and
transfer momentum between the two electron populations. Thus the jet
plasma propagates unperturbed. Where there is no parallel magnetic
field or only a weak magnetic field
($\omega_{pe}/\omega_{ce}=1500$), we observe how the jet and ambient
plasma is heated and how momentum is transferred between the two
populations.

\subsection{Perpendicular Magnetic Field}
We have performed experiments with an ambient magnetic field
perpendicular to the jet flow (Fig.\ \ref{fig:setup}), with field
strengths corresponding to $\omega_{pe}/\omega_{ce}$=1500, 40, 20
and 5. By including the displacement current, one can derive the
relativistic Alfv\'en speed
$v_A^{-2}=c^{-2}+(v_{A}^{non-rel.})^{-2}\simeq
c/[1+(\omega_{pe}/\omega_{ce})^2 (m_i/m_e)\gamma_{jet}^{-2}]^{1/2}$,
where $v_{A}^{non-rel.}=B/[\mu_0 n (m_i+m_e)]^{1/2}$ is the
non-relativistic counterpart. From this we calculate the
corresponding relativistic Alf\'ven Mach numbers
$\gamma_{jet}v_{jet}/v_A=$ 6572, 175, 88, and 22.

Again, the $\omega_{pe}/\omega_{ce}=1500$ run has been chosen
because it resembles the typical density and microgauss magnetic
field strength of the ISM. We find that the magnetic field generated
by the two-stream instability dominates the ambient magnetic field,
and the result resembles the case with no ambient magnetic field.
Furthermore, as expected in both the parallel and perpendicular
cases, the electron momentum distributions (Fig.\ \ref{fig:pdf}) are
very similar, except for a weak merging between the ambient and jet
electrons in the perpendicular case.
\begin{figure}[!ht]
\begin{center}
\epsfig{figure=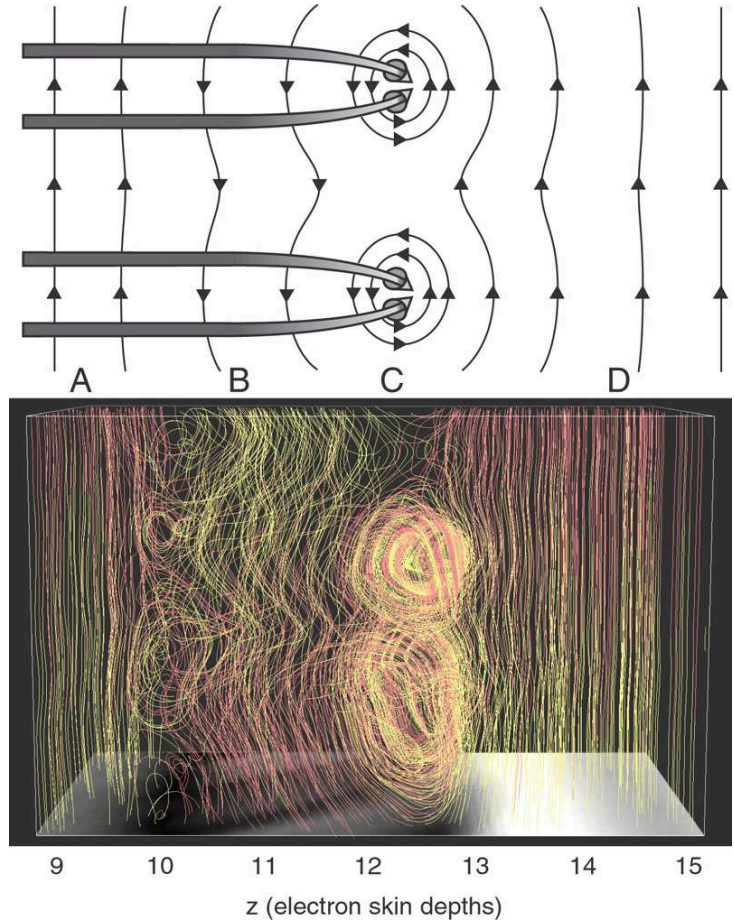,width=.7\textwidth} \caption[3D snapshot
of the complicated magnetic field structure for shocks with a
perpendicular ambient magnetic field.]{Snapshot at
$t=16\omega_{pe}^{-1}$ of the highly complicated magnetic field
topology resulting when a jet is injected into a plasma with an
ambient magnetic field transverse to the jet flow. The bottom panel
shows magnetic field lines in a subsection of the computational box
(from $z=9\lambda_e$ to $z=15\lambda_e$. The top panel refers to a
schematic explanation in the $x-z$ plane: Jet electrons are bent by
the ambient magnetic field (region A). As a result of the Weibel
instability, the electrons bundle into current beams (region C) that
in turn reverse the field topology (region B). This will eventually
bend the jet beam in the opposite direction.} \label{fig:step1}
\end{center}
\end{figure}

\begin{figure}[!h]
\begin{center}
\epsfig{figure=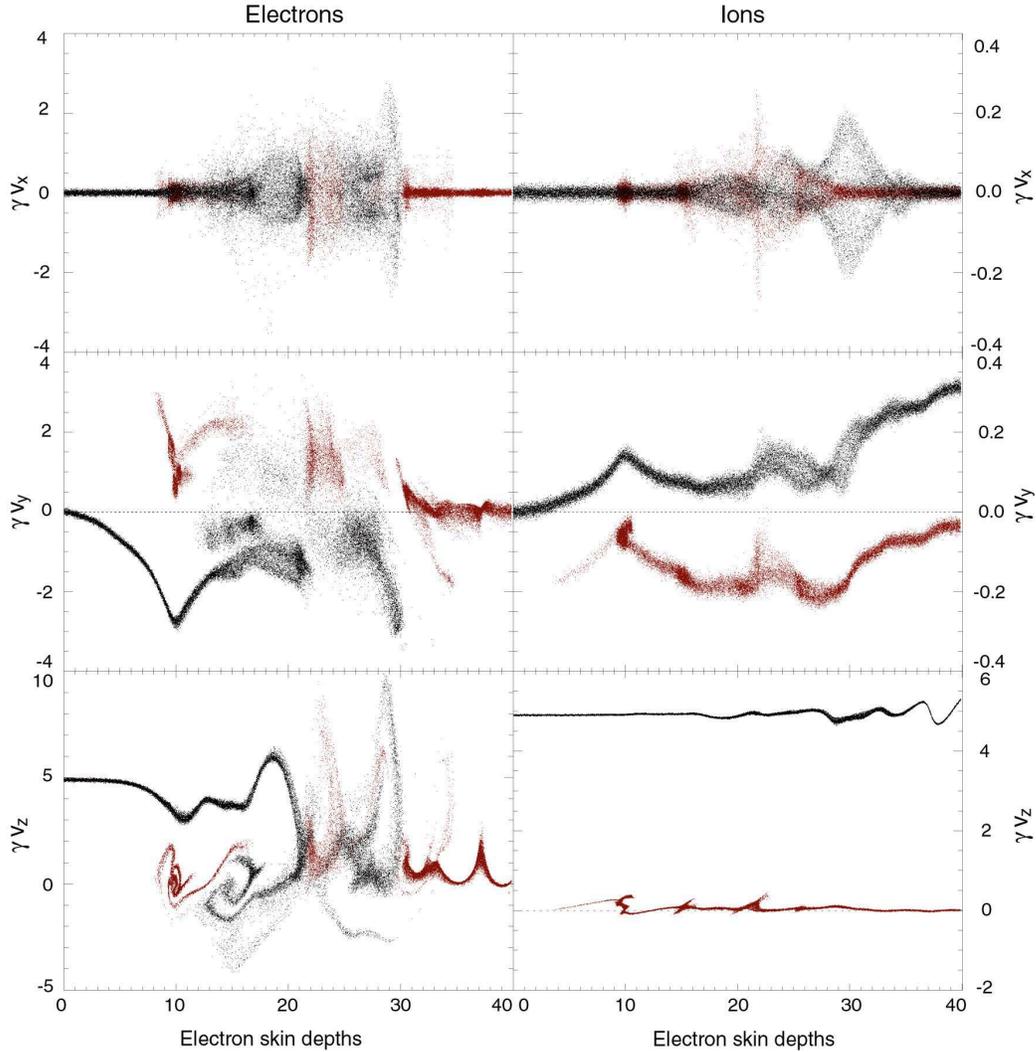,width=\textwidth}
\caption[Phase-space plot of electrons and ions in a shock with
transverse ambient magnetic field]{Phase-space plot of electrons
({\it left column}) and ions ({\it right column}) at time
$t=30\omega_{pe}^{-1}$. The data is from a simulation of a shock
with transverse ambient magnetic field corresponding to
$\omega_{pe}/\omega_{ce}=20$. The three rows correspond to $\gamma
v_x$, $\gamma v_y$ and $\gamma v_z$. The jet plasma ({\em black
dots}) are injected at $z=0$ with $\gamma=5$. The ambient electrons
({\em red dots}) are initially at rest but are strongly accelerated
by the jet.} \label{fig:scatter}
\end{center}
\end{figure}

In the run with $\omega_{pe}/\omega_{ce}=20$, the result differs
substantially from the previous cases. With reference to Fig.\
\ref{fig:step1}, we describe the different stages of the evolution:
Initially, the injected particles are deflected by the ambient
magnetic field. The magnetic field is piled up behind the jet front,
and the enhanced magnetic fields bend jet electron trajectories
sharply. This has two implications. First, the ions, being more
massive, penetrate deeper than the deflected electrons. This creates
a charge separation near the jet head that effectively accelerates
both ambient and injected electrons behind the ion jet front as
shown in Fig.\ \ref{fig:scatter}. Second, the deflected electrons
eventually become subject to the Weibel two-stream instability. This
forms electron current channels at some angle to the initial
direction of injection, as shown in the upper panel of Fig.\
\ref{fig:step1}. Around these current channels, magnetic loops are
induced (Fig.\ \ref{fig:step1}, region C). Magnetic islands are
formed and the ambient magnetic field is reversed behind the loops
(region B). In this region, we find acceleration of electrons in the
$x$-direction. The activity in this region has similarities to
reconnection, but it is beyond the scope of this thesis to
investigate this topic.

\begin{figure}[htb]
\begin{center}
\epsfig{figure=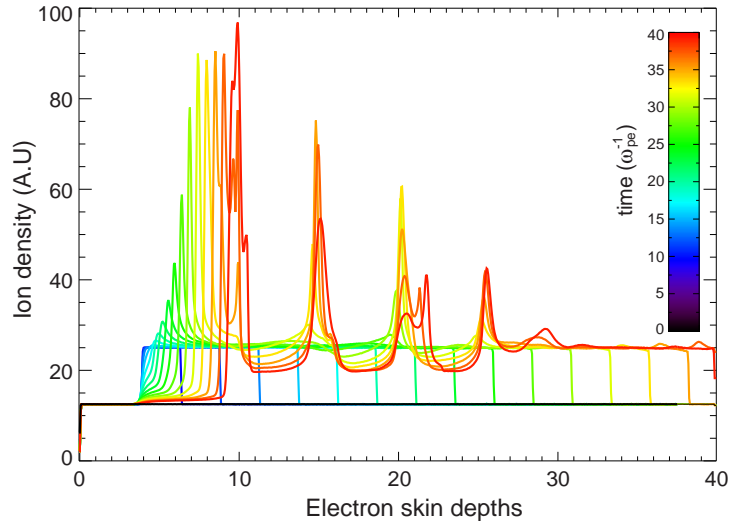,width=.7\textwidth} \caption[Ion density
as function of shock depth]{The ion density averaged over the box
width as a function of shock depth. The different colors correspond
to different times. The simulation is for
$\omega_{pe}/\omega_{ce}=20$. The distance between each peak is
comparable to the ion skin depth.} \label{fig:dni_time}
\end{center}
\end{figure}

In other regions, the ambient magnetic field is strongly compressed,
and this amplifies the field strength up to 5 times the initial
value. As a result, parts of the jet electrons are reversed in their
direction. This may be seen from Fig.\ \ref{fig:scatter}, which
shows a phase-space plot of both ambient and jet electrons at $t =
30\omega_{pe}^{-1}$. We see several interesting features here. In
the region $z=(15-21)\lambda_e$, we observe how ambient electrons
are swept up by the jet. Behind the jet front, both ambient and jet
electrons are strongly accelerated since the jet ions, being
heavier, take a straighter path than the jet electrons, and this
creates a strong charge separation. The excess of positive charge at
the very front of the jet head is very persistent and hard to shield
since the jet ions are moving close to the speed of light. Thus,
there is a continuous transfer of $z$-momentum from the jet ions to
the electrons. In the case of the perpendicular ambient field, more
violent processes take place than in the case of the parallel field,
which may be seen in Fig.\ \ref{fig:pdf} ({\it right panel}). Here
we see that mixing of the two plasma populations is much more
effective for the perpendicular case. However, the spectrum of the
electron's momentum is highly nonthermal, with strong acceleration
of both jet and ambient electrons. The cutoff in electron
acceleration depends on the magnetic field strength. The maximum
($\gamma v_{\parallel} \approx 10$) at $z = 20\lambda_{e}$ in Fig
\ref{fig:scatter} corresponds to the cutoff shown by the dotted line
in the right panel in Fig.\ \ref{fig:pdf}. It should be noted that
the current channels that are caused by the bent jet electron
 trajectories
at the early time, as shown in Fig.\ \ref{fig:step1}, are also seen
in Fig.\ \ref{fig:scatter}. The first current channels have moved to
around $z = 20\lambda_{e}$. At $z = 15 \lambda_{e}$, a second
current channel is created by the deflected jet electrons. This
periodic phenomenon involves the ions (Fig.\ \ref{fig:dni_time}) in
a highly nonlinear process but is beyond the scope of this thesis
and will be investigated in subsequent work.

\section{Conclusions}
Using a three-dimensional relativistic particle-in-cell code, we
have investigated how an ambient magnetic field affects the dynamics
of a relativistic jet in the collisionless shock region. We have
examined how the different ambient magnetic topology and strength
affect the growth of the electron Weibel two-stream instability and
the associated electron acceleration. This instability is an
important mechanism in collisionless shocks.  It facilitates
momentum transfer between colliding plasma populations
\citep{bib:Kazimura,bib:medvedevloeb,bib:nishikawa,bib:nishikawa2004,
bib:silva,bib:frederiksen2004} and can accelerate electrons to
nonthermal distributions \citep{bib:hoshino2002, bib:hededal2004,
bib:saito2003}. Collisionless shocks are found in the interface
between relativistic outflows (e.g., from gamma-ray bursts and
active galactic nuclei) and the surrounding medium (e.g., the ISM).

We find substantial differences between the cases of ambient
magnetic fields transverse and parallel to the jet flow. However,
common to both cases is that it takes an ambient magnetic field
strength much stronger than the strength of the magnetic field
typically found in the ISM to effectively suppress the Weibel
two-stream instability. In the case of a parallel magnetic field,
$\omega_{pe}/\omega_{ce}$ must be smaller than 5 to effectively
suppress the instability. This result is in good agreement with
two-dimensional simulations by \cite{bib:saito2003}, and thus this
limit seems independent of $\gamma_{jet}$. For a typical ISM density
of $10^6\mathrm{m}^{-3}$, this corresponds to a milligauss magnetic
field. We emphasize the role of $\omega_{pe}/\omega_{ce}$ as an
important parameter for collisionless shocks, as was also pointed
out by \cite{bib:shimada2004}.

In the case of perpendicular injection, the dynamics are different
from the parallel injection. Here, the electrons are deflected by
the magnetic field, and this creates a charge separation from the
less deflected ions. The charge separation drags the ambient and jet
electrons, and consequently they are strongly accelerated along the
$z$-direction. Furthermore, as a result of the Weibel instability,
current channels are generated around the ambient magnetic field,
which is curled and locally reversed.

These simulations provide insights into the complex dynamics of
relativistic jets. Further investigations are required to understand
the detailed physics involved. Larger simulations (above $10^9$
particles) with longer boxes are needed to cover the instability
domain of the ions, to investigate the full evolution of the
complicated dynamics, and to resolve the whole shock ramp.

\chapter{Non--Fermi Acceleration in Plasma
Shocks}\label{sec:acc}




Collisionless plasma shock theory, which applies, for example, to
the afterglow of gamma-ray bursts, still contains key issues that
are poorly understood. In this chapter, I discuss the results of
charged particle dynamics in a highly relativistic collisionless
shock numerically using $\sim 10^9$ particles
\citep{bib:hededal2004}.  We find a power-law distribution of
accelerated electrons, which upon detailed investigation turns out
to originate from an acceleration mechanism that is decidedly
different from Fermi acceleration.
Electrons are accelerated by strong filamentation
instabilities in the shocked interpenetrating plasmas and coincide spatially with the
power-law--distributed current filamentary structures. These structures are an
inevitable consequence
of the now well-established Weibel--like two--stream instability that operates
in relativistic collisionless shocks.
The electrons are accelerated and decelerated instantaneously and locally: a scenery
that differs qualitatively from recursive acceleration mechanisms such as Fermi acceleration.
The slopes of the electron distribution power-laws are in concordance with the
particle power law spectra inferred from observed afterglow synchrotron radiation
in gamma-ray bursts, and the mechanism can possibly explain more generally the
origin of nonthermal radiation from shocked interstellar and circumstellar regions
and from relativistic jets.

\section{Introduction}
Given the highly relativistic conditions in the outflow from
gamma-ray bursts (GRBs), the mean free path for particle Coulomb
collisions in the afterglow shock is several orders of magnitude
larger than the fireball itself.
In explaining the microphysical processes that work to define the
shock, MHD becomes inadequate and collisionless plasma shock theory
stands imperative.
In particular, two key issues remain, namely,
the origin and nature of the magnetic field in the shocked region
and the mechanism by which electrons are accelerated from a thermal
population to a power-law distribution $N(\gamma)d\gamma\propto\gamma^{-p}$.
Both ingredients are needed to explain the
observed afterglow spectra \citep[e.g.][]{2000ApJ...538L.125K, 2001ApJ...560L..49P}.

Regarding the origin of the magnetic field in the shocked region,
observations are not compatible with a compressed interstellar
magnetic field, which would be orders of magnitude smaller than
needed \citep{1999ApJ...511..852G}. It has been suggested that a
Weibel--like two--stream instability can generate a magnetic field
in the shocked region (\citeauthor{bib:medvedevloeb}
\citeyear{bib:medvedevloeb}; \citeauthor{bib:frederiksen2002}
\citeyear{bib:frederiksen2002}; \citeauthor{bib:nishikawa}
\citeyear{bib:nishikawa}; \citeauthor{bib:silva}
\citeyear{bib:silva}). Computer experiments by
\citet{bib:frederiksen2004} showed that the nonlinear stage of a
two--stream instability induces a magnetic field in situ with an
energy content of a few percent of the equipartition value,
consistent with that required by observations.

Fermi acceleration \citep{1949PhRv...75.1169F} has, so far,
been widely accepted as the mechanism that provides the inferred electron
acceleration.
It has been employed extensively in Monte Carlo simulations
(e.g.~\citet{bib:Niemiec} and references therein),
where it operates in conjunction with certain assumptions about the scattering of particles
and the structure of the magnetic field.
The mechanism has, however,
not been conclusively demonstrated to occur in ab initio particle simulations.
As pointed out by \citet{bib:Niemiec},
further significant advance in the study of relativistic shock particle acceleration is
unlikely without understanding the detailed microphysics of collisionless shocks. Also,
recently \citet{bib:baring} found that particle distribution functions (PDFs)
inferred from GRB observations are in contradistinction with standard
acceleration mechanisms such as diffusive Fermi acceleration.

In this chapter, we study ab initio the particle dynamics in a
collisionless shock with bulk Lorentz factor $\Gamma=15$. We find a
new particle acceleration mechanism, which we present in Section
\ref{sec:accmech}. Detailed numerical results are presented and
interpreted in Section \ref{sec:expacc}, while Section
\ref{sec:acc-conc} contains the conclusions.
\begin{figure*}[!th]
\begin{center}
\epsfig{figure=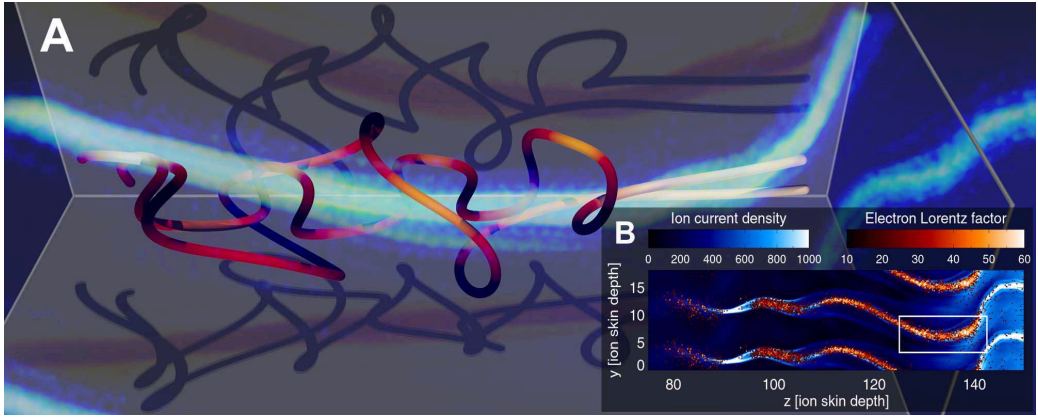,width=\textwidth} \caption[Ray-traced 3D
figure of the current density in a $\Gamma=15$ shock]{(A) Ray-traced
electron paths ({\em red}) and current density ({\em blue}). The
colors of the electron paths reflect their four-velocity according
to the color table in the inset (B). The shadows are equivalent to
the $x$ and $y$ projections of their paths. The ion current density
is shown with blue colors according to the color table in the inset.
The inset also shows the ion current density ({\em blue}) integrated
along the $x$-axis with the spatial distribution of fast-moving
electrons ({\rm red}) over plotted. The main figure is an
enlargement of the white square in the inset.}
\label{fig:acceleration}
\end{center}
\end{figure*}

\section{A new acceleration mechanism}\label{sec:accmech}
We have performed a series of numerical experiments where
collisionless shocks are created by two colliding plasma
populations.  These experiments are described in more detail below,
but a common feature is that the electron PDF has a high-energy tail
that is power-law distributed.  By carefully examining the paths of
representative accelerated electrons, tracing them backward and
forward in time, we have been able to identify the mechanism
responsible for their acceleration. The acceleration mechanism,
which as far as we can tell has not been discussed in the literature
previously, works as follows:

When two nonmagnetized collisionless plasma populations
interpenetrate, current channels are formed through a Weibel--like
two--stream instability (\citeauthor{bib:medvedevloeb}
\citeyear{bib:medvedevloeb}; \citeauthor{bib:frederiksen2002}
\citeyear{bib:frederiksen2002}; \citeauthor{bib:nishikawa}
\citeyear{bib:nishikawa}; \citeauthor{bib:silva}
\citeyear{bib:silva}). In the nonlinear stage of evolution of this
instability, ion current channels merge into increasingly stronger
patterns, while electrons act to Debye shield these channels, as
shown by \citet{bib:frederiksen2004}. That work further showed that
a Fourier decomposition of the transverse ion current filaments
exhibits power-law behavior, which has been recently confirmed by
\citet{bib:medv}.

\begin{figure}[!ht]
\begin{center}
\epsfig{figure=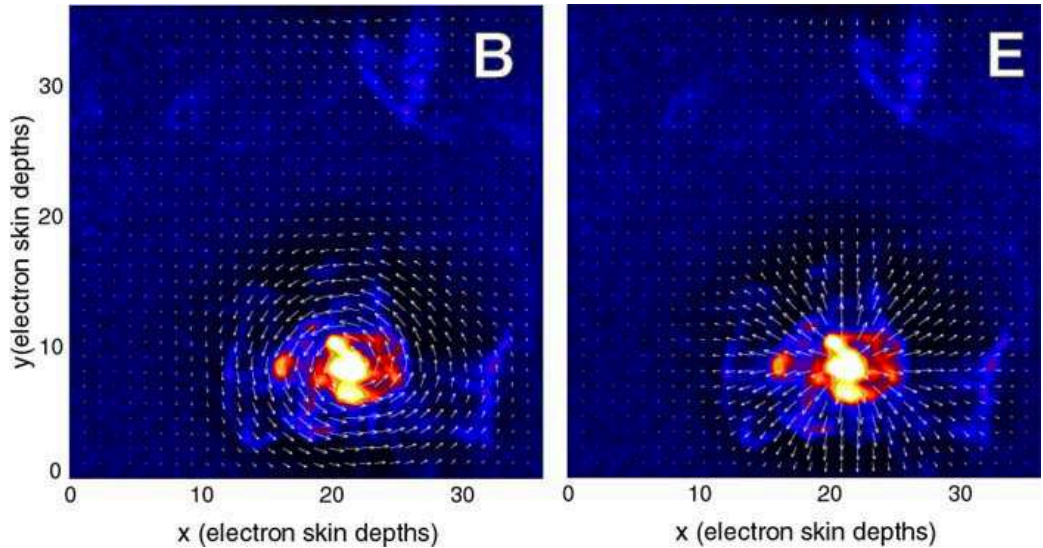,width=\textwidth} \caption[The
electric and magnetic field in the vicinity of an ion current
channel]{A contour plot of the ion current strength with arrows
overplotted to represent the magnetic field ({\it left panel}) and
the electric field ({\it right panel}). } \label{fig:acc_arrow}
\end{center}
\end{figure}

Figure \ref{fig:acc_arrow} presents the electric and magnetic field
in the vicinity of an ion current channel, generated by the Weibel
two-stream instability. At distances less than the Debye length, the
ion current channels are surrounded by transverse electric fields
that accelerate the electrons toward the current channels. However,
the magnetic fields that are induced around the current channels act
to deflect the path of the accelerated electrons, boosting them
instead in the direction of the ion flow. Since the forces working
are caused by quasi--stationary fields, the acceleration is a simple
consequence of potential energy being converted into kinetic energy.
Therefore, the electrons are decelerated again when leaving the
current channel and reach their maximal velocities at the centers of
the current channels. Hence, as illustrated by
\fig{fig:acceleration}B, the spatial distribution of the high-energy
electrons is a direct match to the ion current channels and the
properties of the accelerated electrons depend primarily on the
local conditions in the plasma.

One might argue that the near--potential behavior of the electrons, where they essentially
must lose most of their energy to escape from the current channels, would make the
mechanism uninteresting as an acceleration mechanism since fast electrons cannot easily
escape.  However, this feature may instead be a major advantage, since it means that
energy losses due to escape are small and that the electrons remain trapped long enough
to have time to lose their energy via a combination of bremsstrahlung and synchrotron
or jitter radiation.
We observe that only a very small fraction of the electrons manage to escape,
while still retaining most of their kinetic energy.
This happens mainly at sudden bends or mergers of
the ion channels, where the electron orbits cannot be described in terms of
a particle moving in a static electromagnetic field.

\begin{figure}[!ht]
\begin{center}
\epsfig{figure=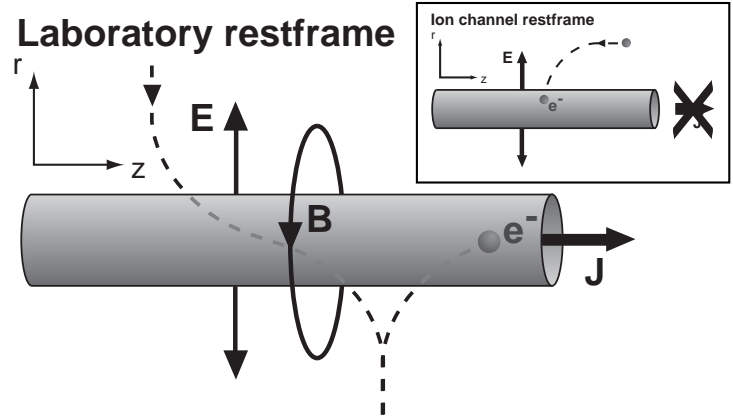,width=.7\textwidth} \caption[Schematic of a
non-fermi acceleration mechanism]{Ion current channel surrounded by
an electric and a magnetic field. A schematic drawing of the
scenario in Fig.\ \ref{fig:acc_arrow}, seen from the side. Electrons
in the vicinity of the current channels are thus subject to a
Lorentz force with both an electric and magnetic component, working
together to accelerate the electrons along the ion flow. Crossing
the center of the channel, the process reverses, leading to an
oscillating movement along the channel.} \label{fig:current_acc}
\end{center}
\end{figure}

To analyze the acceleration scenario quantitatively, we use the
sketch in \fig{fig:current_acc}. We assume that the ion current
channel has radius $R$, that the total charge inside the cylinder
per unit length is $\rho_c$, and that the ions stream with velocity
$u$ in the laboratory rest frame (see \fig{fig:current_acc} and
inset for definition of rest frames). Consider an electron with
charge $-q$ and mass $m$ at a distance $r$ from the center of the
channel, initially having no velocity components perpendicular to
the cylinder. If we analyze everything in the ion channel rest
frame, the problem reduces to electrostatics and it is possible to
analytically calculate the change in four-velocity of the electron
when it reaches the surface of the cylinder. Since the electric
force only works along the $r$-axis, the four-velocity along the
$z$--axis of the electron is conserved in the ion channel rest
frame. Hence, we can calculate both the total change in energy and
the change in the different velocity components. Returning to the
laboratory rest frame, we find
\begin{align}\label{eq:acc}
\Delta\gamma_{electron} &= \frac{q \rho_c}{2 \pi m c^2 \epsilon_0}
             \ln \frac{r}{R} \\
\Delta(\gamma v_z )_{electron} &= u \Delta\gamma_{electron}\, .
\end{align}
The change in the Lorentz boost is directly proportional to
the total charge inside the channel
and inversely
proportional to the electron mass. Debye shielding
reduces the electric field further away from the ion channel, so the estimate
above is only valid for distances smaller than a Debye length.

\section{Computer Experiments}\label{sec:expacc}
The experiments were performed with the three-dimensional
relativistic kinetic and electromagnetic particle--in--cell code
described by \citet{bib:frederiksen2004}. The code works from first
principles, by solving Maxwell's equations for the electromagnetic
fields and solving the Lorentz force equation of motion for the
particles.

Two colliding plasma populations are set up in the rest frame of one
of the populations (downstream, e.g., a jet). A less dense population (upstream,
e.g. the interstellar medium) is continuously injected at the left boundary with a relativistic velocity corresponding
to a Lorentz factor $\Gamma=15$. The two populations initially differ in density by a factor of 3.
We use a computational box with $125\times125\times2000$ grid points and a
total of $8\times10^8$ particles. The ion rest-frame plasma frequency in the downstream
medium is $\omega_{pi}=0.075$, rendering the box 150 ion skin depths long.
The electron rest-frame plasma frequency is $\omega_{pe}=0.3$ in order to resolve
also the microphysics of the electrons.
Hence, the ion-to-electron mass ratio is $m_i/m_e = 16$. Other mass ratios and
plasma frequencies were used in complementary experiments.
Initially, both plasma populations are unmagnetized.

The maximum experiment duration has $t_{max} =$ 340
$\omega_{pi}^{-1}$, which is sufficient for the continuously
injected upstream plasma ($\Gamma = 15$, $v\sim c$) to travel 2.3
times the length of the box. The extended size and duration of these
experiments enable observations of the streaming instabilities and
concurrent particle acceleration through several stages of
development \citep{bib:frederiksen2004}. Momentum losses to
radiation (cooling) are presently not included in the model. We
have, however, verified that none of the accelerated particles in
the experiment would be subject to significant synchrotron cooling.
The emitted radiation may thus be expected to accurately reflect the
distribution of accelerated electrons.

When comparing numerical data with eq. \ref{eq:acc}, we take $r$ to
be the radius where Debye shielding starts to be important. Using a
cross section approximately in the middle of \fig{fig:acceleration},
we find $\Delta(\gamma v_z)_{electron} = 58 \ln (r/R)$. It is hard
to determine exactly when Debye shielding becomes effective, but
looking at electron paths and the profile of the electric field, we
estimate that $\ln (r/R) \approx 1.3$. Consequently, according to
eq. \ref{eq:acc}, the maximally attainable four-velocity in this
experiment is in the neighbourhood of $(\gamma v_z)_{max}=75$. This
is in good agreement with the results from our experiments, where
the maximum four-velocity is $(\gamma v_z)_{max}\simeq80$.

\begin{figure}[!t]
\begin{center}
\epsfig{figure=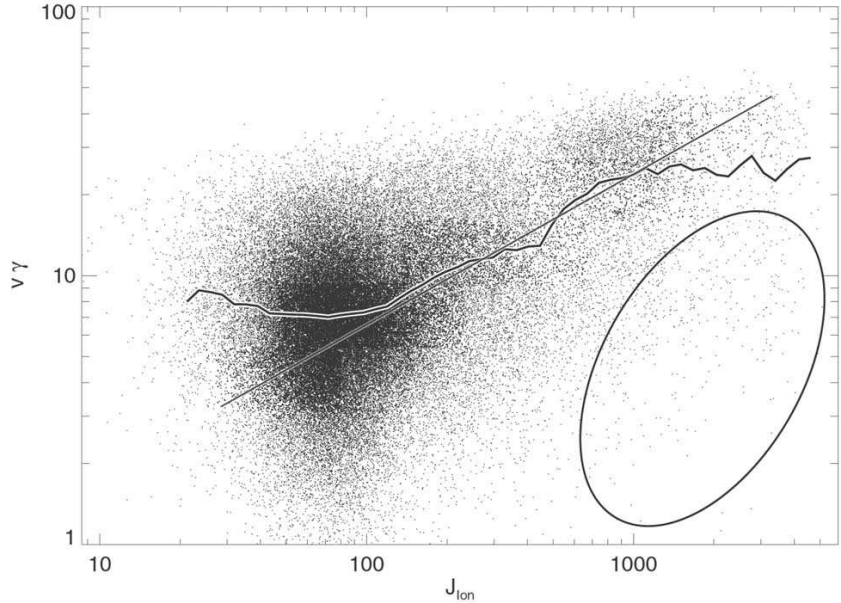,width=12cm} \caption[Electron scatter plot
connecting ion current density and electron velocity]{A scatter plot
of the local ion current density $J_{Ion}$ versus the four velocity
of the electrons in a region downstream of the shock. Overplotted is
a line (thin) showing the average four velocity as a function of
$J_{Ion}$, and a line (thick) showing a straight line fit. Because
'cold' trapped thermal electrons (indicated with the ellipse) exist
inside the ion current channel they count towards lowering the
average four velocity at high $J_{Ion}$. If we cleaned our scatter
plot, statistically removing all thermal electrons we would see a
much tighter relation. Such cleaning, though, is rather delicate and
could introduce biases by itself. The trend is clearly there though
even for the 'raw' data.} \label{fig:jvg}
\end{center}
\end{figure}

The theoretical model does of course not cover all details of the experiment.
For example, in general the
electrons also have velocity components parallel to the magnetic field; instead of
making one-dimensional harmonic oscillations in the plane perpendicular to the current
channel, the electrons will describe ellipsoidal paths.
\fig{fig:acceleration} shows the path of two electrons in the vicinity of an ion channel. But, overall, the electrons behave as expected from the model considerations. Consequently, high-speed electrons are tightly coupled to the ion channels,
as clearly illustrated by \fig{fig:acceleration}B.

\begin{figure}[!t]
\begin{center}
\epsfig{figure=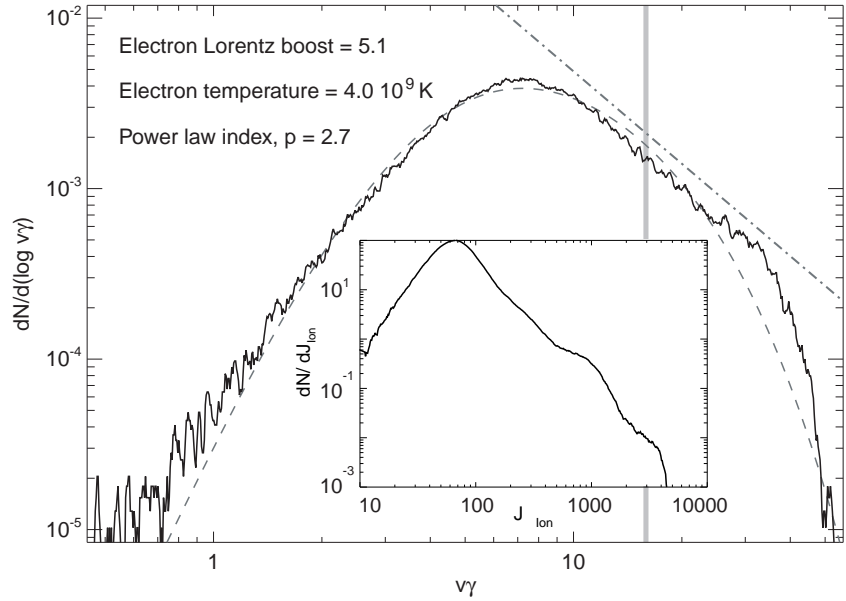,width=12cm} \caption[Non-thermal electron
distribution function in a $\Gamma=15$ shock]{The normalized
electron particle distribution function downstream of the shock. The
dot--dashed line is a power law fit to the non--thermal high energy
tail, while the dashed curve is a Lorentz-boosted thermal electron
population. The histogram is made from the four velocities of
electrons in a thin slice in the $z$--direction of the computational
box.  The grey vertical line at $\gamma v=15$ indicates the
injection momentum. The inset shows a similar histogram for ion
current density sampled in each grid point in the same slice. The
bump in the inset is a statistical fluctuation caused by a single
ion channel. } \label{fig:pdfpower}
\end{center}
\end{figure}

\Fig{fig:pdfpower} shows that the electrons are power-law--distributed at
high energies, with index $p=2.7$.
The electrons at the high gamma cutoff are found where the ion current peaks,
as may be seen from \fig{fig:jvg}. The maximum ion current is limited
by the size of our box; larger values would probably be found if the
merging of current channels could be followed further downstream.
The PDF is not isotropic in any frame of reference because of the high
anisotropy of the Weibel-generated electromagnetic field.
The power-law in the electron PDF is dominant for $10<\gamma<30$.
Likewise, a power law dominates the ion current channel strength, $J_{Ion}$, for
$100<J_{Ion}<1000$ ({\em inset}).
A relation between the power-law distributions of these two
quantities on their respective intervals is provided with
\fig{fig:jvg}: we see that the average four-velocity is proportional
({\em straight line fit}) to a power of the local ion current
density on their respective relevant intervals, $10<\gamma<30$ and
$100<J_{Ion}<1000$. Their kinship stems from the fact that
acceleration is local. The value $J_{Ion}$ has a power-law tail, and
its potential drives the high energy distribution of the electrons
according to eq. \ref{eq:acc}, thus forming a power-law--distributed
electron PDF.

Measuring the rate at which the in--streaming ions transfer momentum to the
ion population initially at rest allows us to make a crude estimate of the length scales
over which the two--stream instability in the current experiment would saturate owing to ion
thermalization. A reasonable estimate appears to be approximately 10 times
the length of the current computational box, or about 1500 ion skin depths. Assuming that
the shock propagates in an interstellar environment with a plasma density of
$\sim 10^6$ m$^{-3}$, we may calculate a typical
 ion skin depth. Comparing this value with the upstream ion skin depth from our experiments,
we find that the computational box corresponds to a scale of the
order  of $10^7$ m, or equivalently that the collisionless shock
transition region of the current experiment corresponds to about
$10^8$ m. For an ion with a Lorentz factor $\gamma=15$, this length
corresponds roughly to 40 ion gyro radii in the average strength of
the generated magnetic field. But we stress that the in--streaming
ions do not really gyrate, since they mainly travel inside the ion
current channels where the magnetic field, by symmetry, is close to
zero. Also, the strong electromagnetic fields generated by the
Weibel instability and the nonthermal electron acceleration, which
is crucial from the interpretation of GRB afterglow observations,
emphasize the shortcoming of MHD in the context of collisionless
shocks.

By rescaling the experiment to physical units we find that the
electrons are accelerated to maximum energies in the neighbourhood
of 5 GeV. Even further acceleration may occur as ion channels keep
growing down stream, outside of our computational box.

The scaling estimates above depend, among other things, on plasma
densities, the bulk Lorentz factor, and the mass ratio ($m_i/m_e$).
A parameter study is necessary to explore these dependencies, but
this is beyond the scope of this thesis. We thus stress that the
 extrapolations performed here are speculative and that
unresolved physics could influence the late stages
of the instability in new and interesting ways.

When the in--streaming ions are fully thermalized, they can no
longer support the magnetic field structures. Thus, one might
speculate that the radiating region of the GRB afterglow is very
thin, as  suggested by \citet{bib:RossiRees}. Furthermore,
traditional synchrotron radiation theory does not apply to an
intermittent magnetic field generated by the two--stream
instability, since the electron gyroradii often are larger than the
scales of the magnetic field structures. We emphasize the importance
of the theory of jitter radiation \citep{bib:Medvedev_jitter}.

\section{Conclusions}\label{sec:acc-conc}
We have proposed a new acceleration mechanism for electrons in
collisionless shocks. The theoretical considerations were suggested
by particle--in--cell computer experiments, which also allowed
quantitative comparisons with the theoretical predictions. We have
shown that the nonthermal acceleration of electrons is directly
related to the ion current channels in the shock transition zone.
The results are applicable to interactions between relativistic
outflows and the interstellar medium. Such relativistic outflows
occur in GRB afterglows and in jets from compact objects
\citep{2004Natur.427..222F}. The suggested acceleration scenario
might overcome some of the problems pointed out by
\citet{bib:baring} regarding the apparent contradiction between
standard Fermi acceleration and spectral observations of GRBs.

The mechanism has important implications for the way
we understand and interpret observations of collisionless shocks:
\begin{itemize}
\item The acceleration mechanism is capable of creating a power-law
electron distribution in a collisionless shocked region. In the
computer experiment presented here, a bulk flow with $\Gamma=15$
results in a power-law slope $p=2.7$ for the electron PDF.
Additional experiments are needed to disentangle which parameters
determine the exact value of the slope.

\item The acceleration is local; electrons
are accelerated to a power law in situ. Therefore, the observed
radiation field may be tied directly to the local conditions of the
plasma and could be a strong handle on the physical processes.

\item Our results strengthen the point already made by
\citet{bib:frederiksen2004} that the fractions of the bulk kinetic
energy that go into in the electrons and the magnetic field,
$\epsilon_e$ and $\epsilon_B$, respectively, are not free and
independent parameters of collisionless shock theory. Most likely,
they represent interconnected parts of the same process.

\item In the case of a weak or no upstream magnetic field, the Weibel--like two--stream
instability is able to provide the necessary electromagnetic fields.
We have shown here that the collisionless shocked region is
relatively thin, and we suggest that the nonthermal radiation
observed from GRB afterglows and relativistic jets in general is
emitted from such a relatively thin shell.

\end{itemize}

It is clear that the non-thermal electron acceleration, the ion
current filamentation, the magnetic field amplification/generation,
and hence the strong non-thermal radiation from the shock, is beyond
the reach of MHD to explain. Whether the relativistic MHD jump
conditions become valid on any larger scale is not possible to
decide from the simulations presented in this chapter.

\chapter{Radiation from Collisionless Shocks}\label{sec:rad}
In this chapter I investigate the radiation emitted from
collisionless plasma shocks. I construct a novel tool that generates
radiation spectra directly from PIC simulations. This is an
important step in order to link simulations with observations.
 Before
utilizing this powerful tool, I perform numerous tests to support
its credibility. I then investigate the nature of 3D jitter
radiation in various setups, including snapshots from the PIC code.

\section{Introduction}
 We first derive\index{derive} an
expression for the frequency spectrum of the radiation emitted from
an accelerated relativistic charged particle. We initially follow
\cite{bib:rybicki}.

From Eq.\ \ref{eq:dpdomega} we have the total energy $W$ emitted per
unit solid angle per unit time from an accelerated charge
\begin{equation}
\frac{dP}{d\Omega}=\frac{d^2W}{d\Omega
dt}=\frac{\left|R\vect{E}(t)\right|^2}{\mu_0 c}\label{eq:dwdOmega}.
\end{equation}
The total energy $W$ radiated per unit solid angle per unit
frequency is then
\begin{equation}
\frac{d^2W}{d\Omega
d\omega}=4\pi\frac{\left|R\vect{E}(\omega)\right|^2}{\mu_0
c}\label{eq:dwdOmegadomega},
\end{equation}
where $2\pi$ comes from Parseval's theorem for Fourier transforms
and the extra factor $2$ comes from a symmetry argument of
$\vect{E}(\omega)$ (no negative frequencies) \citep{bib:rybicki}.
$\vect{E}(\omega)$ is the Fourier transform of $\vect{E}(t)$, which
we define as
\begin{equation}
\vect{E}(\omega)=\frac{1}{2\pi}\int_{-\infty}^{\infty}\vect{E}(t)e^{i\omega
t}dt\label{eq:a_omega}.
\end{equation}
Combining Eq.\ \ref{eq:e_rad_field}, Eq.\ \ref{eq:dwdOmegadomega}
and Eq.\ \ref{eq:a_omega} we find that
\begin{eqnarray}
\frac{d^2W}{d\Omega
d\omega}&=&\frac{1}{4\pi^2}\left|\int_{-\infty}^{\infty}R\vect{E}(t)e^{i\omega
t}dt\right|^2\nonumber\\
&=&\frac{\mu_0 c q^2}{16\pi^3}\left|\int_{-\infty}^\infty
{\left[\frac{\vect{n}\times[(\vect{n}-\vect{\beta})\times\vectdot{\beta}]}{(1-\vect{\beta\cdot
n})^3}\right]_\mathrm{ret} \ e^{i\omega t}dt}
\right|^2\label{eq:dwdw_nonretard}.
\end{eqnarray}

We now expand this into the retarded time frame: We need an
expression for $dt$. From Section \ref{sec:retard} we recall that
the retarded time is given by
\begin{equation}
t'=t-R(t')/c\label{eq:ret_t}.
\end{equation}
Taking the derivative of this expression with respect to $t$ gives
\begin{equation}
\frac{dt'}{dt}=1-\frac{1}{c}\frac{dR(t')}{dt}=1-\frac{1}{c}\frac{dR(t')}{dt'}\frac{dt'}{dt}.
\end{equation}
We may obtain an expression for $\dot R(t')$ by expanding derivative
of the identity $R^2=\vect{R}^2$ into $2R\dot
R=2\vect{R}\cdot\vectdot{R}=-2\vect{R}\cdot\vect{v}$ since
$\vect{R}\equiv\vect{r}-\vect{r}_0\Rightarrow\vectdot{R}=-\vectdot{r}_0\equiv-\vect{v}$.
Thus $\dot R(t')=-\vect{n}\cdot\vect{v}$ and we have
\begin{equation}
dt=(1-\vect{n}\cdot\vect{\beta})dt'\label{eq:ret_dt}.
\end{equation}
With Eq.\ \ref{eq:ret_t} and \ref{eq:ret_dt} we may write Eq.\
\ref{eq:dwdw_nonretard} as
\begin{eqnarray}
\frac{d^2W}{d\Omega d\omega}&=&\frac{\mu_0 c
q^2}{16\pi^3}\left|\int_{-\infty}^\infty
{\frac{\vect{n}\times[(\vect{n}-\vect{\beta})\times\vectdot{\beta}]}{(1-\vect{\beta\cdot
n})^2} \ e^{i\omega(t'-\vect{n\cdot r}_0(t')/c)}dt'}
\right|^2\label{eq:retard_fourier}.
\end{eqnarray}
In the exponential function, we have used that
$R(t')\sim|\vect{r}|-\vect{n\cdot r_0}$. This is valid when
$R(t')\gg r_0(t')$ \citep{bib:jackson}, which is always the case for
observations at cosmological distances. The $|\vect{r}|$ can be
neglected since in the exponential function, this corresponds only
to an overall phase factor.

In theory, this expression gives us the energy radiated per unit
solid angle per unit frequency interval. However, two obstacles
remain: We must know $\vect{r}_0$, \vect{\beta} and \vectdot{\beta}
as functions of time, and we must perform the time integration in
Eq.\ \ref{eq:retard_fourier}. Regarding the first obstacle, we need
an analytical expression for the particle orbit as a function of
time. This may be found for a particle moving in a homogeneous
electromagnetic field in the non-relativistic limit. However, even
here we are left with an integration that cannot be performed
analytically. To make further progress, we assume that the external
electromagnetic field is purely magnetic.
\begin{figure}[htb]
\begin{center}
\epsfig{figure=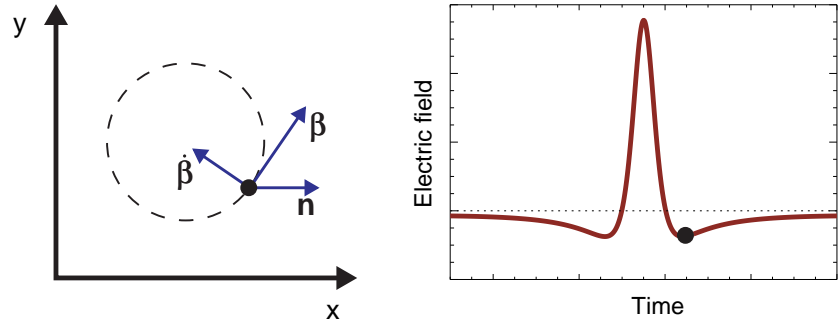,width=0.8\linewidth}
\caption[Electric field from a gyrating charged particle] {The path
of a charged particle moving in a homogenous magnetic field ({\it
left panel}). The particle  radiates a time dependent electric
field. An observer situated at great distance along the
\vect{n}-vector sees the retarded electric field from the gyrating
particle ({\it right panel}). As a result of relativistic beaming,
the field is seen as pulses peaking when the particle moves directly
towards the observer.} \label{fig:eretarded}
\end{center}
\end{figure}

The derivation from here to obtaining an expression for the emitted
radiation as a function of frequency is rather extensive
\citep{bib:jackson}. In appendix \ref{app:spectrum}, we follow this
derivation with the extension of including the full angular
dependency and we find the following result
\begin{eqnarray}
\frac{d^2W_\perp}{d\omega d\Omega}&=&\frac{\mu_0 c q^2
\omega^2}{12\pi^3}\left(\frac{r_L
\theta_\beta\sin\theta}{c}\right)^2
\frac{\left|K_\frac{1}{3}\left(\chi/\sqrt{\cos\theta\beta^3}\right)\right|^2}{(\cos\theta\beta^3)}\label{eq:dwperp}\\
\frac{d^2W_\parallel}{d\omega d\Omega}&=&\frac{\mu_0 c q^2
\omega^2}{12\pi^3}\left(\frac{r_L \theta_\beta^2\beta^2}{c}\right)^2
\frac{\left|K_\frac{2}{3}\left(\chi/\sqrt{\cos\theta\beta^3}\right)\right|^2}{(\cos\theta\beta^3)^2}\label{eq:dwpar},
\end{eqnarray}
where $\theta$ is the angle between $\vect{n}$ and the orbital plane
$\theta_\beta^2\equiv2(1-\beta\cos\theta)$, $\chi=\omega r_L
\theta_\beta^3/(3c)$ and $r_L$ the gyro-radius $\gamma m v/(q B)$.
For $\beta\to1$ and $\theta\to0$, this expression converges toward
the solution one normally find in text books
\citep{bib:jackson,bib:rybicki}.

 We restrict the solution to $\cos\theta>0$. This is justified in the relativistic limit since
the radiation from even mildly relativistic particles is beamed into
a narrow cone around the velocity vector. Equation \ref{eq:dwperp}
and Eq.\ \ref{eq:dwpar} describe the synchrotron radiation spectrum
as observed in the orbital plane. At low frequencies it has a
characteristic power-law slope of $\omega=2/3$. Above some critical
frequency it drops of as $\omega e^{-k \omega}$.

Figure \ref{fig:eretarded} shows the electrical field $\vect{E}(t)$
from a particle that gyrates in a homogenous magnetic field as
measured by an observer situated at great distance in the direction
along \vect{n}.

\section{Obtaining spectra from PIC simulations}
In general, the approximations above will not be entirely accurate.
In the PIC simulations (and in the "real world"), the morphology of
the electromagnetic field is often complicated, with a non-vanishing
electric field. In this case, the paths of the particles are neither
circular nor necessarily periodic. To obtain spectra from the
simulations we need to take another approach. By sampling positions,
velocities and accelerations from the simulation over a suitable
time interval we can numerically evaluate the Fourier integral.
First, we have to choose in which time frame we perform this
integration: "proper" time $t$ or retarded time $t'$. The
corresponding integrals are given in Eq.\ \ref{eq:dwdw_nonretard}
and Eq.\ \ref{eq:retard_fourier}. At a first glance, Eq.\
\ref{eq:dwdw_nonretard} appears the most appealing since here we can
use the standard Fast Fourier Transform (FFT) method, which is
numerically faster. However, the term $(1-\vect{\beta\cdot n})$ is
numerically ill-behaved, and Eq.\ \ref{eq:dwdw_nonretard} has a
higher power dependency on this term than Eq.\
\ref{eq:retard_fourier}. Furthermore, for a relativistic particle
moving towards the observer, all features in the emitted radiation
appear strongly contracted in the $t$-frame and are thus
considerably harder to resolve numerically. Finally, in order to use
an FFT, the numerical representation of $\omega$ needs to have
uniform spacing, which is an inconvenient limitation. .

Before we proceed, several other numerical issues need to be
considered.

\subsection{Resolution constraints}\label{sec:res_con}
\begin{figure}[htb]
\begin{center}
\epsfig{figure=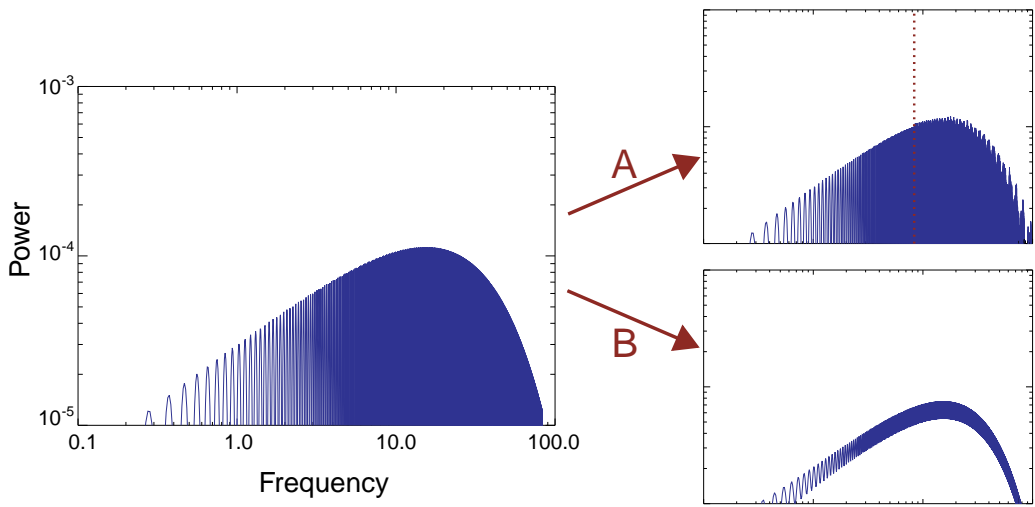,width=\linewidth}
\caption[Numerical distortion of the Fourier spectrum]{The observed
power spectrum from a single charged particle, gyrating in a
magnetic field ({\it left panel}). {\it Case A}: Reducing the sample
rate distorts the frequencies above the Nyquist frequency, marked by
a vertical dotted line. {\it Case B}: If the total time interval
becomes too small (smaller than the orbit period) the main frequency
lobes become smeared out. The units on the axis are arbitrary.}
\label{fig:distortion}
\end{center}
\end{figure}

\begin{figure}[htb]
\begin{center}
\epsfig{figure=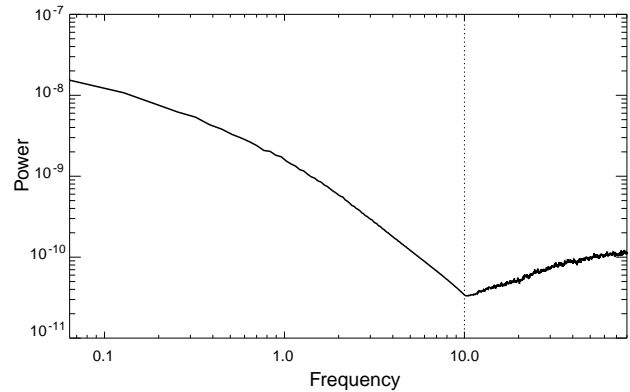,width=.6\linewidth} \caption[Numerical
artifacts above the Nyquist frequency]{An example of a numerical
artifact introduced when trying to extend the spectrum to above the
Nyquist frequency ({\it dotted line}). The effect is know as
aliasing. The example spectrum is from electrons participating in
the Weibel two-stream instability.} \label{fig:aliasing}
\end{center}
\end{figure}

Several issues should be kept in mind when implementing the time
integration of the retarded signal. Constraints exist that are tied
to the uncertainty relation
\begin{equation}
\Delta\omega\Delta t > 1.
\end{equation}
As explained by \cite{bib:rybicki} "this uncertainty relation is not
necessarily quantum in nature, but is a property of any wave theory
of light". As a result of this, the maximum frequency that can be
represented in the resulting spectrum is limited by the number of
points that one samples in a given time interval. This maximum
frequency is known as the Nyquist frequency and equals half the
sampling frequency (e.g.\ \citec{bib:press}). It corresponds to a
rapid oscillating signal that repeats once per two sample points in
the time domain. Above the Nyquist frequency, the solution becomes
aliased, noisy, and in general invalid (see Fig.\
\ref{fig:distortion} and Fig.\ \ref{fig:aliasing}). Thus, we set an
upper limit on the frequency band to
\begin{equation}
\omega_{N}=\frac{1}{2 \Delta t},
\end{equation}
where $\Delta t$ is the time between each sample point. The low end
of the frequency band is also limited since the lowest frequency
that can be represented is that of a wave that repeats only once in
the whole time interval. In fact, we must sample for long enough to
detect not only low frequencies in the signal, but also small
differences between frequencies (the resolution). Thus, the length
of time for which we sample the signal ($T_s$) determines our
ability to resolve differences between frequencies. This defines the
frequency resolution
\begin{equation}
\Delta\omega=\frac{1}{T_s}.
\end{equation}

The effect of a too short total time interval can be seen from Fig.\
\ref{fig:distortion}. Here we see that the individual spikes in the
synchrotron spectrum are broadened out. So as a rule of thumb, the
total length of the time sample should be at least one over the
lowest gyro-frequency in the particle ensemble for which we wish to
obtain a spectrum. For an ensemble of, say, thermal particles this
poses a lesser problem, since the individual frequency peaks become
averaged out to a continuum anyway.

\subsection{Time domain windowing}
To determine the radiation power spectrum from an accelerated
charged particle we sample the position, velocity and acceleration
over a given time interval $T_s=t_2-t_1$. However, even for periodic
orbits we cannot presume that this time interval is an integer
number times the orbital period. And in the general case of sampling
many particles over the time interval, in an inhomogeneous
electromagnetic field, there will most certainly be a high level of
a non-periodicity. However, the theory behind Fourier
transformations assumes that the signal being processed is a
periodic waveform. Connecting the ends of non-periodic signal
introduces a discontinuity in the signal. Such a discontinuity in
the time-domain can only be represented by a superposition of all
frequencies in the frequency domain. The result is that any peak
frequency ({\it main lobe}) is smeared out over a frequency range
and {\it side lobes} occur around the main lobe. This is known as
spectral leakage (e.g. \citec{bib:press}). From experiments we have
found that this leakage can alter the characteristic $\omega^{2/3}$
slope in the spectrum from a single synchrotron particle.

\begin{figure}[htb]
\begin{center}
\epsfig{figure=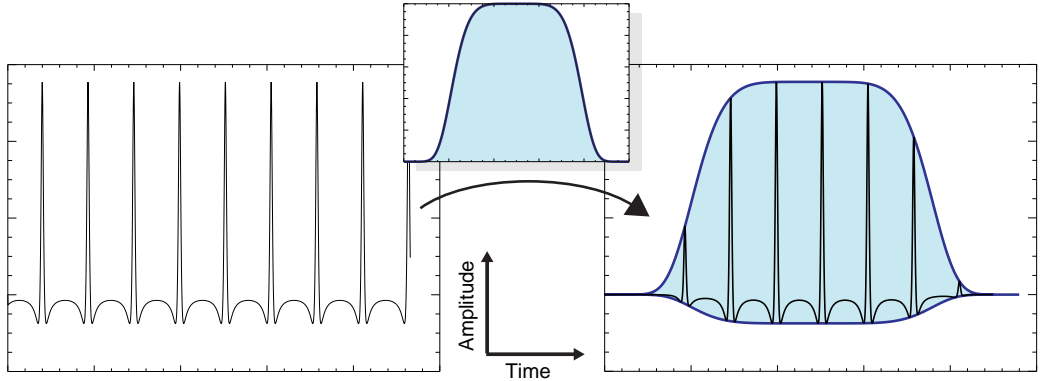,width=\linewidth} \caption[Windowing of a
sampled signal]{Windowing of a signal sampled over a time block
ensures that the signal becomes periodic and continuous in the end
points. This greatly reduces the spectral leakage.}
\label{fig:windowing}
\end{center}
\end{figure}

To solve this problem, we apply a filtering function on the time
dependent signal, to ensure that the signal becomes exactly zero in
the end points of the chosen time sample block (and thus periodic),
so that they match up and the transition is continuous (Fig.\
\ref{fig:windowing}). The technique is known as windowing and is a
standard tool in discrete signal analysis. The choice of window
function depends on the problem. Several standard functions exist
(Hanning, Hamming, Kaiser-Bessel...). After trying several of these,
as well as other functions on the synchrotron radiation case we have
found a good {\it ad hoc} window function
\begin{equation}
W(t)={\mathrm{Exp}}\left[-\left(k\frac{t-{\small
\frac{1}{2}}(t_1+t_2)}{t_2-t_1}\right)^m\right]\label{eq:window}
\end{equation}
with the parameters $k=3$ and $m=6$.

After applying a window function to the signal it is important to
renormalize the sample in order to retain the amplitude of the
spectrum. Thus we multiply the final spectrum with the factor
$A_W^2$ where
\begin{equation}
A_W=\frac{(t_2-t_1)}{\int_{t1}^{t2}W(t)dt}.
\end{equation}

With the window function applied on the sampled data, the expected
$\omega^{2/3}$ synchrotron slope is effectively restored.

\begin{figure}[!h]
\begin{center}
\epsfig{figure=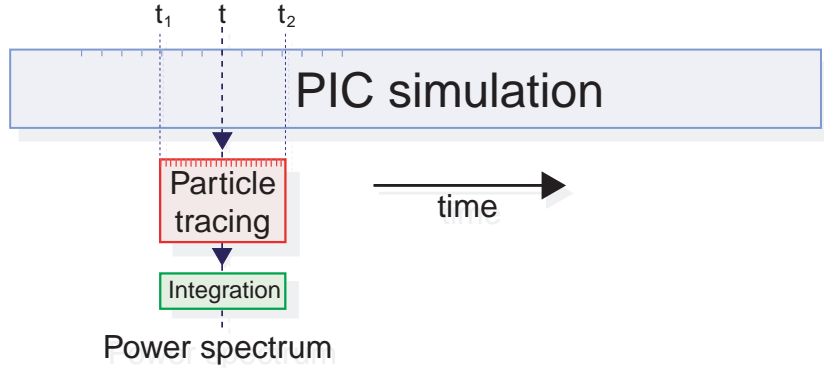,width=0.8\linewidth} \caption[The
scheme for obtaining synthetic spectra]{To obtain a spectrum from
the PIC simulation at time $t$, particles are traced over a time
interval around $t$ with high temporal resolution. Positions,
velocities, and accelerations are sampled and from equation
\ref{eq:retard_fourier} we find the radiation spectrum.}
\label{fig:spectrum_scheme}
\end{center}
\end{figure}

\subsection{Implementation and testing}
The procedure above has been implemented in two steps. The first
part is to sample position, velocity and acceleration for each
particle. In general, one could dump particle data directly from the
simulations. This, however, requires not only vast amount of disk
storage, but one also needs to run the simulations with a time step
much smaller than otherwise needed. The reason for this is discussed
in Section \ref{sec:res_con}. Typically, the time step in the
spectrum tracer is of the order of 100 times smaller than in the PIC
code. Since the simulations already take several weeks to run, we
choose to take another approach (Fig.\ \ref{fig:spectrum_scheme}):
When a simulation has finished, we choose a snapshot for which we
want to obtain the radiation spectrum. Loading fields and particles
from the chosen time step, we trace a given number of the particles
back and forth in a time segment with very high temporal resolution,
while keeping the electromagnetic fields frozen. In this sense, the
particles can be seen as test particles. At each small time step we
then store the particle positions, velocities and accelerations in a
file. This data is read by another program that evaluates the
Fourier integral (in the retarded time frame) given a direction
\vect{n} towards the observer. Finally we add the spectrum of each
particle into a total spectrum. Adding the spectra from each
particle linearly is valid as long as the phase of each contribution
is completely uncorrelated to the others. This is clearly an
assumption, which would be violated for example in a maser
arrangement. The spectrum may finally be Lorentz/Dobbler boosted to
any observer rest frame.

\begin{figure}[!h!]
\begin{center}
\epsfig{figure=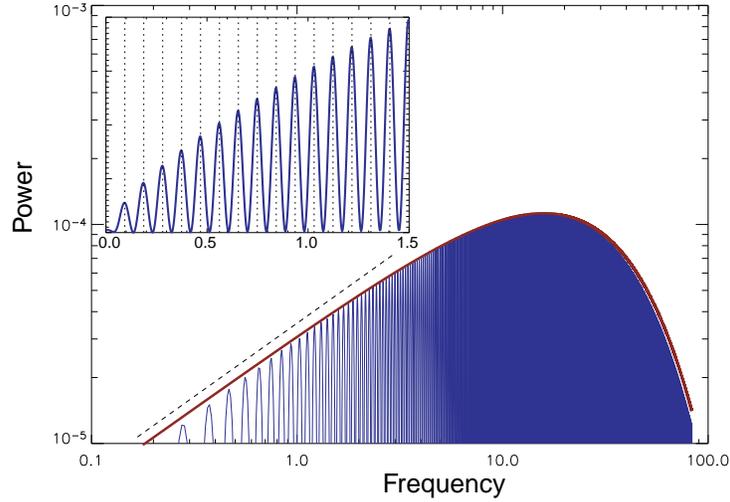,width=0.7\linewidth}
\caption[Single particle synchrotron spectrum obtained from
simulations]{The spectrum obtained from a single particle gyrating
in a magnetic field from the code ({\it blue}). The thick red line
shows the theoretical synchrotron spectrum. The dashed line
indicates the power-law slope $2/3$. The inset in the upper left
corner shows that the spectrum consists of a discrete set of spikes
that are integer overtones of the gyrofrequency, in this case
$\omega_B=0.094$ (arbitrary units) ({\it dotted vertical lines}).
The particle has a Lorentz factor $\gamma=5$. All numbers are in
simulations units.} \label{fig:spectrum_single}
\end{center}
\end{figure}

\subsubsection{Test: Synchrotron radiation}
To verify the ability of radiation code to produce correct spectra
from accelerated charges, a series of tests have been performed. The
first test is that of synchrotron radiation. We place a single
relativistic particle in a homogeneous magnetic field (Fig.\
\ref{fig:eretarded}). The theory predicts that the emitted energy
spectrum should consist of a collection of peaks, positioned at the
gyro-frequency and higher harmonics. The amplitude of the individual
peaks follow the continuum profile given by Eq.\ \ref{eq:dwpar}.
 Fig.\ \ref{fig:spectrum_single} shows the spectrum we obtain
from the code for a single relativistic charged particle in a
homogenous magnetic field. We find very good agreement between the
spectrum obtained from the simulation (blue line) and the
theoretical synchrotron spectrum from Eq.\ \ref{eq:dwpar} (red
line). As expected, the peaks are found at harmonics of the gyro
eigenfrequency and the low energy part follows a power-law slope
$\omega^{2/3}$.

\begin{figure}[htb]
\begin{center}
\epsfig{figure=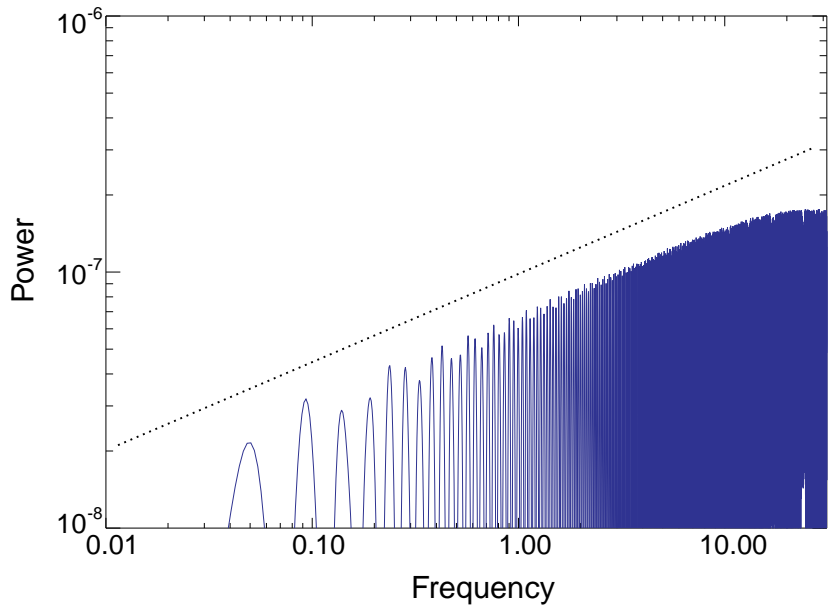,width=0.8\linewidth} \caption[Spectrum
from a particle ensemble with an isotropic mono-energetic
distribution function]{The low energy part of the spectrum from an
ensemble of electrons with $\gamma=10$ and random orientation of the
momentum vector. The ensemble is placed in a homogeneous magnetic
field. The dotted line indicates a power law slope of 1/3. 512
particles were traced for this spectrum.}\label{fig:synch_iso}
\end{center}
\end{figure}

If one observes not a single synchrotron particle, but rather a
whole ensemble of particles, the characteristics of the spectrum
change. If all the particles in the ensemble have the same energy,
but have momentum vectors that are isotropic distributed, the
characteristic low-energy power-law slope of the integrated spectrum
is altered. Integrating Eq.\ \ref{eq:dwperp} and \ref{eq:dwpar} over
all pitch angles yields a spectrum with a low-energy part of the
spectrum that has a power-law slope of $1/3$ rather than $2/3$
\citep{bib:rybicki}. The results from the simulations are in
excellent agreements with this. Figure \ref{fig:synch_iso} shows the
spectrum from 512 particles, sampled from a mono-energetic
distribution function ($\gamma=10$).

\subsubsection{Test: Bremsstrahlung}
Another important test that our radiation tool must pass is to
reproduce the theoretically predicted spectrum of bremsstrahlung.
Bremsstrahlung is emitted when the path of a charged particle is
perturbed by an electric field (e.g. an ion). In the present version
of the PIC code (and in most other PIC codes), the particles cannot
interact directly but only through the grid. For simulations where
each cell in the computational domain is quasi-neutral,
bremsstrahlung is not included. The next generation of out PIC code
will be able to include particle-particle interactions to some
extent (see Chapter \ref{chap:photonplasma}). We can, however, still
test the radiation module by placing a single ion/electron pair in a
simulation box.

Where synchrotron radiation is a periodic phenomenon tied to a
magnetic field, bremsstrahlung is non-periodic and connected to
electric deflections. So in a sense, this test complements the
synchrotron test very nicely.
\begin{figure}[htb]
\begin{center}
\epsfig{figure=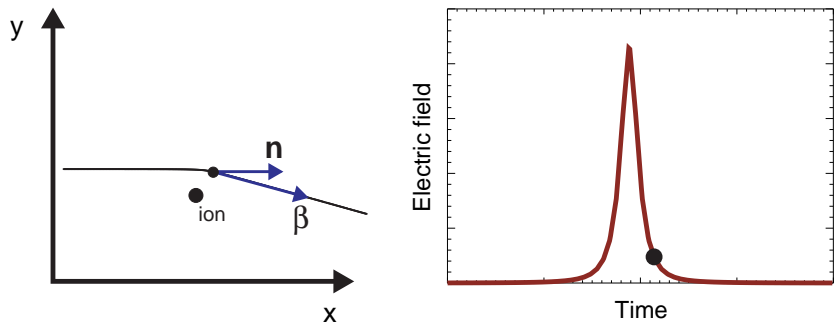,width=0.8\linewidth}
\caption[Bremsstrahlung test setup]{A relativistic electron passes
an ion (left frame). The electrical far-field peaks as the electron
is accelerated by the ion (right frame).} \label{fig:bremss_setup}
\end{center}
\end{figure}
Figure \ref{fig:bremss_setup} shows the example setup where an
electron passes an ion and becomes subject to acceleration. In the
relativistic limit the theoretical spectrum for this process may be
derived using the beautiful theory of virtual photons
\citep{bib:rybicki}: In the rest frame of the electron, the electric
field from the ion appears as a short, Lorentz contracted,
electromagnetic pulse that can be view as a virtual photon. This
virtual photon Compton scatters of the electron to produce the
emitted radiation. Transforming back to the lab frame of the ion
gives the relativistic bremsstrahlung emission from the electron
\begin{equation}
\frac{W_\perp}{d\omega
}=\frac{q^6Z^2}{24\pi^4\epsilon^3c^3m^2v^2b^2}\left(\frac{\omega
b}{v\gamma^2}\right)^2K_1^2\left(\frac{\omega b}{v\gamma^2}\right)
\label{eq:bremss}.
\end{equation}
Here, $Z$ is the number of charges in the ion, $b$ is the minimum
distance between that ion and the electron and $K_1$ is the modified
Bessel function of the second kind. We note that a factor of $K_1$
is missing in Eq.\ 5.23 on page 164 in \cite{bib:rybicki}. In Eq.\
\ref{eq:bremss}, the $\perp$ indicates that we have neglected the
contribution from the component of the ions electric field that is
parallel to the path of the incoming electron. This approximation is
valid in the ultra relativistic limit ($\gamma\gg1$). Figure
\ref{fig:spectrum_bremss} shows the simulation results ({\it thick
blue}) compared with Eq.\ \ref{eq:bremss} ({\it thin red}) for
different electron energies ($\gamma v$=0.5, 2.5, 5.0, 7.5, 10.0).
We find good agreement between simulations and the theoretical
result for $\gamma v>1$. The deviation for $\gamma\to1$ is because
of the approximation in Eq.\ \ref{eq:bremss} above.

\begin{figure}[htb]
\begin{center}
\epsfig{figure=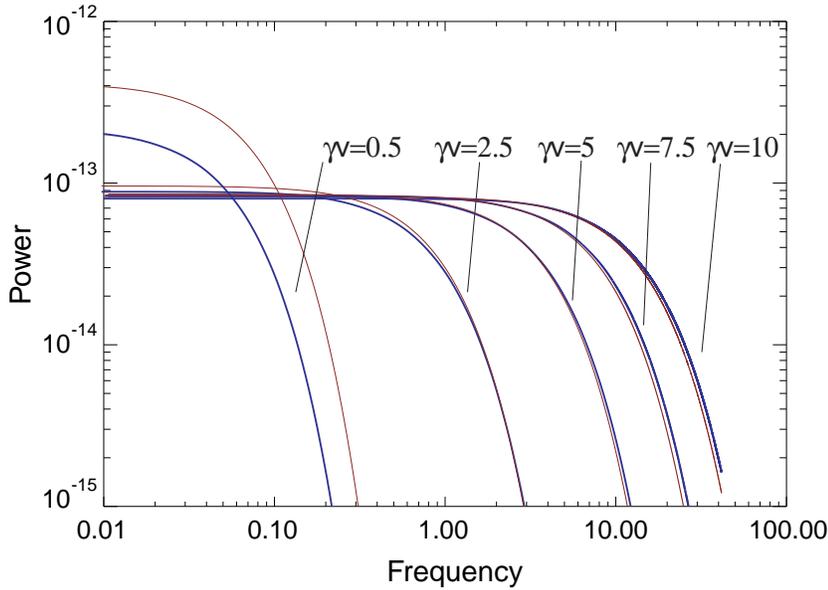,width=0.8\linewidth}
\caption[Single particle bremsstrahlung spectrum obtained from
simulations]{The theoretical ({\it thin red}) and simulated ({\it
thick blue}) bremsstrahlung spectrum from an electron passing an
ion. Four cases have been tested, differing only in initial electron
energy (relativistic momentum $\gamma v$=0.5, 2.5, 5, 7.5, 10).}
\label{fig:spectrum_bremss}
\end{center}
\end{figure}

\subsubsection{Test: Undulator radiation}
\begin{figure}[htb]
\begin{center}
\epsfig{figure=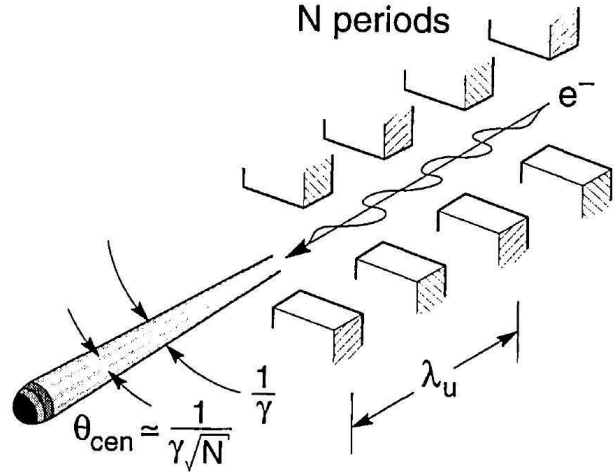,width=0.6\linewidth} \caption[Schematic
view of an undulating electron]{A schematic view of a undulator
experiment in a cylotron. A cluster of magnets are aligned along the
path of the electron. When the electron passes the array of
magnetic, it radiates at the undulator frequency.
 From \cite{bib:attwood1993}}
\label{fig:undulator}
\end{center}
\end{figure}

A final but crucial test is whether the radiation code can reproduce
the spectrum from undulator radiation. Undulator radiation is
emitted when a relativistic electron traverses a periodic magnetic
field so weak that the deflection angle stays within the
relativistic beaming cone. The phenomenon is well known, and widely
used, in synchrotron storage ring experiments. Figure
\ref{fig:undulator} shows a schematic view of the generation of
undulator radiation \citep{bib:attwood1993}. A full introduction to
the field of undulator and wiggler radiation is given by
\cite{bib:attwood}.

The spectrum from a relativistic electron moving in a transverse
periodic undulator field (with magnetic wavelength $\lambda_u$ and
maximum strength $B_0$), is characterized by a well confined peak
near $\omega_u$, determined by the undulator equation
\citep{bib:kincaid}
\begin{equation}
\omega_u=\frac{2 c
\gamma^2}{\lambda_u(1+K^2/2+\gamma^2\theta_u^2)}\label{eq:undulator}.
\end{equation}
Here, $\theta_u$ is the angle between the path of the electron and
the line of sight and $K$ is the periodic magnet parameter, defined
as
\begin{equation}
K=\lambda_u q B_0/(2\pi m_e c)\label{eq:magnet}.
\end{equation}
In the weak-field case ($K<1$) the deflection angle of the
undulation is smaller than the beaming opening angle $1/\gamma$ and
the radiation is confined into the angle $1/\gamma$. For $K>1$ the
radiation (called wiggler radiation) is emitted into a cone of
half-angle $K/\gamma$. In this section we focus on $K<1$. The width
of the peak is determined by how many periods $N$ the electron path
is sampled over. The more undulations, the better defined the peak
becomes.

\begin{figure}[htb]
\begin{center}
\epsfig{figure=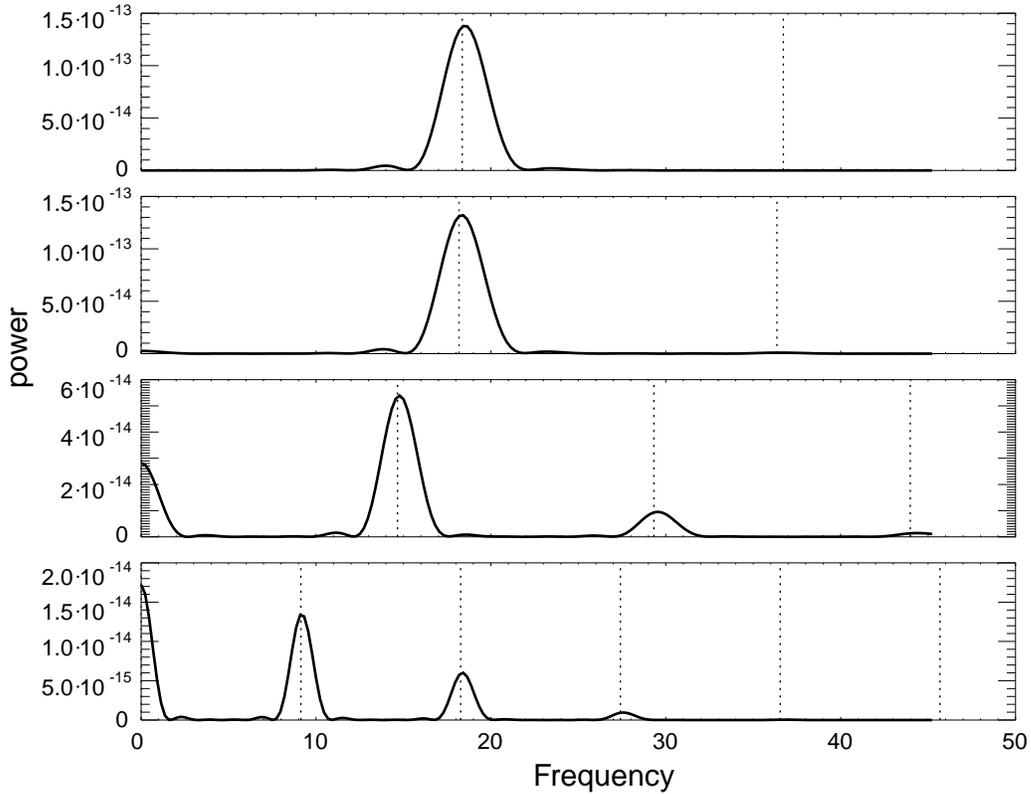,width=\linewidth} \caption[Single
particle Undulator spectrum obtained from simulations]{Undulator
spectra generated when a $\gamma=10$ electron traverses a periodic
magnetic field. The spectra are obtained from simulations with
different viewing angles $\theta_u$= 0, 0.01, 0.05, and 0.1 (top to
bottom panel). The dotted lines show the expected undulator
frequency $\omega_u$ and its higher harmonics.}
\label{fig:undulator_spec}
\end{center}
\end{figure}

Figure \ref{fig:undulator_spec} shows the result of simulations
where a single, relativistic electron ($\gamma=10$) moves in a
periodic magnetic field. The undulator wavelength is $\lambda_u=11$
grid-zones and the periodic magnet parameter $K=0.0015$ (Eq.\
\ref{eq:magnet}). The four panels in the figure represent different
angles between the electron path and the line of sight
($\theta_u$=0, 0.01, 0.05, and 0.1). As expected \citep{bib:attwood}
the spectrum peaks at the frequency given by Eq.\ \ref{eq:undulator}
(dotted lines). As $\theta_u$ is increased, higher harmonics of
$\omega_u$ becomes visible. We thus find excellent agreement between
the undulator equation Eq.\ \ref{eq:undulator} and the simulations.

\section{Jitter radiation}
After having tested the radiation generation module thoroughly, we
now focus on some more scientific issues. The PIC plasma simulation
code, in combination with the radiation generation module, provides
us with a powerful tool for testing various non-linear problems.

\begin{figure}[htb]
\begin{center}
\epsfig{figure=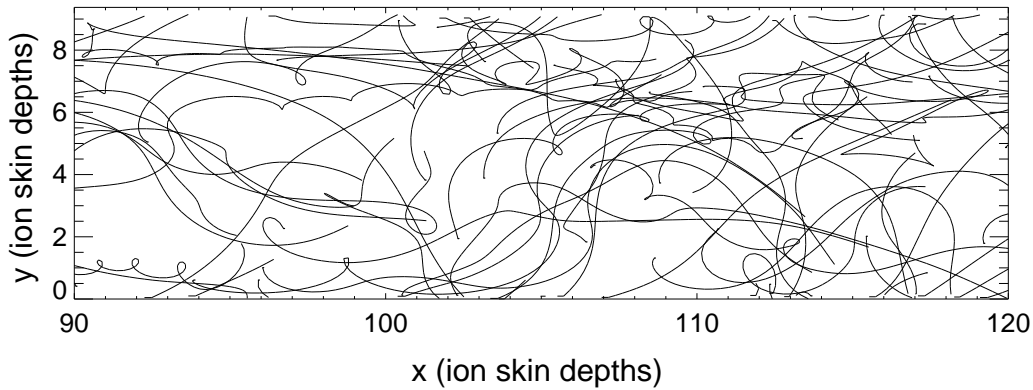,width=\textwidth} \caption[Traced path of
electrons from the PIC simulations]{An example of the complicated
paths of the electrons under the influence of electromagnetic field
generated by the Weibel two-stream instability.}
\label{fig:tracerpar}
\end{center}
\end{figure}

We choose to address several problems, all connected to the theory
of 3D jitter-radiation from GRB plasma shocks. Jitter radiation was
introduced to the GRB community by \cite{bib:Medvedev_jitter}. It
covers the regime where the magnetic field is inhomogeneous on
scales smaller than the Larmor radius (see Fig.\
\ref{fig:tracerpar}) and the electron's transverse deflections in
these fields are much smaller than the relativistic beaming angle.
The foundation and motivation for such a theory was based on the
small-scale nature of the magnetic fields generated by the
two-stream instability. In the one-dimensional analytical approach,
\cite{bib:Medvedev_jitter} found that the low frequency slope of the
spectrum could be steeper than the $1/3$ slope of synchrotron
radiation.

In this section, we first perform a parameter study to investigate
the nature of the radiation emitted from an ensemble of relativistic
electrons in different small-scale magnetic configurations (all with
$K<1$).

Secondly, we compare the radiation that is emitted from a electron
population in a generic, turbulent magnetic field (power-law
distributed power-spectrum) with a two-stream generated magnetic
field obtained from a snapshot from the PIC simulations. The
question is: Is it reasonable to apply the theory of
jitter-radiation directly on GRB shocks, assuming some generic
structure of the magnetic field or must one include the complicated
spatial details of a two-stream generated magnetic field? Our
concern is that the ordered sub-structures in the two-stream
generated magnetic field may affect the radiation spectrum.


\begin{figure}[htb]
\begin{center}
\epsfig{figure=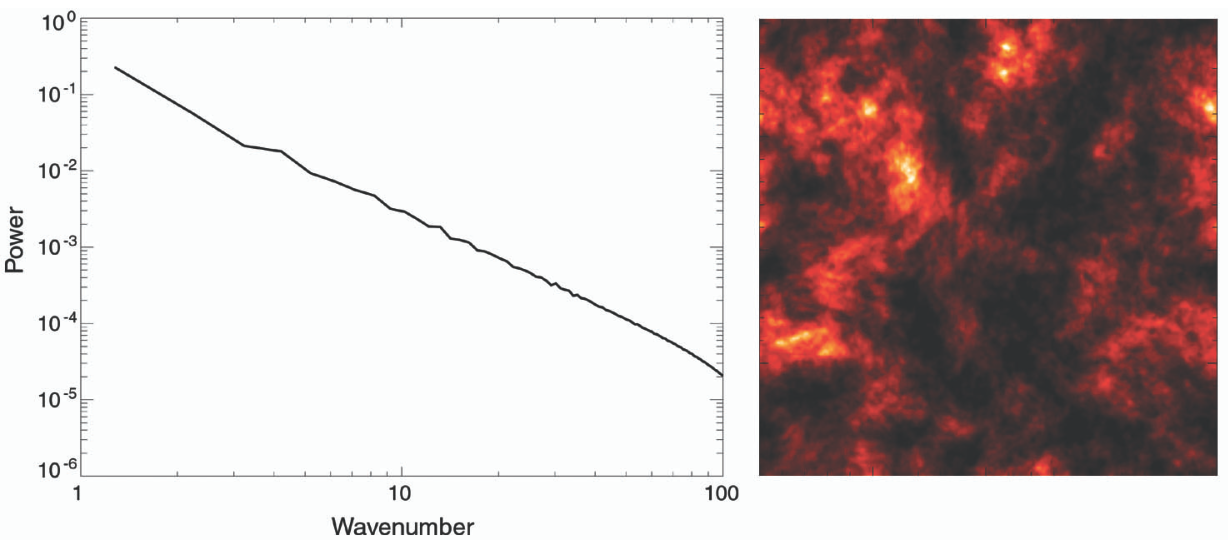,width=0.75\linewidth}
\epsfig{figure=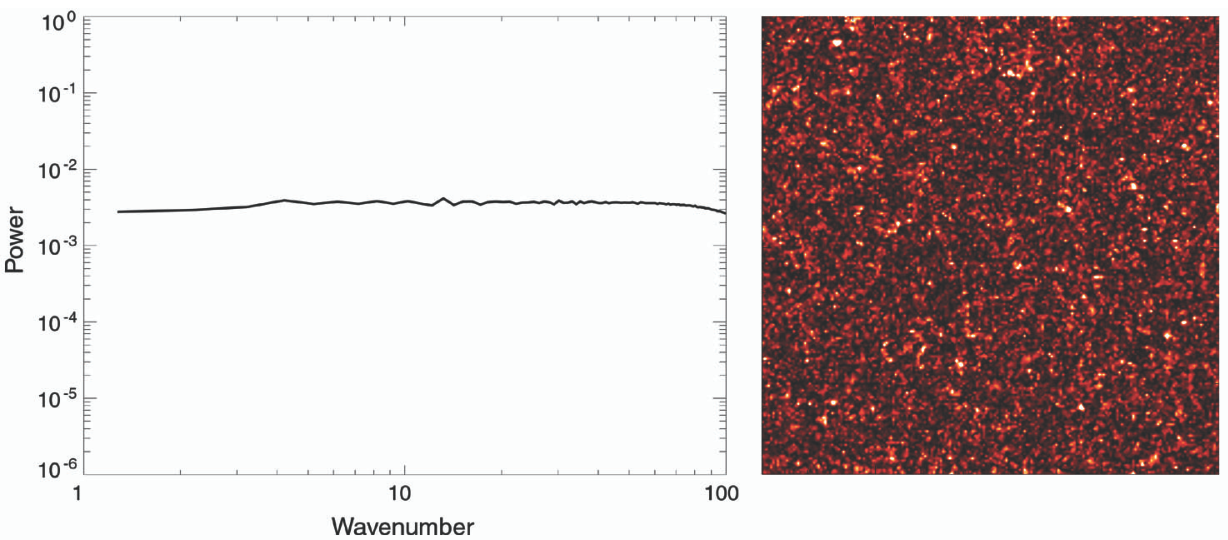,width=0.75\linewidth}
\epsfig{figure=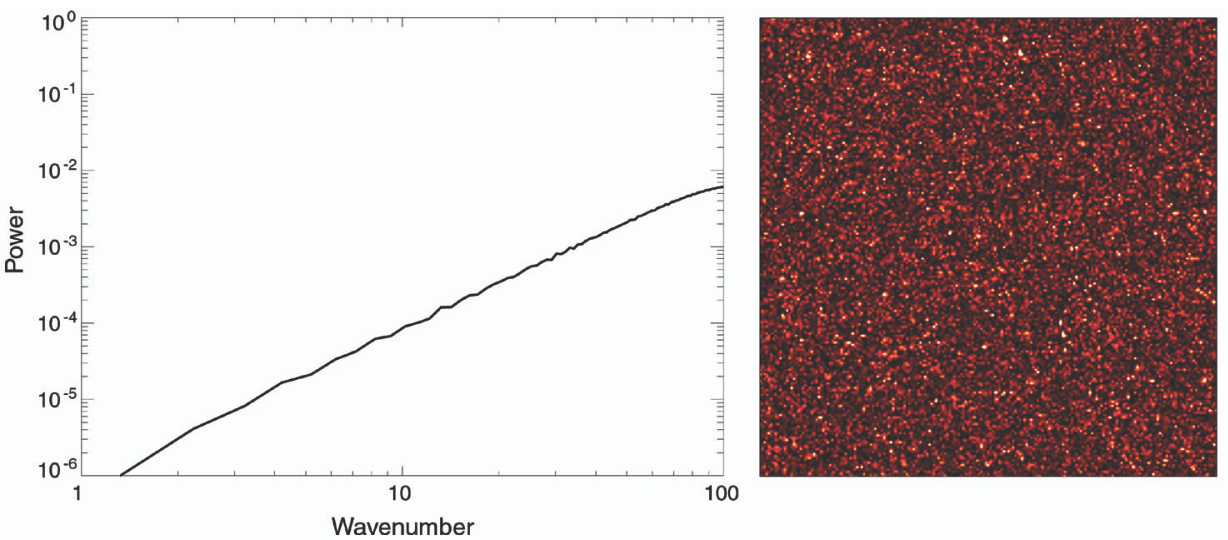,width=0.75\linewidth} \caption[2D slices
of a power-law distributed turbulent magnetic field for different
values of $\mu$]{2D-slices of a turbulent magnetic field for three
different values of the spectral power law index $\mu$: $\mu=-2$
(red noise, {\it top}), $\mu=0$ (white noise, {\it middle}) and
$\mu=2$ (blue noise, {\it bottom}). }\label{fig:div_mu_cont}
\end{center}
\end{figure}
\subsection{The 3D jitter spectrum}\label{sec:jitter}
Jitter radiation is in family with undulator radiation, except that
the magnetic field has power in many Fourier harmonics rather than
only one, and in general follows a power-law in Fourier space. We
define the power-spectrum of the magnetic field
\begin{equation}
P_B(k)=C_B k^{\mu}\label{eq:b_k}
\end{equation}
 (note that \cite{bib:Medvedev_jitter} defines the spectrum as $B_k=C_B
 k^{\mu}$,
 which means that there is a factor 2 difference on $\mu$ compared to our definition).
In the original calculations, \cite{bib:Medvedev_jitter} assumed
that $\mu\ge1$. However, large-scale PIC-simulations (e.g.\
\cite{bib:frederiksen2004}) have shown that the magnetic field
generated by the non-linear Weibel two-stream instability is indeed
power-law distributed but has $\mu<0$.

\begin{figure}[htb]
\begin{center}
\epsfig{figure=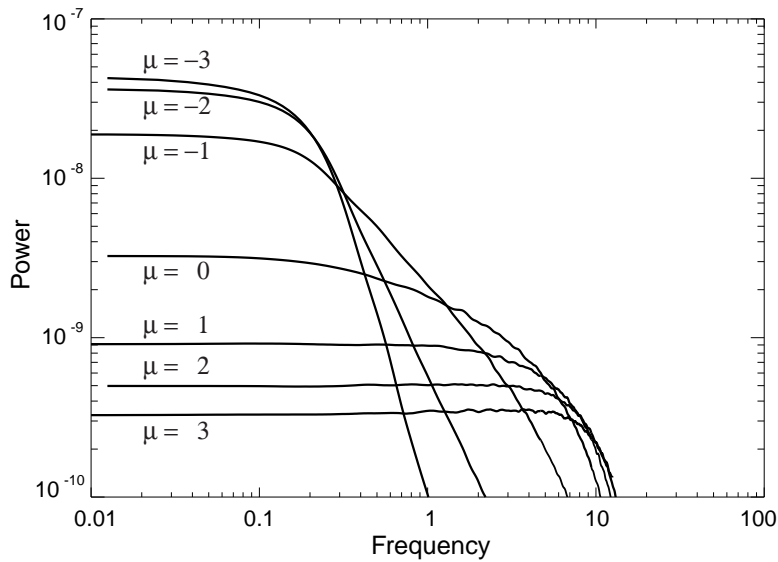,width=0.75\linewidth} \caption[Jitter
spectra from electrons, moving in a turbulent magnetic field, for
different values of $\mu$]{The jitter spectra from a mono-energetic
ensemble of electrons ($\gamma=3$) moving in a turbulent magnetic
field. The different graphs are for simulations with different
values of $\mu$. The amplitude of the magnetic field is set so that
the periodic magnet parameter $K<1$ for all $k$-nodes (Eq.\
\ref{eq:magnet}).}\label{fig:div_mu}
\end{center}
\end{figure}

In this section we investigate the radiation from an ensemble of
relativistic electrons, moving in a random magnetic field, for
different values of the spectral power law index $\mu$ (see Fig.\
\ref{fig:div_mu_cont}). The result is a three-dimensional
generalization of the jitter-theory that differs substantially from
the one-dimensional result found by \cite{bib:Medvedev_jitter}. In
this thesis, we do not include the regime where an additional
large-scale amplitude field is present \citep{bib:Medvedev_jitter}.

We have performed simulations of an isotropic mono-energetic
ensemble of electrons ($\gamma=5$). The isotropic distributions may
be seen as an integration over all angles of the undulator equation.
The electrons are placed in a simulation box with
$200\times200\times800$ grid-zones with a turbulent magnetic field
that has a Fourier distribution as in Eq.\ \ref{eq:b_k}. Seven
different values of $\mu$ have been examined. The amplitude of the
magnetic field is set so that the mean magnetic field energy is the
same in each simulation. The periodic magnet parameter $K$ is less
than 1 for all Fourier nodes. The range of the power-law is limited
to $k_{min}<k<k_{max}$ where $k_{min}$ and $k_{max}$ correspond to
minimum and maximum frequencies that can be represented by the
simulation grid (Fig.\ \ref{fig:div_mu_cont}).
\begin{figure}[htb]
\begin{center}
\epsfig{figure=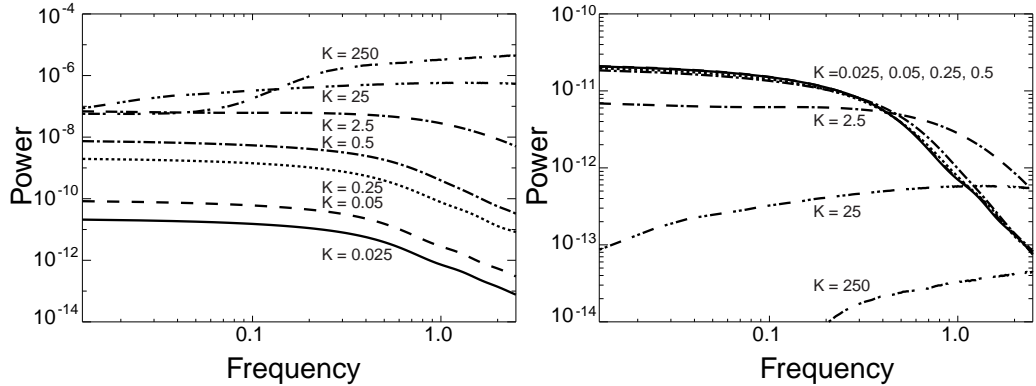,width=\linewidth} \caption[The
transition from undulator radiation to wiggler radiation]{Increasing
the magnetic field increases the magnet parameter $K$ (Eq.\
\ref{eq:magnet}). The left panel shows how the amplitude of the
undulator spectrum increases with $\langle B^2\rangle $. When $K$
becomes larger than 1, a transition to the wiggler radiation
spectrum occurs. The right panel shows the same spectra divided with
$\langle B^2\rangle$.}\label{fig:undulator1}
\end{center}
\end{figure}

The resulting 3D jitter radiation spectrum from the electrons can be
seen in Fig.\ \ref{fig:div_mu}. Quite interestingly, all the spectra
have similarity with the bremsstrahlung spectrum for low energies
(flat spectrum). At higher energies, the $\mu>0$ cases drop off very
sharply -- faster than the bremsstrahlung spectrum. For $\mu<0$ the
high-energy part of the spectrum is found empirically to be a
power-law with a slope of $\omega^{\mu-1}$. The reason why the
$\mu>0$ peaks at so much higher frequencies than for $\mu<0$ can be
understood from Eq.\ \ref{eq:undulator}, which shows that the
characteristic frequency scales as $k$ ($\sim 1/\lambda_u$).

We have also tested how the spectrum depends on the magnetic field
amplitude. From Fig.\ \ref{fig:undulator1} we find that the
amplitude scales with $B^2$ as expected from Larmor's radiation
formula. Above the limit where the deflection angle becomes
comparable to, or larger than the beaming angle ($K>1$), the
spectrum enters the wiggler-domain. The wiggler radiation spectrum
is close to the synchrotron spectrum \citep{bib:attwood}.

\begin{figure}[htb]
\begin{center}
\epsfig{figure=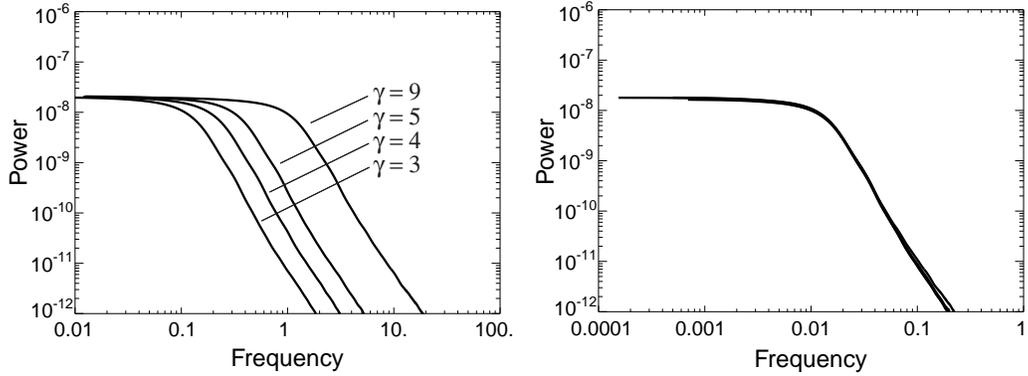,width=\linewidth} \caption[Jitter
spectra from electrons moving in a turbulent magnetic field with
$\mu<0$]{The radiation spectrum from four different mono-energetic
ensembles of electrons ($\gamma$=3, 4, 5, and 9) with isotropic
distributed momentum-vectors, placed in a turbulent magnetic field
with spectral slope $\mu=-3$ ({\it left panel}). The right panel
shows the same four graphs but shifted in frequency with a factor
$\gamma^{-2}$ (renormalized).}\label{fig:jitter_n3}
\end{center}
\end{figure}

\begin{figure}[htb]
\begin{center}
\epsfig{figure=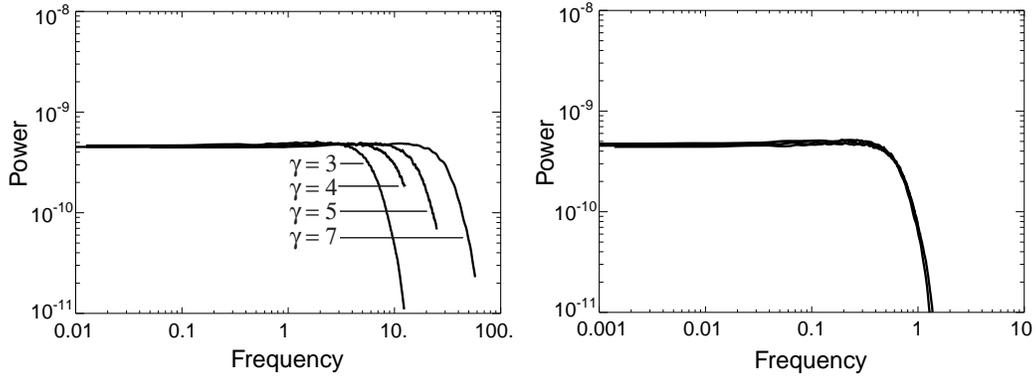,width=\linewidth} \caption[Jitter
spectra from electrons moving in a turbulent magnetic field with
$\mu>0$]{The radiation spectrum from four different mono-energetic
ensembles of electrons ($\gamma$=3, 4, 5, and 7) with isotropic
distributed momentum vectors, placed in a turbulent magnetic field
with spectral slope $\mu=+3$ ({\it left panel}).  The right panel
shows the same four graphs but shifted in frequency with a factor
$\gamma^{-2}$ (renormalized).}\label{fig:jitter_p3}
\end{center}
\end{figure}

We finally test the $\gamma$-dependency. Figure \ref{fig:jitter_n3}
and Fig.\ \ref{fig:jitter_p3} show the jitter radiation spectra for
$\mu<0$ and $\mu>0$ for electron ensembles with different energies.
Even though the spectra in the two cases differ, they both scale the
same way: The spectra are shifted with $\gamma^2$ in frequency in
agreement with Eq.\ \ref{eq:p_rad_a} (and Eq.\ \ref{eq:undulator}).

\subsection{Jitter radiation from collisionless plasma shocks?}
In this section, we compare the radiation spectrum from an ensemble
of electrons, moving in two different magnetic topologies: 1) the
magnetic field generated by the Weibel two-stream magnetic field
$\vect{B}_0$ (see Section \ref{sec:weibel} in this thesis,
\citec{bib:medvedevloeb} and \citec{bib:frederiksen2004}) and 2) a
randomized magnetic field $\vect{B}_1$ that has the same average
energy density and spectral power spectrum as the two-stream
generated field. The purpose of this exercise is to test if there
are any differences in the resulting radiation spectra. $\vect{B}_0$
is readily available from the PIC simulations. Constructing
$\vect{B}_1$ is more tricky. To make sure that $\vect{B}_1$ has
exactly the same average energy density and power spectrum
properties as the PIC-field $\vect{B}_0$, we generate the random
field in the following way:

The basic idea is to Fourier transform the magnetic field from
$\vect{B}_0$, make a random phase shift of all $k$-nodes in the
Fourier domain and then transform back to real-space. This is easily
done but has one serious flaw: we cannot be sure that the resulting
magnetic field obeys $\nabla\cdot\vect{B}_1=0$ even though
$\vect{B}_0$ does. The solution is to work on the vector potential
rather than directly on
 $\vect{B}_0$.
The vector potential $\vect{A}_0$ is a vector field that satisfies
$\vect{B}_0=\nabla\times\vect{A}_0$. Inverting this equation
requires $\vect{B}_0$ to be periodic. In the PIC simulations,
$\vect{B}_0$ is already periodic in the two-dimensions perpendicular
to the jet flow ($x,y$). To make the magnetic field periodic in the
$z$-direction we apply a windowing function that connects the
magnetic field-lines through the $z$-boundary. Away from the
boundary, the magnetic field remains unchanged. We have tested that
this filter does not alter the power spectrum of $\vect{B}_0$ except
for a few percent in the very highest frequencies. Ensured that
$\vect{B}_0$ is periodic, the Fourier transform of $\vect{A}_0$ into
$\hat{\vect{A}}_0$ is
\begin{equation}
\hat{\vect{A}}_0(\vect{k})\equiv\mathrm{FFT}(\vect{A}_0)=i\vect{k}\times\mathrm{FFT}(\vect{B}_0)/|\vect{k}|^2.
\end{equation}
$\vect{k}$ is the wave vector and FFT represent a fast Fourier
transformation.

\begin{figure}[thb]
\begin{center}
\epsfig{figure=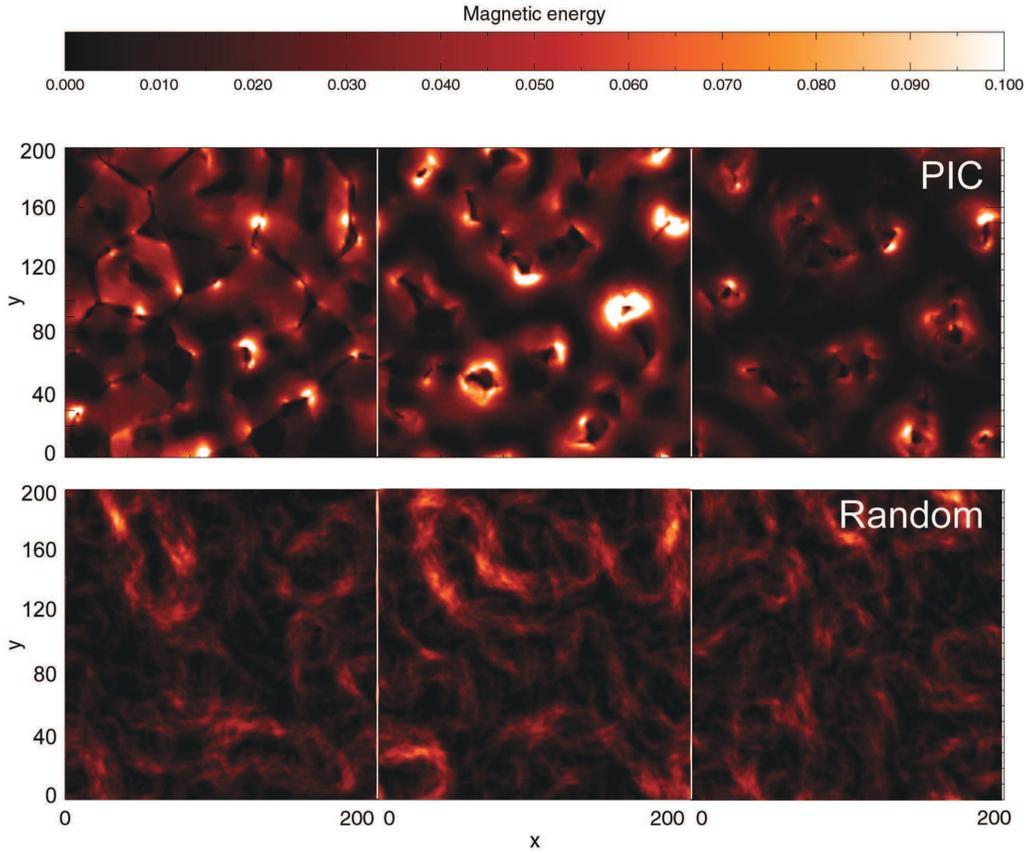,width=\linewidth} \caption[Slices
of the two-stream generated magnetic field compared with a random
field]{The spatial distribution of magnetic energy in three slices
sampled at different depth in the plasma shock. The top panel shows
$\vect{B}_0$ generated by the two-stream instability. The lower
panel shows a random magnetic field $\vect{B}_1$ with the same
spectral power distribution as $\vect{B}_0$.
}\label{fig:b_shift_cont}
\end{center}
\end{figure}

We then scramble the vector potential by a random phase shift of all
the Fourier harmonics
\begin{equation}
\hat{\vect{A}}_1(\vect{k})=\hat{\vect{A}}_0(\vect{k})
e^{i2\pi\vect{k}\tilde{f}},
\end{equation}
where $\tilde{f}$ is a matrix of random numbers in the range
$[0;1]$. The size of $\tilde{f}$ matches the total number of data
points in the three-dimensional ${\vect{B}}_0$-array. $\vect{B}_1$
is then given as $\vect{B}_1=\nabla\times\vect{A}_1$. The
transformation is performed in IDL (Interactive Data Language by
Research Systems Inc.) and this adds a few complications to the
process, mainly connected to the representation and renormalization
of the Fourier transforms. To ensure that no errors have been
introduced in the process of obtaining $\vect{B}_1$, we check and
find that
\begin{itemize}
\item $\vect{B}_0$ and $\vect{B}_1$ have identical Fourier transforms
and equivalently, identical spectral power distribution,
\item $\vect{B}_1$ is real if $\vect{B}_0$ is real,
\item the magnetic energy is conserved through the transformation
$\int\vect{B}_1^2dV=\int\vect{B}_0^2dV$, and
\item $\vect{B}_1$ is divergence free ($\nabla\cdot\vect{B}_1=0$).
\end{itemize}

\begin{figure}[thb]
\begin{center}
\epsfig{figure=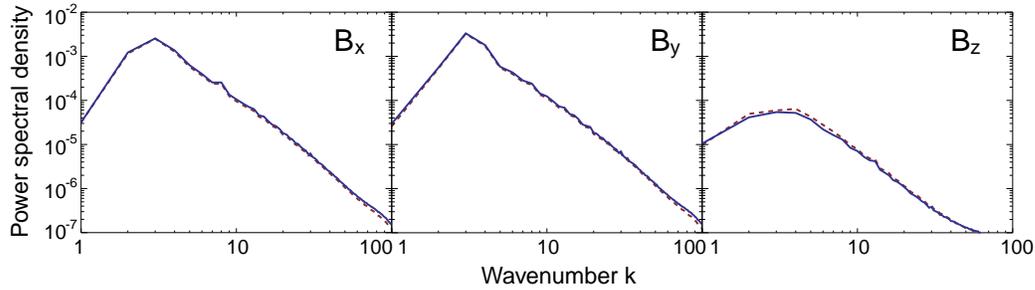,width=\linewidth} \caption[Power spectrum
of the magnetic field generated by the Weibel two-stream
instability]{Power spectrum of the magnetic field generated by the
Weibel two-stream instability $\vect{B}_0$ ({\it full line}) and the
random phase shifted field $\vect{B}_1$ ({\it dotted
line}).}\label{fig:power_b}
\end{center}
\end{figure}

Figure \ref{fig:b_shift_cont} shows the input magnetic field
$\vect{B}_0$ and the resulting $\vect{B}_1$. Three slices are shown
for each field, sampled at different depths in the shock. Figure
\ref{fig:power_b} shows the power spectrum of $\vect{B}_0$ and
$\vect{B}_1$. Clearly, the randomized magnetic field $\vect{B}_1$
has the same statistical characteristics as the magnetic field from
the simulations.

\begin{figure}[thb]
\begin{center}
\epsfig{figure=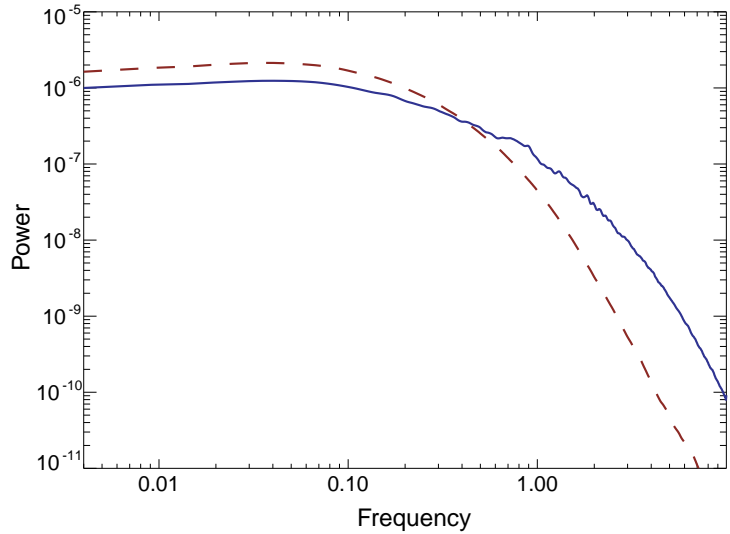,width=.7\linewidth} \caption[Radiation
from PIC simulations compared to radiation from a random
field]{Radiation from PIC simulations ({\it solid blue}) compared to
radiation from a random field with the same statistical properties
({\it dashed red}). Both spectra are generated by using the same
ensemble of 20000 electrons.}\label{fig:pic_random}
\end{center}
\end{figure}

In Fig.\ \ref{fig:pic_random} we compare the radiation spectrum from
a thermal electron ensemble in  $\vect{B}_0$, with the spectrum of
the same ensemble in $\vect{B}_1$. $\vect{B}_0$ is from PIC
simulations of a plasma shock ($\Gamma=3$ simulation from Chapter
\ref{sec:weibel}). $\vect{B}_1$ is created as described above. Both
spectra are generated by tracing 20000 electrons. To make the
spectra as realistic as possible, the electron positions and
velocities are sampled from a snapshot of the PIC simulation.

We find that the two spectra peak at the same frequency but diverge
considerably for high frequencies where the radiation spectrum from
the random field $\vect{B}_1$ has a harder cutoff. The reason for
this is because of the higher peak-values in the magnetic field from
the PIC simulations (see Fig.\ \ref{fig:b_shift_cont}). Even though
$\vect{B}_0$ and $\vect{B}_1$ have the exact same power-spectrum,
there exists phase-correlations in $\vect{B}_0$ that give the field
a more coherent structure ($\vect{B}_1$ has a larger volume filling
factor). We know also that the frequency where the synchrotron
spectrum peaks $\nu_c$ scales linearly with the magnetic field
strength. Finally, the high-energy electrons are primarily found
near ion-current channels where the magnetic field peaks (see Fig.\
\ref{fig:Slice}, \ref{fig:acceleration} and \ref{fig:jvg}). We
conclude that it would appear that one may use a random magnetic
field for analytical calculations, but only as a first
approximation. We also stress that the simulation results we have
used are from simulations that do not include the full shock ramp.
The spectra may diverge even more for radiation from a full shock.

\subsubsection{Radiation due to Electric Fields}
\begin{figure}[htb]
\begin{center}
\epsfig{figure=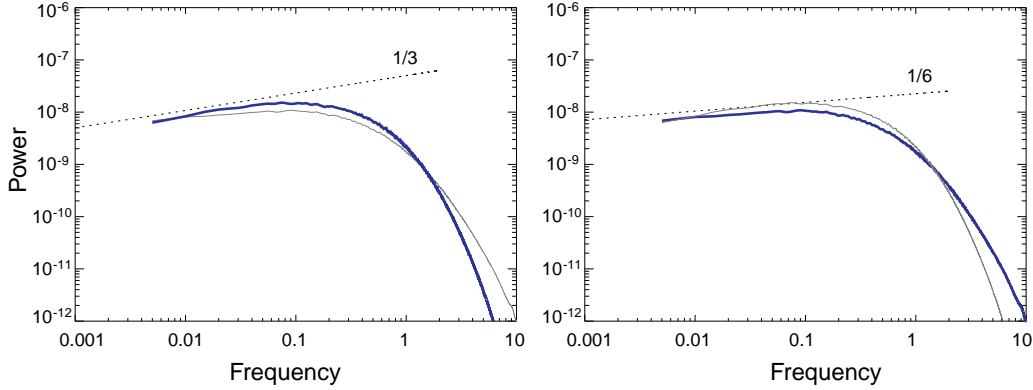,width=\linewidth} \caption[Influence
on a magnetic field on the radiation spectrum]{The spectrum of a
mono-energetic ($\gamma=3$), isotropic electron population in the
electromagnetic field generated by the Weibel two-stream
instability. In the left panel, only the magnetic field is included
whereas the right panel includes both the magnetic and electric
field. The grey shaded line in the left panel is a copy of the
spectrum in the right panel and visa verse. 20000 particles were
traced in each spectrum. Both axis are in arbitrary
units.}\label{fig:spec_b-e}
\end{center}
\end{figure}

Electric fields are most often neglected in generation of radiation
from astrophysical shocks. In PIC simulations we have found that for
collisionless shocks where the magnetic field is generated by the
Weibel two-stream instability, electric fields exist with an energy
density of the order of  $10\%$ of the generated magnetic field.
Figure \ref{fig:spec_b-e} shows two radiation spectra emitted from a
collisionless shock. The electron distribution and electromagnetic
field that lie behind the radiation are taken directly from
self-consistent PIC code simulations. The left panel of
\ref{fig:spec_b-e} shows a spectrum where only the magnetic field is
taken into account, while the right panel shows the spectrum when
the electric field is also included. For a pure magnetic field we
immediately identify the wiggler/synchrotron signature from the
$\omega^{1/3}$ low energy slope. The main effect from including the
electric field is that the low energy slope is flattened to
$\sim\omega^{1/6}$ and the high-energy exponential cut-off is softer
and goes to higher energies. From these plots we conclude that it is
important to also include the electric field generated in the Weibel
two-stream instability.

\section{Spectra from the PIC simulations rescaled into 'real space'}
The ultimate goal is to obtain spectra from the PIC shock
simulations that can be compared directly to observations. Even
though it is beyond the scope of this thesis, this will eventually
allow us to put constraints on the physics of GRB afterglows.

In the PIC simulations, several parameters have been scaled to
simulation units. To make the simulation results more directly
comparable to real observations, it is necessary to correct for
this. Rescaling the radiation spectra obtained from simulations
involves several effects:

\begin{description}
\item[Simulation time-scales:] To rescale the time-scale of the simulations to real space,
we must divide all frequencies with the plasma frequency in the
upstream part of the simulation box $\omega_p'$, and multiply with
the real space (ISM) plasma frequency $\omega_p$. That is, we have
to multiply with a factor $R_1=\omega_p/\omega_p'$. In the
non-linear phase of the Weibel two-stream instability, the ions are
dominating the evolution of the instability. Therefore, we set
$\omega_p'$ equal to the upstream ion plasma frequency from the
simulations ($\omega_p'=0.025$). Since we are focusing on the
afterglow domain in the simulations, we choose the ion plasma
frequency of the inter-stellar medium for the real space plasma
frequency, ($\omega_p\simeq1300\ \mathrm{s}^{-1}$). In this case,
$R_1=1300/0.025\simeq5\cdot10^4$.
\item[Ion to electron mass ratio:] For numerical convenience, the ion (proton) to electron mass ratio are rescaled
(see Section \ref{sec:pic}). The ion/electron charge ratio remains
unity. From the simulations it is found that the ions dominate the
Weibel two-stream instability. Therefore, we can see the mass
rescaling as an unphysical rescaling of the electron mass/charge
ratio. In the code, the ion/electron mass ratio is typically 16.
Using the real mass ration 1836 would mean that the electrons become
accelerated to higher Lorentz factors (but the same energy) by a
factor 1836/16. In Section \ref{sec:jitter} we found that the peak
frequency in the spectra scales with $\gamma^2$. This is consistent
with synchrotron radiation \citep{bib:rybicki}: In synchrotron
radiation, the peak frequency is proportional to $\gamma^2/m_e$.
Thus we must effectively shift the radiation spectrum up in
frequency by a factor $R_2=(1836/16)^3=1.5\cdot10^6$
\item[Relativistic Doppler shift:] All the simulations are carried out in the rest frame of the shock, moving with a bulk lorentz factor
$\Gamma$ towards the observer (only GRB jets moving directly towards
us will trigger the space telescopes as a result of the relativistic
beaming). We must therefore shift the simulation radiation spectra
with a relativistic Doppler correction term $R_3\simeq2\Gamma$.
\end{description}

\begin{figure}[htb]
\begin{center}
\epsfig{figure=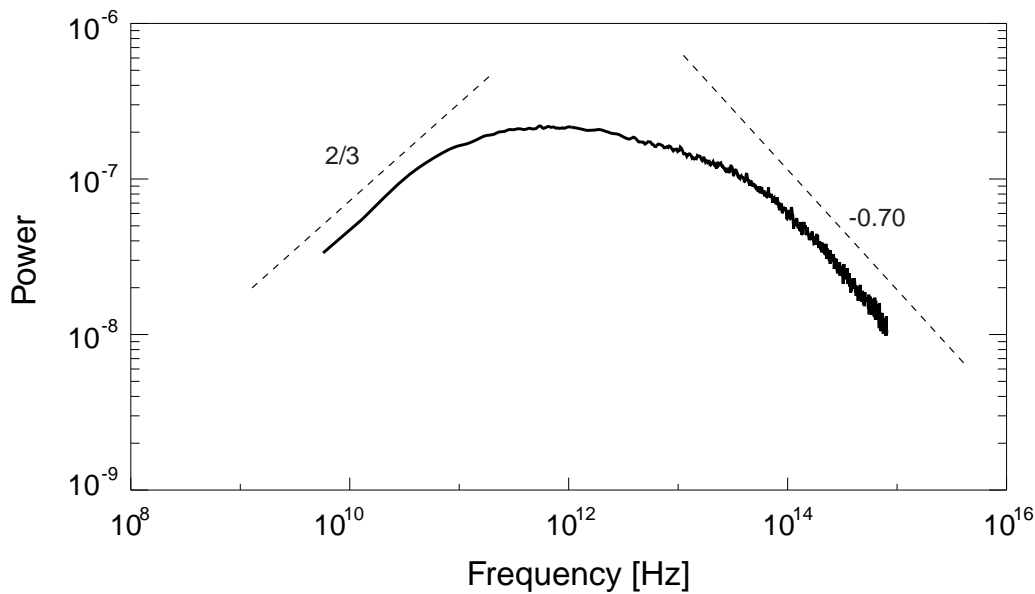,width=\linewidth} \caption[A
synthetic spectrum from a gamma-15 shock wave]{The spectrum from a
$\Gamma=15$ plasma shock propagating into an ISM-like medium. The
frequency scale is boosted in accordance with the physical arguments
given in the text. The spectrum peaks in the far infrared. Below the
peak, we find a power-law segment with slope
$P(\omega)\propto\omega^{2/3}$. For frequencies above the peak
frequency, the spectrum continues into the near infrared/optical
band, following a power-law with slope
$P(\omega)\propto\omega^{-\beta}$, where $\beta=0.7$.
}\label{fig:spec_g15}
\end{center}
\end{figure}

In summary, to compare the spectrum we have obtained from the
simulations of a $\Gamma=15$ shock propagating through the ISM, we
must shift the frequency axis with a factor $R=R_1 R_2
R_3\simeq2\cdot10^{12}$. This puts us somewhere near the optical
frequency band. In Fig.\ \ref{fig:spec_g15} we show the radiated
spectrum, sampled from 20.000 electrons. After applying the
rescaling, we find that the spectrum peaks in the far infrared area.
Below the peak, we find a power-law segment with slope
$P(\omega)\propto\omega^{2/3}$. This is interesting because it is
steeper than the synchrotron $1/3$ slope and thus might shed light
on the issue about "the line of death" \citep{bib:preece1998} for
many bursts. For frequencies above the peak frequency, the spectrum
continues into the near infrared/optical band, following a power-law
with slope $P(\omega)\propto\omega^{-\beta}$ with $\beta\simeq0.7$.

From the power-law slope in the spectrum with ($\beta=0.7$), the
standard procedure in the GRB community would find that the
electrons have a power-law distribution, with slope $p=2.4$
(determined by solving $(p-1)/2=\beta$). The standard conclusion
would be that this is consistent with Fermi acceleration. However,
we strongly emphasize that the electrons that produced the spectrum
in Fig.\ \ref{fig:spec_g15} are not Fermi accelerated. The electrons
are instantaneously accelerated and decelerated in the highly
complicated electric and magnetic field near the ion current
channels under emission of strong radiation (see an example of the
complicated electron-paths in Fig.\ \ref{fig:tracerpar}). This
differs substantially from the iterative acceleration in Fermi
acceleration.

We note that the box of the PIC simulation behind the spectrum
presented in Fig.\ \ref{fig:spec_g15} is not large enough to cover
the whole shock ramp (ions are not fully thermalized when they leave
the simulation domain). This means that additional features might be
added to the spectrum for even larger simulations. Moreover, the
spectrum does not include synchrotron self-absorbtion.

\section{Summary}

I have developed, and tested, a tool can create radiation spectra
directly from particle-in-cell simulations. By tracing particles,
the code performs the exact radiation field Fourier integration
(Eq.\ \ref{eq:retard_fourier}) without having to make assumptions
about the magnetic field, particle orbit, beaming angle, and so
forth.

The radiation tool has been thoroughly tested and successfully
reproduces synchrotron radiation, bremsstrahlung, and undulator
radiation from small-angle deflections.

I have used the tool to investigate the properties of
three-dimensional jitter radiation in magnetic fields with different
turbulent configurations. By tracing isotropic momentum
distributions of electrons through a random magnetic field with a
power spectrum that follows a power-law distribution in the Fourier
domain $P_B(k)\propto k^{\mu}$, I have determined the resulting
spectrum. I have focused on the weak field limit (where the
deflection angle is less than the beaming cone angle, $K<1$). For
all values of $\mu$, the radiation spectrum has a flat low-energy
part (similar to bremsstrahlung). For $\mu<0$, the high-energy part
of the spectrum follows a power-law with a slope $\alpha\simeq\mu-1$
independent of the electron energy. For $\mu>0$, the flat spectrum
continues to higher frequencies than for $\mu<0$. The spectrum has a
hard cut-off for high frequencies. With the present simulations, it
is not possible to determine if the spectrum has peaks for even
higher frequencies.

 For all values
of $\mu$ and $K<1$, the spectrum shifts in frequency with $\gamma^2$
and scales with the amplitude of the magnetic field squared $B^2$,
in good agreement with Larmor's formula for radiated power. In the
limit of large deflections ($K>1$) the spectrum eventually converges
through the wiggler spectrum to the ordinary synchrotron spectrum.

For all the spectra presented in this chapter, I have checked for
convergence: For each spectrum obtained I have made a second run
with twice the number of particles. If the two spectra have
different shape, the process is repeated by doubling the particle
number until the result converges. This process is important since
radiation from high-energy particles is highly beamed. This is prone
to distort the high-energy radiation spectrum because of bad
statistic.

Computationally, the generation of spectra from simulations is not
an easy task. One may say that it is an example where {\it nature is
not nice} to scientists. Especially, obtaining a radiation spectrum
from high-energy particles is not easy since it is necessary to
integrate for a time period that grows linearly with $\gamma$ in
order to sample at least a couple of orbits. But at the same time it
is crucial to sample a very high number of time-steps per time unit,
to resolve the peak in the electrical field, which is very narrow
($\gamma^{-2}$) because of relativistic beaming. So in total, the
computational cost grows with $\gamma^3$. The good news is that each
particle and frequency bin may be treated independently, which makes
the problem {\em embarrassingly parallel} (in the jargon of High
Performance Computing).

The method of tracing particles in an electromagnetic field that is
fixed to a single snapshot is clearly an approximation. For high
frequency spectra, this is not a problem, since we sample for a very
short period of time. For low frequencies, it is a potentially more
severe approximation since it doesn't make sense to trace the
particle for too long time in a field that would have changed
significantly during this process. Improvements can be made in
several ways. One method is to pick out a number of particles before
the large simulation starts and then integrate these particles with
a high number of sub-steps. Another approach would be to interpolate
field values to the particles position not only in the spatial
dimensions (which is already implemented) but also in time from
different field snap shots.

An interesting step for the future is to determine the polarization
in the synthetic spectra from the PIC simulations. Even though time
did not allow for it in this thesis, it should not be difficult with
the tools that are already developed. One could then investigate how
the radiation spectrum, including polarization change as functions
of the viewing angle measured from the jet propagation direction.

In the near future, the computational resources will reach a level
where one can resolve fully three-dimensional collisionless shocks.
It then becomes possible to make composite spectra from different
simulations with varying bulk Lorentz factor as a function of
viewing angle. This will allow us to test many interesting aspects
with regard to jet structure and polarization predictions from
shock-generated electromagnetic fields.

I end this chapter by emphasizing that strong magnetic field
generation, particle acceleration and emission of non-thermal
radiation are unavoidable consequences of the collision of two,
initially unmagnetized, plasma shells.

\chapter{A next generation PIC code}\label{chap:photonplasma}
\section{Introduction}
Over the last couple of years the Copenhagen group has been using
PIC models that include electromagnetic fields and charged particles
to understand the plasma microphysics of collisionless shocks
\cite{bib:frederiksen2002,bib:frederiksen2004,bib:hededal2004,bib:hededal2005}.
It has turned out to be a very successful tool, but it is still
limited in the scope of phenomena that can be addressed. Even though
a large class of astrophysical environments are indeed
collisionless, scattering and collision processes do play an
important role in several key scenarios. Examples are given below.
Another key ingredient, which has been missing in charged particle
simulations, is a full treatment of photon propagation. It can be
argued that photons are represented directly on the mesh by
electromagnetic waves, which certainly is correct. But the mesh can
only represent waves with frequencies smaller than the Nyquist
frequency. The physical length of a typical cell has in our
applications typically been $10^5-10^6\ \textrm{cm}$ and hence it is
clear that only low frequency radio waves can be represented.
High-frequency photons have to be implemented as particles that
propagate through the box and interact, either indirectly through
messenger fields on the mesh, or directly with other particles. A
valuable consequence of modelling the detailed photon transport is
that extraction of electromagnetic spectra is trivial. Even in cases
where the photon field is only a passive participant, this fact
should not be underestimated as it enables direct comparison with
observations.

There exists Monte Carlo based particle codes (see
e.g.~\cite{bib:stern95} and references therein) that address various
particle interactions, but one of their main shortcomings is the
poor spatial resolution. This makes it impossible to couple the
particle aspects to a self-consistent evolution of the plasma.

Our goal has been to develop a framework where both
electromagnetic fields and scattering processes are included in a consistent way. We can then
correctly model the plasma physics and the radiative dynamics. The
scattering processes include, but are not limited to, simple
particle-particle scattering, decay and annihilation/creation
processes. Our new code is not limited in any way to charged
particles, but can also include neutrals such as photons and neutrons.

In the next section we describe some of the physics that can be
addressed with this new code. In section \ref{sec:NGPimplementation}
we discuss how the code has been implemented, the general framework
and in detail, which physical processes that are currently
implemented. In section \ref{sec:NGPresults} we present the results
of a preliminary toy experiment that we have performed to validate
the code. In the last section \ref{sec:NGPdiscussion} we summarize.

\subsection{Motivation}
Before we continue and describe in detail the methods, physics and
test problems we have implemented and used, it is important to
consider the general class of scenarios we have had in mind as
motivation for developing the code. There are several key objects,
where only the bulk dynamics is understood, and we are lacking
detailed understanding of the microphysics.

\subsubsection{Internal shocks in Gamma-Ray Bursts}
In the internal/external GRB shock model, the burst of gamma-rays is
believed to be generated when relativistic shells collide and dissipate
their relative bulk energy \cite{bib:rees1992,bib:meszaros1993}.
The nature of the radiation is presumably inverse Compton scattering and
synchrotron radiation. Particle/photon
interactions might also play an important role in the very early
afterglow as suggested by
\cite{bib:thompson2000,bib:beloborodov2002}: Even though the medium
that surrounds the burst (ISM or wind) is optically very thin to
gamma-rays, a tiny fraction of the gamma-rays will Compton
scatter on the surrounding plasma particles. This opens up for the
possibility of pair-creation between back scattered and outgoing
gamma-rays. The creation of pairs may increase the rate of back
scattered photons in a run-away process \cite{bib:stern2003}.
The Compton scattering may accelerate the pair-plasma through the surrounding medium with many
complicated and non-linear effects, including streaming plasma
instabilities and electromagnetic field generation. Hence, it is
crucial that plasma simulations of internal GRB plasma shocks
include lepton-photon interactions.

\subsubsection{Solar corona and the solar wind}
Space weather (defined as the interaction of the solar wind on the
Earth) is in high focus for several reasons. Not only is the Sun our
closest star, providing us with invaluable data for stellar
modelling, but also coronal mass ejections from the Sun potentially
have impact on our every day life. The strong plasma outflows from
the sun can induce large electrical discharges in the Earths
ionosphere. This may disrupt the complex power grids on Earth,
causing rolling blakcouts such as the one in Canada and North
America in 1989. Also high-energy particles can be hazardous to
astronauts and airline passengers. Computer simulations have
provided a successful way of obtaining insight in these complex
plasma physical processes. However, in the solar coronal and in the
solar wind plasma out to distances beyond the earth orbit,
difficulties arise in finding the right formalism to describe the
plasma. Neither a collisionless model based on the Vlasov equation
nor an MHD fluid model provides a adequate framework for
investigation. The problem has already been studied using three
dimensional PIC simulations but without taking collisions into
account (e.g. \cite{bib:buneman1992,bib:hesse2001}).

\subsubsection{The corona of compact objects}
The bulk dynamics of accreting compact objects have been modelled
for many years using fluid based simulations (e.g.
\cite{bib:balbus2003} and references therein). Nevertheless, it has
been a persistent problem to extract information about the radiating
processes. Furthermore in the corona the MHD approximation becomes
dubious, just as in the solar corona. The environment around a
compact object is much more energetic than the solar corona, and
therefore radiative scattering processes play an important role.
Pair production is also believed to be abundant. Using our new code
it would be possible to model a small sub box of the corona. The
main problem here -- as in most numerical implementations --  is to
come up with realistic boundaries for the local model. A shearing
box approach may be appropriate, but in fact we can do even better.

The size of a stellar mass black hole is around $10^6\ \textrm{cm}$.
In a fluid simulation we want to model the accretion disk--compact
object system out to hundreds of radii of the compact object. The
normal approach is to use a non-uniform mesh. Nonetheless, the
Courant criteria, which determines the time step, is still limited
by the sound crossing time of the compact object. I.e.~the time step
is limited by the size of the innermost (and smallest) cells in the
mesh. The very small time step corresponds to those found in a
typical particle simulation, where the strict time step arises from
the need to resolve plasma oscillations. Hence data from an MHD
simulation could provide temporally well-resolved fluxes on the
boundaries of the much smaller sub box containing the particle
simulation.

In this sense the particle simulation will act as a probe or
thermometer of the fluid model. The particle model includes the
full microphysics in a realistic manner and most importantly
includes photon transport. Realistic spectra
could be obtained indirectly from the fluid model, testing
fluid theory against observations. We have already started
preliminary work on interfacing fluid models with the old PIC code.

\subsubsection{Pre-acceleration in Cosmic Ray acceleration}
Accepting Fermi acceleration as a viable mechanism for accelerating
electrons and creating the non-thermal cosmic ray spectrum is still
left with some big unanswered questions. One is that the Fermi
mechanism requires injection of high-energy electrons while still
keeping a large, low-energy population to sustain the magnetic
turbulence. Hence, a pre-acceleration mechanism needs to be
explained.

The shocks in supernova remnants are believed to be cosmic ray
accelerators. However, the Fermi acceleration process in shocks is
still not understood from first principles but rely on assumptions
on the electromagnetic scattering mechanism. PIC codes would seem
ideal in exploring the mechanism from first principles, since they
include field generation mechanisms and the back-reaction that the
high-energy particles have on this scattering agent. In Supernova
remnants however, the mean free path for Coulomb collisions are
comparable to the system and particle-particle interactions cannot
be fully neglected.

\section{Implementation}\label{sec:NGPimplementation}
Implementing any state-of-the-art large-scale numerical code is a
big undertaking, and can easily end up taking several man years. We
estimate that the final version of the next generation code will
contain more than 50.000 lines of code. Starting in February this
year, it has taken us three man months to implement the current
incarnation of the code, which has already grown to approximately
10.000 lines. Besides T.~Haugb{\o}lle and C.~B.~Hededal, the
development is done together with {\AA}.~Nordlund and
J.~T.~Frederiksen. Fortunately we have a good tradition and
expertise for numerical astrophysics in Copenhagen and we have been
able to port different technical concepts and solutions from our
suite of fluid codes and to a lesser extent from the old PIC code.
The aim is to build an extremely scalable code that is able to run
on thousands of CPUs on modern cluster architectures and utilize MPI
as the inter node communication protocol. In this chapter we will
not go further into technical details. Instead we will put emphasis
on the important concepts and physics and how we have implemented
these.

\subsection{Concepts}
The two fundamental objects in a particle-in-cell code are the mesh
and the particles. We have adopted the solver and interpolation
routines from the old PIC code to solve the Maxwell equations and
find fluxes and densities on the mesh. The mesh is used to
distribute messenger fields -- such as the electromagnetic fields --
and to calculate volume averaged fluxes and densities of the
particles The latter are used as source terms in the evolution of
the messenger fields. The particles really represent an ensemble of
particles and are often referred to as \emph{pseudoparticles}
\cite{bib:birdsall} or {\em large particles}. A so-called smoothing
kernel describes the density distribution of a single pseudoparticle
on the mesh. In our implementation the volume of a particle is
comparable to a cell in the mesh.

\subsubsection{Pseudoparticles with variable weights}
The concept of
pseudoparticles is introduced since the ``real space'' particle
density easily exceeds any number that is computationally reasonable
(i.e. of the order of a billion particles). The pseudoparticle charge to mass ratio
is kept the same as the ratio for a single particle.

In ordinary PIC codes the weight of each pseudoparticle of a given
species is kept constant throughout the simulation. The benefit is a
simple code and an unique identity for each particle. The first is a
convenience in the practical implementation, the second important
when understanding the detailed dynamics and history of a single
particle.

Notwithstanding possible conveniences, as detailed below in section
\ref{scat}, we have decided to improve this concept to a more
dynamical implementation where each pseudoparticle carries a
individual weight. Particles are then allowed to merge and split up
when a cell contains too many/few particles, or when particles are
scattered. The concept is sometimes used in smooth particle
hydrodynamics (SPH), where different techniques have been proposed
for the splitting and merging of particles. It is both used to
adjust the density of individual particles \cite{bib:trulsen2001}
and in the conversion of gas-- to star particles in galaxy formation
models \cite{bib:governato2004}. An important quality of SPH is its
adaptive resolution capabilities. These are important in the
description of collapsing self-gravitating systems, ranging from
core collapse supernovae to the formation of galaxy clusters,
scenarios where matter is collapsing many orders of magnitude, and
therefore the smoothing length or volume of the individual particles
is readjusted accordingly. Consequently when splitting particles, or
adjusting the weights in an SPH code it is important to match
precisely the spatial density distribution of the parent particle to
the spatial distribution of the child particles. In PIC codes though
the spatial size or smoothing parameter of an individual particle is
determined beforehand by the mesh spacing. This is reasonable since
we are not interested in adaptive resolution but rather a kinetic
description of the plasma dynamics. Splitting a {\it parent}
particle with weight $w_p$ into {\it child} particles with weights
$w^i_c$ is therefor trivial. The requirements of conservation of
mass and four velocity together with conservation of the density and
flux distribution in the box, can all be satisfied by setting
\begin{align}
w_p &= \sum^n_{i=1} w^i_c & e_p &= e^i_c &
\gamma_p\vec{v}_p &= \gamma^i_c\vec{v}^{\, i}_c
\end{align}
since the smoothing kernel is determined by the mesh spacing, not the
mass of the individual particle.

The merging or renormalization of pseudoparticles requires a much more thorough
analysis. Up to now we have investigated two schemes, one that respects conservation
of mass, energy and four velocity by merging three particles into two at a time,
and one where only mass, energy and average direction is conserved by merging two particles into one
particle. While these schemes probably
work well for approximately thermal distributions, it will easily give
rise to a large numerical heating when considering head on beam collisions.
We believe it can be improved by first selecting a random ``merger particle'' and
then find other particles in the local cell, that are close to the merger
particle in momentum space. A more radical approach is to resample
the full phase distribution in a cell every time the number density becomes
above a certain threshold. Nevertheless, it requires testing of different extreme
situations to find the optimal method to merge particles, and it is still a work in progress.

To obtain the results, that we present in section \ref{sec:NGPresults}, we
ran the code without merging of the pseudoparticles activated.

\subsubsection{Scattering processes and splitting of particles}\label{scat}
In Monte Carlo based particle codes the generic way to compute an
interaction is first to calculate the probability for the interaction
$P_S$, then compute a random number $\alpha$. If $\alpha \le P_S$ then
the full pseudoparticle is scattered otherwise nothing happens. This
probabilistic approach is numerically rather efficient and simple to implement, but
it can be noisy, especially when very few particles are present in a cell.
In large particle Monte Carlo codes the typical cell contains up to $10^4$
particles per species per cell (hence ``large particle''). In our
PIC code typical numbers are $10^1-10^2$ particles per species per cell, since
we need many cells to resolve the plasma dynamics. For our requirements the
probabilistic approach would result in an unacceptable level of noise. For example, in a beam
experiment the spectra of the first generation of scattered particles may come
out relatively precise, but the spectra of higher generation scattered particles
(i.e.~particles that are scattered more than once) will come out with
poor resolution or require an excessive amount of particles. Another well known
consequence of the probabilistic approach is that for a given experiment
the precision goes in the best case inversely proportional to
\emph{the square root} of the number of particles used in the experiment.
\begin{figure}[htb]
\begin{center}
\epsfig{figure=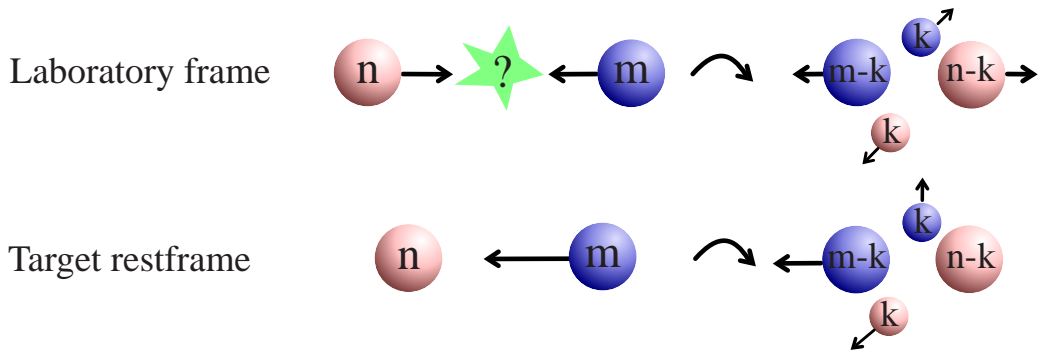,width=\textwidth}
\caption[Schematics of a generic scattering process]{To implement the
scattering of
two pseudoparticles we transform to the rest frame of the target particle
(shown as red/light gray) and computes the probability $P(n)$ that a single
incident particle (shown as blue/dark gray) during a timestep $\Delta t$ is
scattered on the $n$ target particles. If the incident particle has weight
$m$, then $k=P(n) m$ particles will interact and two new pseudoparticles are
created.}
\label{fig:splitting_schematic}
\end{center}
\end{figure}
To increase effective spectral resolution we have instead decided to take a
more direct approach. For simplicity we will here describe the method for a
two-particle interaction, and disregard
all factors converting code units to physical units. For example, the weight
of a pseudoparticle is proportional to the number of physical particles in
the
pseudoparticle. Although, these prefactors all represent trivial conversions of units,
they must be taken into account in the real code.

Consider a single cell containing a single pseudoparticle (red) with weight $w_t=n$ and a single pseudoparticle (purple) with weight
$w_i=m$, where
$n>m$ (see Fig.~\ref{fig:splitting_schematic}). We first select the red particle
as the \emph{target}, since $n>m$, and the purple as the \emph{incident}
particle. We then transform the four velocity of the incident particle to the rest
frame of the target particle, and calculate the total cross section
$\sigma_t$ of the interaction. Conceptually we consider the process as a single incident particle
approaching a slab of the target particle. The number density of target
particles in the slab can be calculated from the weight $w_t$ as
$\rho_t = w_t/\Delta V$, where
$\Delta V = \Delta x \Delta y \Delta z$ is the volume of a single cell. Given the number density
the probability that a single incident particle is scattered
\emph{per unit length} is
\begin{equation}
P_l = \rho_t \sigma_t = \frac{w_t \sigma_t}{\Delta V}
\end{equation}
During a time step $\Delta t$ the incident particle travels
$\Delta l =v_{i} \Delta t$,
and the probability that a single incident particle is scattered then becomes
\begin{align}\nonumber
P_S &= 1 - \exp\left[ - P_l \Delta l\right] \\ \label{eq:scat}
    &= 1 - \exp\left[ - \frac{w_t \sigma_t v_i \Delta t}{\Delta V}\right]
\end{align}
The weight of the incident pseudoparticle is $w_i=m$. Pseudoparticles
represent an ensemble of particles. Therefore
$P_S$ is the fraction of incident particles that are scattered on the
target. To model the process we create two new particles with weight
$w_{new} = w_i P_S = k$. Given the detailed interaction, we can calculate
the theoretical angular distribution of scattered particles in accordance with the
differential scattering cross section. Drawing from this distribution we
find the momentum and energy of the new scattered particles. The
weights of the target and incident particles are decreased to $w_t=n-k$ and
$w_i=m-k$ respectively (see Fig.~\ref{fig:splitting_schematic}).

Our method faithfully represents the actual physics even for small
cross sections. However, if all the particles are allowed to
interact, the number of particles in the box will increase at least
proportionally to the total number of particles squared. This is
potentially a computational run away. Normally we will have on the
order of up to $100$ particles per species per cell, but to be
computationally efficient we only calculate interactions for a
subset of the particles in a cell. This subset is chosen at random
according to an arbitrary distribution we are free to select. If the
probability that two particles are selected for scattering in a
given timestep is $Q$ then the travelling length $\Delta l$ simply
has to be adjusted as $\Delta l/Q$. If this arbitrary distribution
is chosen cleverly, the particles with the largest cross section are
actually the ones selected most often for scattering, and everything
ends up as a balanced manner: We only calculate the full cross
section and scattering as often as needed, and the computational
load that is given to a certain particle is proportional to the
probability of that particle to scatter. We rely on the merging of
particles as described above to avoid the copious production of
pseudoparticles. Every time the number of pseudoparticles in a given
cell crosses a threshold, pseudoparticles are merged and this way
the computational load per cell is kept within a given range.

\subsection{Neutron decay}
Free neutrons not bound in a nucleus will decay with a
half-life a little longer than ten minutes. The neutron
decays into an electron and a proton and an electron antineutrino
to satisfy lepton number conservation
\begin{equation}
n \to p + e^{-} + \bar{\nu}_e
\end{equation}
The rest mass difference of the process (0.78 MeV) goes into kinetic energy
of the proton, electron and neutrino. Let the neutron lifetime be $\tau$ in
code units. If $\tau$ is comparable to or less than a typical timestep, then
practically all neutrons decay in one iteration, and it is irrelevant to
include them. If $\tau$ is much larger than the total runtime, the
neutron can be considered a stable particle (unless the
neutron density in the box is much larger than the proton-- or electron density). If instead $\tau \simeq \alpha
\Delta t$ where $\alpha \sim 100$, then we can select a fraction $f$ of the
pseudoparticle neutrons in each cell and let them decay. This is done in an
analogous manner to the generic scattering process described above in
section \ref{scat}. The weight of the selected neutron
is decreased with a factor
\begin{equation}
\exp\left[-\frac{f\Delta t}{\gamma \tau}\right]\,,
\end{equation}
where $\gamma$ is the Lorentz boost of the neutron pseudoparticle and $f$
is chosen to give reasonable values for the decrease in the weight. At the
same time a pair of electron
and proton pseudoparticles is created with the same weight.
The generated particles share the excess mass of the process (where the
neutrino is neglected for now, but could be included in the future).
The momenta are selected to give an isotropic distribution in the rest frame of the
decaying neutron.

\subsection{Compton scattering}
Here we briefly describe a specific physical scattering mechanism,
which have already been implemented in the code, namely Compton
scattering.

\begin{figure}[htb]
\begin{center}
\epsfig{figure=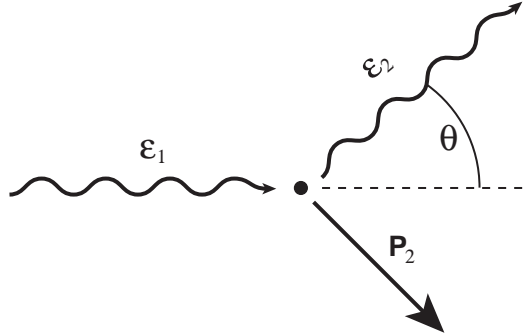,width=.5\textwidth}
\caption[Schematic picture of Compton scattering] {Schematic view
of the Compton scattering process.
Impinging on the electron, an incoming photon with energy
$\epsilon_1$ is scattered into the angle $\theta$ with
energy $\epsilon_2$. In the initial rest-frame of the electron, the
electron will be recoiled to conserve energy and momentum.}
\label{fig:compton_schematic}
\end{center}
\end{figure}

Compton scattering is the relativistic generalization of the
classical Thompson scattering process, where a low energy photon
scatters of on a free electron. In the rest frame of the electron,
the photon changes direction and loses energy to the electron, which
is set in motion. The cross section for Thompson scattering is
\cite{bib:rybicki}
\begin{equation}
\sigma_T=\frac{8\pi}{3}r_0^2
\end{equation}
where $r_0\equiv e^2/(m c^2)$ is called the {\it classical electron
radius}. The Thompson scattering approximation is valid as long as
the photon energy is much lower than the electron rest mass $h\nu\ll
m_ec^2$ and the scattering can be regarded as elastic. For photon
energies comparable to, or larger than, the electron rest mass,
recoil effects must be taken into account.
Measured in the electron rest frame we define $\epsilon_1$ as the
photon energy before the scattering,
$\epsilon_2$ as the photon energy after the scattering and $\theta$
the photon scattering angle (\ref{fig:compton_schematic}).
By conservation of energy and momentum one can show
(e.g. \cite{bib:rybicki}) that
\begin{equation}\label{eq:comptonenergy}
\epsilon_2=\frac{\epsilon_1}{1+\frac{\epsilon_1}{m_e c^2}
               (1-\cos\theta)}
\end{equation}
The differential cross section as a function of scattering angle is
given by the Klein-Nishina formula
\cite{bib:klein1929,bib:heitler1954}
\begin{equation}\label{eq:kn}
\frac{d\sigma_C}{d\Omega}=\frac{r_0^2}{2}
                          \frac{\epsilon_2^2}{\epsilon_1^2}
              \left(\frac{\epsilon_1}{\epsilon_2}
+\frac{\epsilon_2}{\epsilon_1}-\sin^2\theta\right)
\end{equation}

The Klein-Nishina formula takes into account the relative intensity
of scattered radiation, it incorporates the recoil factor, (or
radiation pressure) and corrects for relativistic quantum mechanics.
The total cross section is then
\begin{equation}
\sigma_C=\sigma_T\frac{3}{4}\left[\frac{1+x}{x^3}
                 \left\{\frac{2x(1+x)}{1+2x}-\mathrm{ln}(1+2x)
\right\}+ \frac{1}{2x}\mathrm{ln}(1+2x)-\frac{1+3x}{(1+2x)^2}\right]
\end{equation}
where $x\equiv h\nu/(mc^2)$.

\section{Preliminary results}\label{sec:NGPresults}
To test the new code and it capabilities in regard to the inclusion
of collisions, we have implemented and tested a simple scenario
involving Compton scattering.
\begin{figure}[htb]
\begin{center}
\epsfig{figure=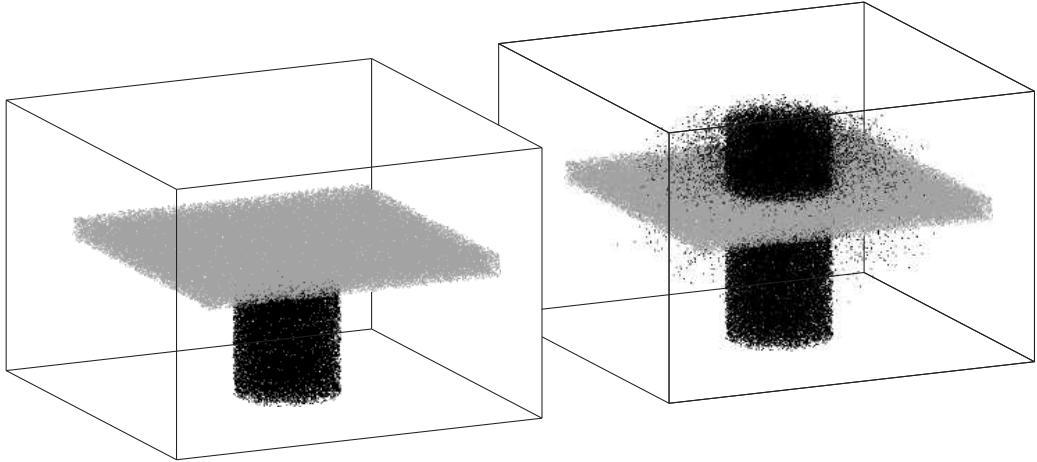,width=1.\textwidth}
\caption[3D
scatter plot of a photon beam passing through a pair plasma] {3D
scatter plot of a photon beam ({\it black}) passing through a cold
pair plasma ({\it gray}). Left panel show initial setup where a
photon beam is injected in the upward direction. Right panel shows
how photons are scattered on the electron-positron pairs}
\label{fig:compt_scatterplot}
\end{center}
\end{figure}

In the test setup, we place a thin layer of cold electron-positron
pair plasma in the computational box. From the boundary, we inject a
monochromatic beam of photons all travelling perpendicular to the
pair-layer (Fig. \ref{fig:compt_scatterplot} left panel). As the
beam passes through the plasma layer, photons are scattered (Fig.
\ref{fig:compt_scatterplot} left panel).

\begin{figure}[htb]
\begin{center}
\epsfig{figure=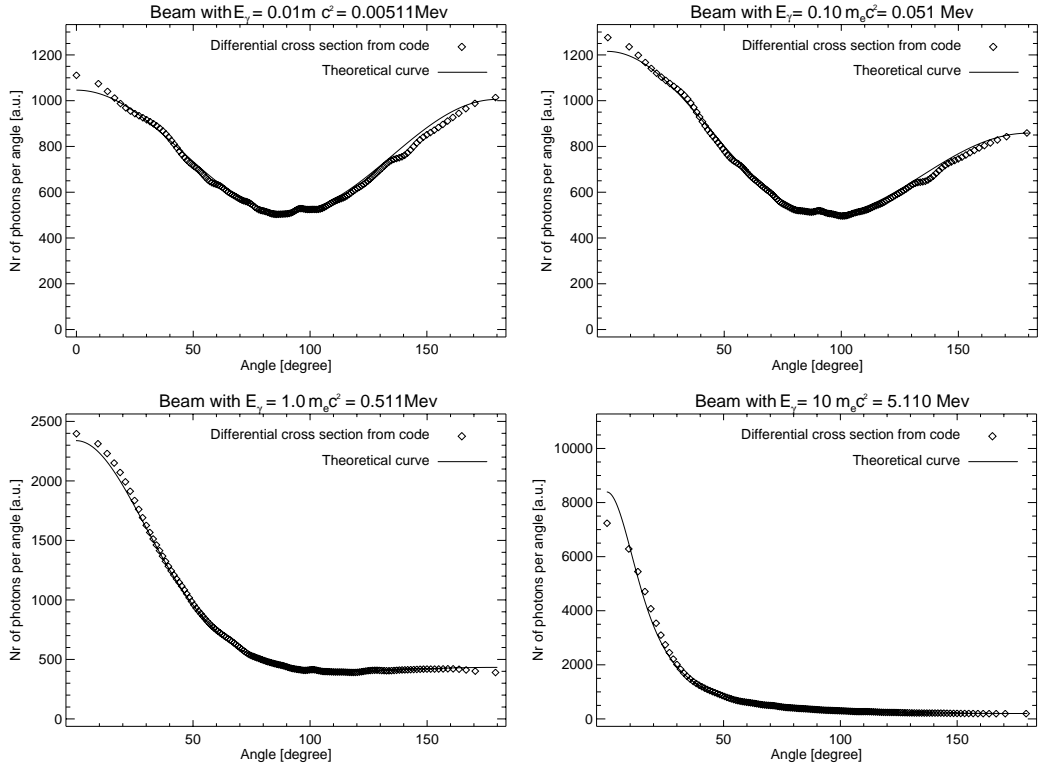,width=1.\textwidth} \caption[The
theoretical Compton scattering cross section compared to data] {The
theoretical Compton scattering differential cross section. We have
performed a test experiment with an incoming laser beam on a very
cold electron population. Over plotted the differential distribution
is the theoretical curve according to Eqs.~(\ref{eq:kn}) and
(\ref{eq:comptonenergy}). } \label{fig:compt_theory}
\end{center}
\end{figure}

For each scattered photon we sample the weight of the photon
and its direction (remembering that all particles are pseudoparticles
that represent whole groups of particles).
Fig. \ref{fig:compt_theory} shows the theoretical cross section as
function of scattering angle compared with the result from the simulations.
Four plots for different energies of the incoming photon
beam are shown. We find excellent agreement between the simulation results and
the theoretical predictions.

\section{Discussion}\label{sec:NGPdiscussion}
A next generation PIC code that includes many different kinds of
scattering process is under development. It will enable us to target
problems that reside in the grey zone between the MHD and
collisionless plasma domains. This domain covers many astrophysical
scenarios of great interest counting internal shocks in gamma-ray
bursts, solar flares and magnetic substorms, compact relativistic
objects, supernova remnants and many more.

The concept of splitting/merging particles and keeping individual
weights of each particle carry many important features. Variable
weights represent the true statistics of a scattering process in an
optimal way compared to the Monte Carlo approach. Also, for
MPI-parallelization it is crucial that the number of particles per
cell is kept more or less constant to ensure an optimal CPU
load-balancing. To localize calculations we are employing a sorting
algorithm that maintains neighbouring particles on the mesh as
neighbours in memory. This is not only good for parallelization, but
also makes all computations very cache efficient; a crucial
requirement on modern computer architectures.

To test the infrastructure of the new code we have implemented
Compton scattering as a simple scattering mechanism. The preliminary
results are very promising in form of excellent agreement with the
theoretical prediction. We note that a recent paper by
\cite{bib:moderski2005} provide an interesting test suite for
various kind of particle-photon interactions that can be tested in
the future. Merging particles has not been satisfactorily
implemented yet. Parallelization of code is still not there yet, and
is necessary to obtain the capability of performing truly
large-scale experiments. In summary: Work has still to be done
before we can start to investigate non-trivial astrophysical
scenarios, nevertheless solid progress has already been made

\vspace{2ex} This chapter has been written jointly by Christian
Hededal and Troels Haugb{\o}lle, reflecting the fact that the
development process of the next generation PIC code has been highly
team based. Essentially everybody has contributed time and effort to
every single source file of the code. It would not make sense to
write the chapter separately, essentially repeating each other and
reusing the same figures.

\chapter{Overall Discussion and Conclusions}\label{sec:concl}

The main source of information we have about gamma-ray bursts (GRBs)
is from multi-wavelength observations of GRB afterglows. The
radiation from GRB afterglows is generated in collisionless shocks
between the external plasma (ISM) and the GRB jet. To make the right
conclusions about burst parameters in these shocks, it is crucial
that we have a firm understanding of how the radiation is generated.
In the current working model, the radiation is believed to be
emitted from power-law distributed electrons in a magnetic field.
Following our lack of knowledge about the plasma-physical details,
it is a general approach to parameterize this ignorance with the
dimensionless parameters $\epsilon_B$, $\epsilon_e$ and $p$. As
usual, $\epsilon_B$ and $\epsilon_e$ express the fraction of the
total internal energy that is deposited in magnetic field and
electrons and $p$ is the slope of the supposedly electron power-law
momentum distribution function.

How the electrons are accelerated and what the origin and nature of
the magnetic field in the afterglow shock is, remain open questions.

The main goal behind the work presented in this thesis has been to
expand our knowledge about the microphysics of collisionless shocks.
Attacking the problem from first principles, I have used
particle-in-cell (PIC) simulations to investigate different aspects
of these shocks. The results may be categorized into three groups:
Magnetic fields in collisionless shocks, non-thermal electron
acceleration and radiation from GRB afterglow shocks. I emphasize
that these three components are highly interconnected and should be
seen as pieces of the same puzzle.

\section{Magnetic fields in collisionless shocks}
Since the main focus is on GRB afterglows, I have investigated only
electron-proton plasma shocks. The three-dimensional plasma PIC
simulations illustrate from first principles a number of fundamental
properties of collisionless plasma shocks.

\begin{itemize}
\item In unmagnetized or weakly magnetized relativistic shocks (below
milligauss), the Weibel two-stream instability will unavoidably
create a magnetic field that amounts to $\sim10\%$ of the
equipartition value ($\epsilon_B\sim0.1$) but varies greatly through
the shock.

\item The nature of the magnetic field is predominantly transverse
to the plasma flow, but a parallel magnetic component is also
present. The parallel magnetic field strength is roughly $10\%$ of
the transverse components. An electric field is also present with an
energy density that amounts to roughly $10\%$ of the magnetic field
energy density.

\item The dominating instability is highly non-linear. The electrons are
rapidly thermalized while the ions form current channels. These
current channels are the main source of the magnetic field.

\item  Being Debye shielded by the hot electrons, the ion current channels are
relatively stable and can penetrate deeply into the shock. Earlier
thinking and concerns were that the generated magnetic field can
only survive over an electron skin depth \citep{bib:gruzinov2001}.
In the simulations presented here, the generated magnetic field is
sustained for several hundred ion skin depths.

\item The structure of the magnetic field is highly patched. The
transverse coalescence scale is comparable to the ion skin depth. A
spatial Fourier decomposition of the magnetic field shows that the
structures follow a power-law distribution with negative slope.

\item I find that in two-dimensional simulations of the Weibel
two-stream instability, both the evolution of the plasma density
profile and the generation of electromagnetic field are in agreement
with three-dimensional simulations.

\item With two-dimensional simulations, it has been able to capture
a full relativistic, collisionless shock. The shock consists of
three segments. 1) An upstream foreshock with great anisotropy,
consisting of instreaming ISM plasma and upscattered shock
particles. In this region, a strong magnetic field generation is
taking place with $\epsilon_B=5-10\%$. 2) A dense thermalization
region where the electrons and ions are close to equipartition
($\epsilon_e=50-80\%$). In this region, the magnetic field is
relatively weak ($\epsilon_B\simeq1\%$) but still strong enough to
account for most GRB afterglow radiation estimates. 3) A hot
downstream region where the magnetic field is of the order percents
of equipartition. This region is clumpy, with large structures that
are of the order $1-20$ ion skin depths in size.

\item If the ejecta is already strongly magnetized with a transverse magnetic field
(e.g.\ carried from the progenitor), the resulting dynamics is very
complex: Particle deflection, charge separation and a mixture of the
electromagnetic Weibel two-stream instability and the electrostatic
Buneman instability make the resulting field very complicated with
local reversal of the ambient field. Electrons are
 accelerated to a non-thermal distribution. In the case of strong
parallel magnetic field, all instabilities are effectively damped.

\item For an ambient magnetic field strength comparable
to the ISM- field, the Weibel instability grows unhindered.
\end{itemize}

\section{Non-thermal electron acceleration}
There has been a general acceptance of Fermi acceleration as the
mechanism that produces the desired electron distribution function.
Nevertheless, this has never been proven in self-consistent
numerical simulations. The idea of  Fermi acceleration in
collisionless shocks is currently facing several problems (see
Chapter \ref{sec:intro} for details):
\begin{itemize}
\item It is assumed that the particles scatter on
electromagnetic waves, but the model does not self-consistently
account for the generation of these waves nor for the back-reaction
that the high-energy particle distribution has on the
electromagnetic field.
\item Relativistic Fermi acceleration requires the downstream
magnetic field to be strongly turbulent on scales comparable to the
gyro-resonant scale (e.g., \citec{bib:ostrowski2002}). This is in
contrast with the standard synchrotron radiation scenario for GRB
afterglows where it is assumed that the magnetic field is nearly
constant on scales much larger than the gyro-radius (see however
\cite{bib:Medvedev_jitter} and Chapter \ref{sec:rad}).
\item Fermi acceleration requires pre-acceleration to above a certain threshold. How
this pre-acceleration works and how large a fraction of the
particles that participate in the non-thermal tail remains
unexplained. In the most well studied mildly relativistic plasma
shock (in the Crab nebula), observations are contradicting what is
normally assumed about this threshold \citep{bib:eichler2005}.
Furthermore, Chandra images of the close by Supernova remnant SN
1006 fail to detect an x-ray halo in front of the shock, which is to
be expected if the standard Fermi diffusive shock acceleration
theory is correct. One expects an X-ray halo around the shock
because higher energy electrons are expected to diffuse further
ahead of the shock \citep{bib:long2003}.
\end{itemize}
I do not claim that Fermi acceleration cannot take place, but urge
the need for a self-consistent treatment of the problem when
computational resources allow this.

I have presented self-consistent, three-dimensional PIC simulations
that have revealed a new non-thermal electron acceleration mechanism
that differs substantially from Fermi acceleration. The acceleration
is a natural consequence of the properties of relativistic
collisionless shocks. Acceleration of electrons is directly related
to the formation of ion current channels that are generated in the
non-linear stage of the Weibel two-stream instability in the shock
transition zone. This links particle acceleration closely together
with magnetic field generation in collisionless shocks.

The resulting electron spectrum consists of a thermal component and
a non-thermal component at high energies. This is in the line with
arguments by \cite{bib:ryde2005} and \cite{bib:rees2005}. In
simulations of mildly relativistic shocks ($\Gamma=3$), the
non-thermal component vanishes in the thermal pool. Computer
experiment with a bulk flow with $\Gamma=15$ results in a power-law
slope $p=2.7$ for the electron distribution function.

\section{Radiation from GRB afterglow shocks}
I have developed a tool that enables us to obtain radiation spectra
directly from PIC simulations. The tool has been thoroughly tested
and successfully reproduces spectra from synchrotron radiation,
bremsstrahlung and undulator radiation from small-angle deflections.
The tool has been used to investigate the properties of
three-dimensional jitter radiation in magnetic fields with different
turbulent configurations. By tracing ensembles of monoenergetic,
isotropic distributions of electrons in a random magnetic field
(whose power spectrum follows a power-law in the Fourier domain
$P_B(k)\propto k^{\mu}$) I have computed and studied the resulting
spectrum. I have focused on the weak field limit (where the
deflection angle is less than the beaming cone angle, $K<1$). For
all values of $\mu$, the radiation spectrum has a flat low-energy
part (like bremsstrahlung). For $\mu<0$, the high-energy part of the
spectrum follows a power-law with a slope $\alpha\simeq\mu-1$
independent of the electron energy. For $\mu>0$, the flat spectrum
continues to higher frequencies than for $\mu<0$. The spectrum has a
hard cut-off for high frequencies.

 For all values
of $\mu$ and $K<1$, the spectrum shifts in frequency with $\gamma^2$
and scales with the amplitude of the magnetic field squared $B^2$,
in good agreement with Larmor's formula for radiated power. In the
limit of large deflections ($K>1$) the spectrum eventually converges
through the wiggler spectrum to the ordinary synchrotron spectrum.

I have furthermore examined the radiation from electrons moving in
1) a magnetic field generated by the Weibel two-stream magnetic
field and 2) a magnetic field that has the same average energy
density and spectral power spectrum as the two-stream generated
field. I find that even though the two spectra peak at the same
frequency, there is a large difference above the peak-frequency. The
reason is that in the magnetic field from the PIC simulations there
exists a phase correlations resulting in higher magnetic peak values
but smaller filling factor. The peak frequency of
synchrotron/wiggler radiation scales linearly with the maximum
magnetic field strength. Furthermore, from other PIC simulations we
know that the high-energy electrons are found near the peaks in
magnetic field.

 In simulations of a collisionless shock that propagates with
$\Gamma=15$ through the interstellar medium, the resulting radiation
spectrum  peaks  around $10^{12}$Hz. Above this frequency, the
spectrum follows a power-law
 $F\propto\nu^{-\beta}$, with $\beta=0.7$. Below the peak frequency, the
spectrum follows a power law  $F\propto\nu^{\alpha}$ with
$\alpha\simeq2/3$. This is steeper than the standard synchrotron
value of $1/3$ and more compatible with observations. Both the slope
and the peak is consistent with observations (e.g.\
\citec{bib:panaitescu2001} who finds $\beta=0.67\pm0.04$, a peak at
$3\times10^{11}$Hz and $\Gamma>10$ for the afterglow of GRB 000301c
after five days).

I stress the following interesting point:
 If one hides
all the real physical details of the magnetic field in the
dimensionless parameters $\epsilon_B$, $\epsilon_e$, and $p$ the
conclusion from standard synchrotron radiation theory would be that
the slope of the electron distribution $N(\gamma)\propto\gamma^{-p}$
is found by solving $-\beta=-(p-1)/2=-0.70\ \to\ p=2.4$ (which is
consistent with standard analysis of the GRB 000301c afterglow,
\citec{bib:panaitescu2001}). This would appear to be consistent with
Fermi acceleration and the supposedly universal $p\simeq2.2\pm0.2$:
Case solved! But a closer look reveals that the acceleration
mechanism is not Fermi acceleration. The electrons are
instantaneously accelerated and decelerated in the highly
complicated electric and magnetic field near the ion current
channels. Strong radiation is produced in this process. The electron
distribution behind the radiation is a mixture of a thermal
component for low energies and a power-law component with $p=2.7$
for high energies \citep{bib:hededal2004} (see also Chapter
\ref{sec:acc}). The paths of the electrons are more random rather
than circular and the electron distribution function varies with
shock depth, as does the magnetic topology and strength. I repeat
and emphasize that the electrons that produced the spectrum in Fig.
\ref{fig:spec_g15} are not Fermi accelerated but a natural
consequence of the Weibel two-stream instability.

\section{Future work}
In this thesis I have shown, from first prinicples, that from the
collision of two plasma shells that are initially cold and
non-magnetized, a strong small scale magnetic is generated
($\epsilon_B\sim0.01-0.1$), particles are accelerated to non-thermal
distributions and the emitted radiation is quite consistent with
observations (comparing Fig. \ref{fig:spec_g15} and data from GRB
000301c, \citec{bib:panaitescu2001}). The next step is to determine
the polarization in the synthetic spectra from the PIC simulations.
Even though time did not allow this in this thesis, it will not be
difficult with the tools that have already been developed. With the
inclusion of polarization it will be possible to investigate how the
radiation spectrum and polarization change as functions of the
viewing angle relative to the jet propagation direction.

In the near future, computational resources will reach a level where
one can resolve full three-dimensional collisionless shocks. This
will enable us to make composite spectra from different simulations
with bulk Lorentz factor varying with viewing angle, and in turn
allow us to test many interesting aspects with regard to the jet
structure and polarization predictions from shock-generated
electromagnetic fields.

Finally, the new generation PIC code that is under development
allows more of physics to be included in the simulations. Some of
the main features are: High-energy photons are treated as particles,
several scattering processes are included (e.g.\ Coulomb and Compton
scattering), parallelization with MPI, pseudo particles with
individual weights, etc. With this code it will become possible to
study also the prompt GRB phase, where Compton scattering and pair
processes are likely to be important.

\pagebreak   \\
\pagebreak   \\
\begin{appendix}
\chapter{The Mean Free Path in a Blast Wave}\label{app:cross_section}
In this appendix I make an estimate of the mean free path for
Coulomb collisions of a relativistic electron with momentum $\gamma
m_e \vect{v}_e$ on ions in a plasma with density $n$.

First we examine the Coulomb collision between an electron and an
ion. Without loss of generality, we may assume that the electron is
travelling in the $yz$-plane along the $z$-axis and that an ion is
positioned in $(x,y,z)=(0,-b,0)$. The ion is surrounded by an
electric field. Because of Lorentz contraction and symmetry
arguments we can assume that the electron will only be affected by
the component that is transverse to $\vect{v}_e$, namely $E_y$. In
the reference frame of the electron, this component is given by
\citep{bib:rybicki}:
\begin{equation}
E_y=\frac{1}{4\pi\epsilon_0}\frac{q\gamma
b}{(\gamma^2v_e^2t^2+b^2)^{3/2}}.
\end{equation}
Here $t$ is the time, centered so that the electron is in $(0,0,0)$
at $t=0$. The force felt by the electron is $\vect{F}=q
E_y\hat{\vect{e}}_y$. The change in the electron's momentum $\delta
p$ is found as:
\begin{equation}
\delta p=\int F_y dt=\int q E_y dt
=\frac{1}{4\pi\epsilon_0}\frac{q^2 \gamma
t}{b\sqrt{\gamma^2v_e^2t^2+b^2}}\label{eq:appdp}.
\end{equation}
The pulse from the ion is felt by the electron in the short time
interval $T\simeq b/(\gamma v_v)$. Inserting this into Eq.\
\ref{eq:appdp} we find
\begin{equation}
\delta p=\frac{1}{4\pi\epsilon_0}\frac{q^2}{\sqrt{2}v_eb}.
\end{equation}
We are interested in collisions that alter the impinging electrons
momentum significantly with $\delta p\simeq\gamma m_e v_e$, and we
can thus find the distance $b_c$ for such a collision:
\begin{eqnarray}
\delta p\simeq\gamma m_e v_e\ \ \ \ \Rightarrow\nonumber\\
b_c\simeq\frac{1}{4\pi\epsilon_0}\frac{q^2}{\sqrt{2}\gamma m_e
v_e^2}.
\end{eqnarray}
Thus, the cross section for the collision is
\begin{equation}
\sigma_c=\pi
b_c^2=\frac{q^4}{16\pi\epsilon_0^2\sqrt{2}\gamma^2m_e^2v_e^4}.
\end{equation}

The collision frequency is $\nu_c=n \sigma_c v_e$ and from this we
find the mean free path for a full collision:
\begin{equation}
\lambda_c\equiv\frac{v_e}{\nu_c}=\frac{1}{n\sigma_c}=\frac{16\pi\epsilon_0^2\sqrt{2}\gamma^2m_e^2v_e^4}{n
q^4}.
\end{equation}

In reality, the mean free path somewhat shorter because of
accumulation of small angle deflections. We can correct for this by
introducing a correction factor $1/\ln \Lambda$, which is of the
order of 0.1 \citep{bib:spitzer}.

For a electron in a blast wave that is expanding with Lorentz factor
$\gamma\simeq\Gamma=10$, into an interstellar medium with density
$n\simeq10^6\mathrm{m}^{-3}$, the mean free path for a collision is
larger than $10^{23}$m. Comparing this number with the typical size
of a GRB blast wave $\sim10^{14}$m we conclude that it is reasonable
to neglect collisions between the ejecta and ISM, and that the shock
between the two is to be regarded as a {\it collisionless shock}.

\chapter{Integration of $\frac{dP}{d\Omega}$}\label{app:dpdomega_integration}
\begin{figure}[!th]
\begin{center}
\epsfig{figure=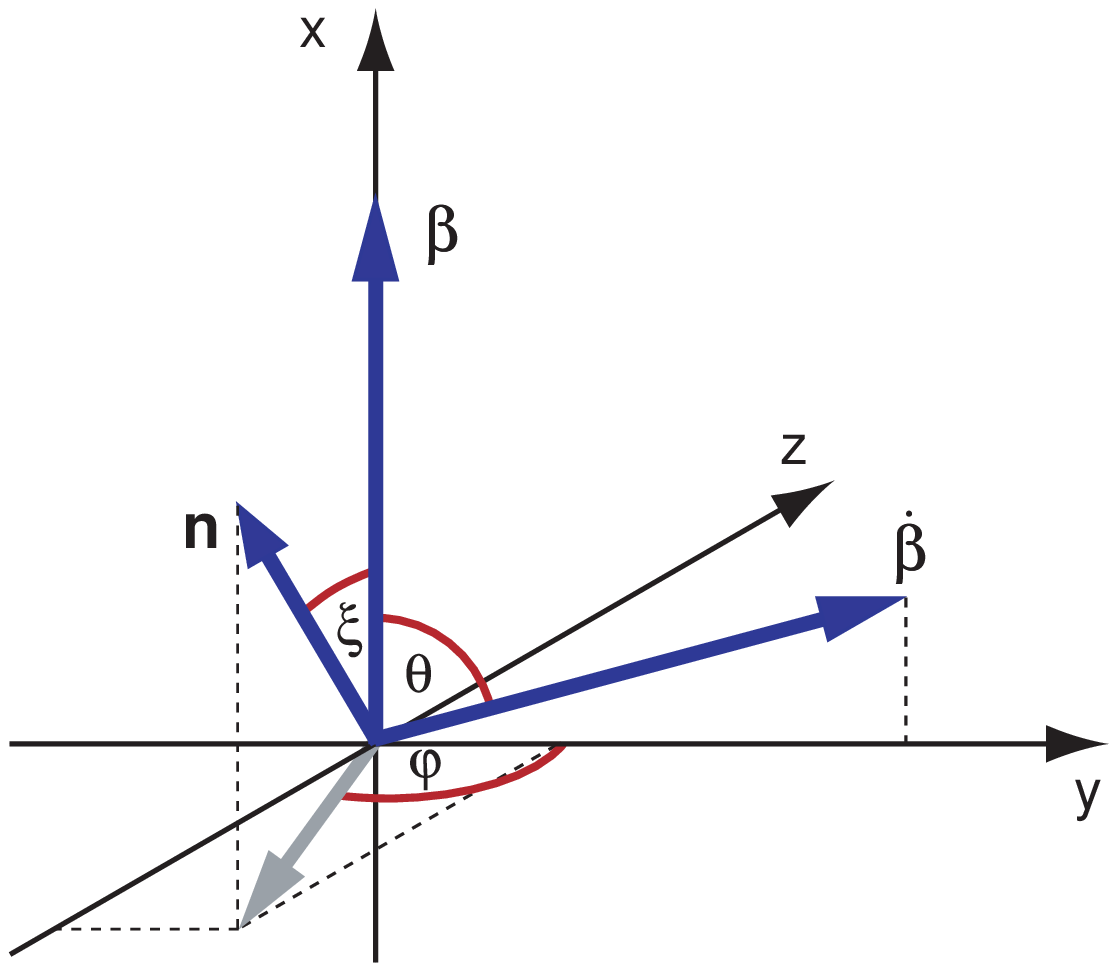,width=0.5\textwidth}
\caption[Coordinate system for deriving $\frac{dP}{d\Omega}$] {We define a
coordinate system with the $x$-axis along the velocity vector
\vect{\beta}. For the integration over the unit solid angle
$d\Omega$ around the unit vector we use the polar coordinates
$(\xi,\varphi)$. The particle is accelerated at some angle $\theta$
to the velocity vector.} \label{fig:app_integration}
\end{center}
\end{figure}

We wish to integrate Eq.\ \ref{eq:P_rad-solid} over all directions
\vect{n} through the solid angle $d\Omega$.
\begin{eqnarray}
\df{P_{rad}}{\Omega}=\frac{\mu_0 q^2
c}{16\pi^2}\frac{\left[\vect{n}\times
\left\{\left(\vect{n}-\vect{\beta}\right)\times\vectdot{\beta}
\right\}\right]^2}{\left(1-\vect{n\cdot\beta}\right)^5}.
\end{eqnarray}
Without loss of generalization we define a Cartesian coordinate
system with the $x$-axis along the particles velocity vector
\vect{\beta} and the acceleration vector \vectdot{\beta} in the
$x$-$y$ plane (Fig.\ \ref{fig:app_integration}).

\noindent Integration over the unit solid angle $d\Omega$ then
yields:
\begin{eqnarray}
P_{rad}=\int\df{P_{rad}}{\Omega}d\Omega=\int_0^{2\pi}\int_0^\pi\df{P_{rad}}{\Omega}\sin\xi
d\xi d\varphi\label{eq:app_p_integrated}.
\end{eqnarray}
With the definition of the polar angles in Fig.\
\ref{fig:app_integration} we can writte out \vect{\beta},
\vectdot{\beta} and \vect{n}:
\begin{equation}
\vect{\beta}=\beta\left(\begin{array}{c} 1\\0\\0\end{array}\right),\
\ \ \ \
\vectdot{\beta}=\dot\beta\left(\begin{array}{c} \cos\theta\\
\sin\theta\\0\end{array}\right),\ \ \ \ \
\vect{n}=\left(\begin{array}{c} \cos\xi\\
\sin\xi\cos\varphi\\ \sin\xi\sin\varphi\end{array}\right)
\end{equation}
\begin{eqnarray}
\df{P_{rad}}{\Omega}=\frac{\mu_0 q^2 c}{16\pi^2}&&\Big\{
\left|\dot{\beta}\sin\xi\left[(\beta-\cos\xi)\cos\varphi\sin\theta+\cos\theta\sin\xi\right]\right|^2\nonumber\\
&&+
\left|\dot{\beta}\sin\xi(\cos\theta\cos\xi+\cos\varphi\sin\theta\sin\xi)\sin\varphi\right|^2\nonumber\\
&&+ \Big|\dot{\beta}\cos\xi((\beta
-\cos\xi)\sin\theta+\cos\theta\cos\varphi\sin\xi)\nonumber\\
&&-\dot{\beta}\sin\theta\sin^2\xi\sin^2\varphi\Big|^2
\Big\}\nonumber\\
&&/(1-|\beta|)^5\label{eq:dpdOmega_expand}.
\end{eqnarray}
Inserting Eq.\ \ref{eq:dpdOmega_expand} into Eq.\
\ref{eq:app_p_integrated} we perform the integration in two steps:
\begin{eqnarray}
P_{rad}&=&\int_0^{2\pi}\int_0^\pi\df{P_{rad}}{\Omega}\sin\xi d\xi
d\varphi\nonumber\\
&=&\int_0^\pi\dot{\beta}^2\pi\big\{(11+6\beta^2+2\beta^2\cos(2\xi)-2\cos(2\theta)\nonumber\\
&&(1+3\beta^2+(3+\beta^2)\cos(2\xi)) -32\beta\cos\xi\sin^2\theta)\nonumber\\
&&\sin\xi-\sin(3\xi)\big\}/(16(1-\beta\cos\xi)^5)\sin\xi d\xi\nonumber\\
&=&\frac{\mu_0q^2c}{12\pi}\gamma^6\dot{\beta}^2\left(2-\beta^2(1-\cos(2\theta)\right)\nonumber\\
&=&\frac{\mu_0q^2c}{6\pi}\gamma^6\dot{\beta}^2\left(1-\beta^2\sin^2\theta\right).
\end{eqnarray}

\chapter{Spectral distribution of Synchrotron Radiation}\label{app:spectrum}
Here we derive the spectral distribution of synchrotron radiation.
We follow the standard derivation found in many textbooks (e.g.\
\cite{bib:rybicki}), but keep the full angular dependency. We start
with the expression for the radiated energy pr. unit frequency pr.
unit solid angle (Eq.\ \ref{eq:retard_fourier})
\begin{equation}
\frac{d^2W}{d\omega d\Omega}=\frac{q^2}{16\pi^3\epsilon_0
c}\left|\int_{-\infty}^\infty
{\frac{\vect{n}\times[(\vect{n}-\vect{\beta})\times\vectdot{\beta}]}{(1-\vect{\beta\cdot
n})^2} \ e^{i\omega(t'-\vect{n\cdot r}(t')/c)}dt'} \right|^2.
\end{equation}
Integration by parts and using the relation
\begin{equation}
\frac{\vect{n}\times[(\vect{n}-\vect{\beta})\times\vectdot{\beta}]}{(1-\vect{\beta\cdot
n})^2}
=\frac{d}{dt}\left[\frac{\vect{n}\times(\vect{n}\times\vect{\beta})}{1-\vect{\beta\cdot
n}}\right],
\end{equation}
we can rewrite into the expression
\begin{equation}
\frac{d^2W}{d\omega d\Omega}=\frac{q^2\omega^2}{16\pi^3\epsilon_0
c}\left|\int_{-\infty}^\infty
\vect{n}\times(\vect{n}\times\vect{\beta})
e^{i\omega(t'-\vect{n\cdot r}(t')/c)}dt \right|^2.
\end{equation}
We now expand the terms in the integrand
\begin{eqnarray}
t'-\vect{n\cdot r}(t')/c)&=&t'-\frac{r_L}{c}\cos\theta\sin(\beta c t'/a)\nonumber\\
&\simeq&(1-\beta\cos\theta)t'+\frac{\beta^3c^2\cos\theta}{6r_L^2}{t'}^3\nonumber\\
&=&\frac{1}{2}\left(\theta_\beta^2t'+\frac{\beta^3c^2\cos\theta}{3r_L^2}{t'}^3\right),
\end{eqnarray}
where $\theta_\beta^2\equiv2(1-\beta\cos\theta)$.
\begin{figure}[!th]
\begin{center}
\epsfig{figure=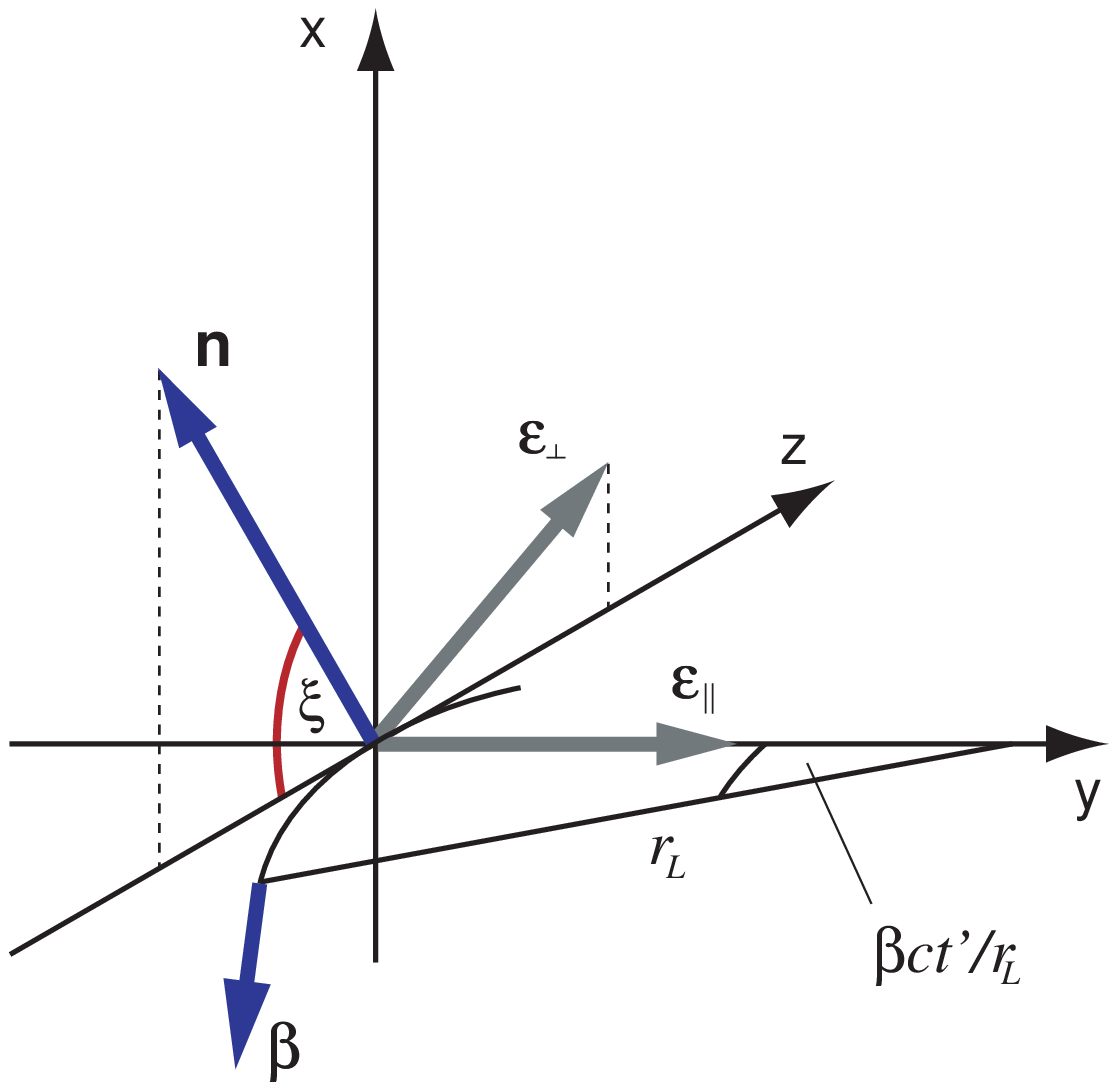,width=0.5\textwidth}
\caption[Coordinate system for deriving $\frac{d^2W}{d\omega
d\Omega}$] {We define a coordinate system with the $x$-axis along
the velocity vector \vect{\beta}. For the integration over the unit
solid angle $d\Omega$ around the unit vector we use the polar
coordinates $(\theta,\varphi)$. The particle is accelerated at some
angle $\theta$ to the velocity vector.} \label{fig:app_integration1}
\end{center}
\end{figure}
\begin{eqnarray}
\vect{n}\times(\vect{n}\times\vect{\beta})&=&
\beta\left[\sin\theta\cos(\beta c t'/r_L)\vect{\epsilon}_\perp-\sin(\beta c t'/r_L)\vect{\epsilon}_\parallel\right]\nonumber\\
&\simeq&\beta\sin\theta\vect{\epsilon}_\perp-\beta^2 c t'/r_L\vect{\epsilon}_\parallel
\end{eqnarray}

The radiation has to polarization components
\begin{eqnarray}
\frac{d^2W}{d\omega d\Omega}=\frac{d^2W_\perp}{d\omega d\Omega}+\frac{d^2W_\parallel}{d\omega d\Omega}
\end{eqnarray}
\begin{eqnarray}
\frac{d^2W_\perp}{d\omega d\Omega}&=&\frac{q^2\omega^2}{16\pi^3\epsilon_0 c}\left|\int_{-\infty}^\infty
\beta\sin\theta
\exp\left[\frac{i\omega}{2}(\theta_\beta^2t'+\frac{\beta^3c^2\cos\theta}{3r_L^2}{t'}^3)\right]dt' \right|^2\\
\frac{d^2W_\parallel}{d\omega
d\Omega}&=&\frac{q^2\omega^2}{16\pi^3\epsilon_0
c}\left|\int_{-\infty}^\infty \frac{\beta^2c t'}{a}
\exp\left[\frac{i\omega}{2}(\theta_\beta^2t'+\frac{\beta^3c^2\cos\theta}{3r_L^2}{t'}^3)\right]dt'
\right|^2.
\end{eqnarray}
We now define
\begin{equation}
K=\frac{q^2\omega^2}{16\pi^3\epsilon_0 c}\ ,\ \ \ \chi\equiv\frac{\omega r_L \theta_\beta^3}{3 c}\ ,\ \ \ y\equiv\frac{c t'}{r_L \theta_\beta}
\end{equation}
\begin{eqnarray}
\frac{d^2W_\perp}{d\omega d\Omega}&=&K\sin^2\theta\left(\frac{r_L \theta_\beta}{c}\right)^2
\left|\int_{-\infty}^\infty
\exp\left[\frac{3}{2}i\chi(y+\frac{1}{3}\cos\theta\beta^3y^3) \right]dy \right|^2\label{eq:dwperp_temp}\\
\frac{d^2W_\parallel}{d\omega d\Omega}&=&K\left(\frac{r_L
\theta_\beta^2\beta^2}{c}\right)^2 \left|\int_{-\infty}^\infty
y\exp\left[\frac{3}{2}i\chi(y+\frac{1}{3}\cos\theta\beta^3y^3)
\right]dy \right|^2\label{eq:dwpar_temp}.
\end{eqnarray}
We now focus solely on the integrals. We hide all the coefficients
by introducing the variables $x$ and $z$.
\begin{eqnarray}
x\equiv\frac{3}{2}\chi\ ,\ \ \
a\equiv\frac{1}{2}\cos\theta\beta^3\chi.
\end{eqnarray}
First, we expand the exponential functions in a geometric
representation,
\begin{eqnarray}
\exp\left[i(x y+ a y^3)\right]&=&\cos(x y+ a y^3)+i\sin(x y+ a y^3).
\end{eqnarray}
The solutions to the integrals of Eq.\ \ref{eq:dwperp_temp} and Eq.\
\ref{eq:dwpar_temp} are expressed by \cite{bib:abramowitz}:
\begin{eqnarray}
\int_{-\infty}^\infty\exp\left[i(x y+ a y^3)\right]dy&=&\int_{-\infty}^\infty\cos(x y + a y^3)dy\nonumber\\
&=&2(3 a)^{-1/3}\pi\mathrm{Ai}\left[(3a)^{-1/3}x\right]\label{eq:ai1}\\
\int_{-\infty}^\infty y\exp\left[i(x y+ a y^3)\right]dy&=&\int_{-\infty}^\infty y\sin(x y + a y^3)dy\nonumber\\
\frac{d}{dx}\int_{-\infty}^\infty\cos(x y + a y^3)dy&=&2(3
a)^{-2/3}\pi\mathrm{Ai}'\left[(3a)^{-1/3}x\right]\label{eq:ai2},
\end{eqnarray}
where $\mathrm{Ai}(z)$ is the Airy function. The Airy function and its derivative can be expressed in
terms of the modified Bessel function of the second kind:
\begin{eqnarray}
K_\frac{1}{3}(\zeta)&=&\pi\sqrt{\frac{3}{z}}\mathrm{Ai}(z)\\
K_\frac{2}{3}(\zeta)&=&-\pi\frac{\sqrt{3}}{z}\mathrm{Ai}'(z),
\end{eqnarray}
where $z\equiv(\tfrac{3}{2}\zeta)^{2/3}$. I our case, $z=(3 a)^{-1/3}x$ and by comparing these two expressions for $z$
and reinstating the definitions of $a$ and $x$ we find the relation:
\begin{equation}
\zeta=\frac{\chi}{\sqrt{\cos\theta\beta^3}}.
\end{equation}
Thus, the Airy function and its derivative in Eq.\ \ref{eq:ai1} and
Eq.\ \ref{eq:ai2} can be replaced with:
\begin{eqnarray}
\mathrm{Ai}\left[(3a)^{-1/3}x\right]&=&K_\frac{1}{3}\left(\chi/\sqrt{\cos\theta\beta^3}\right)\sqrt{\frac{z}{3\pi^2}}\\
\mathrm{Ai}'\left[(3a)^{-1/3}x\right]&=&-K_\frac{2}{3}\left(\chi/\sqrt{\cos\theta\beta^3}\right)\frac{z}{\sqrt{3\pi^2}}
\end{eqnarray}
with $z=(3 a)^{-1/3}x$. With these expressions we arrive at the
final result of Eq.\ \ref{eq:dwperp_temp} and Eq.\
\ref{eq:dwpar_temp}:
\begin{eqnarray}
\frac{d^2W_\perp}{d\omega d\Omega}&=&\frac{q^2\omega^2}{12\pi^3\epsilon_0 c}\left(\frac{r_L \theta_\beta\sin\theta}{c}\right)^2
\frac{\left|K_\frac{1}{3}\left(\chi/\sqrt{\cos\theta\beta^3}\right)\right|^2}{(\cos\theta\beta^3)}\\
\frac{d^2W_\parallel}{d\omega d\Omega}&=&\frac{q^2\omega^2}{12\pi^3\epsilon_0 c}\left(\frac{r_L \theta_\beta^2\beta^2}{c}\right)^2
\frac{\left|K_\frac{2}{3}\left(\chi/\sqrt{\cos\theta\beta^3}\right)\right|^2}{(\cos\theta\beta^3)^2}
\end{eqnarray}

\end{appendix}

\backmatter
\addcontentsline{toc}{chapter}{Bibliography}
\bibliographystyle{astron}
\bibliography{thesis}


\end{document}